\documentclass[12pt]{article}
\pdfoutput=1
\usepackage{tabu}
\usepackage{putex}
\usepackage{amssymb}
\usepackage{amsmath}
\usepackage{empheq}
\usepackage{epsf}%\usepackage{showlabels}
\usepackage{graphicx}
\usepackage{epstopdf}
\usepackage[utf8]{inputenc}
\usepackage{enumerate}
\usepackage{cite}
\usepackage{caption}
\usepackage{subcaption}
\usepackage{hyperref}
\usepackage{braket}
\usepackage{titlesec}
\usepackage{fixltx2e}
\usepackage{cleveref}
\hypersetup{breaklinks}
\usepackage[usenames,dvipsnames,svgnames,table]{xcolor}
\usepackage{comment}
\newlength\mytemplen
\newsavebox\mytempbox

\makeatletter
\newcommand\mybluebox{%
    \@ifnextchar[%]
       {\@mybluebox}%
       {\@mybluebox[0pt]}}

\def\@mybluebox[#1]{%
    \@ifnextchar[%]
       {\@@mybluebox[#1]}%
       {\@@mybluebox[#1][0pt]}}

\def\@@mybluebox[#1][#2]#3{
    \sbox\mytempbox{#3}%
    \mytemplen\ht\mytempbox
    \advance\mytemplen #1\relax
    \ht\mytempbox\mytemplen
    \mytemplen\dp\mytempbox
    \advance\mytemplen #2\relax
    \dp\mytempbox\mytemplen
    {\hspace{1em}\usebox{\mytempbox}\hspace{1em}}}

\makeatother
\def\cline{\centerline{\noindent\rule{10cm}{0.4pt}} \vskip .1 in}
\def\DO{\Delta_{{\cal O}}}
\def\<{\langle}
\def\>{\rangle}
\newcommand   \f  {\phi}
\newcommand{\bea}{\begin{eqnarray}}
\newcommand{\eea}{\end{eqnarray}}
\def\xb{\overline{x}}
\def\Bc{{\cal B}}
\def\Dc{{\cal D}}
\def\pb{\overline p}
\def\Dg{\Delta_{\rm gap}}

\def\foot{\footnote}

\def\o{\over}

\def\d{\delta}

\def\0{{(0)}}
\def\1{{(1)}}
\def\2{{(2)}}
\def\3{{(3)}}
\def\4{{(4)}}

\def\G{\Gamma}

\def\eqr{\eqref}
\def\sec{\section}

\def\l{\lambda}
\def\vs{\vskip .1 in}

\def\a{\alpha}
\def\rar{\rightarrow}

\def\la{\langle}
\def\ra{\rangle}
\def\O{{\cal O}}

\def\ssec{\subsection}
\def\sssec{\subsubsection}
\def\sec{\section}

\def\i{\infty}

\def\O{{\cal O}}
\def\ra{\rangle}
\def\la{\langle}

\def\t{\tau}

\def\s{\sigma}

\def\vs{\vskip .1 in}
\def\D{\Delta}

\def\L{\Lambda}

\def\g{\gamma}
\def\a{\alpha}
\newcommand {\be} {\begin {equation}}
\newcommand {\ee} {\end {equation}}

\newcommand {\bes} {\begin {equation*}}
\newcommand {\ees} {\end {equation*}}

\newcommand{\es}[2] {\begin{equation} \label{#1} \begin{split} #2 \end{split} \end{equation}}
\newcommand{\e}[2] {\begin{equation} \label{#1} #2 \end{equation}}

\newcommand{\Z}{\mathbb{Z}}
\newcommand{\N}{\mathbb{N}}

\newcommand{\beq}{\begin{equation}}
\newcommand{\eeq}{\end{equation}}

\newcommand{\p}{\partial}

\def\be{ \begin{equation} }
\def\ee{ \end{equation} }

\def\half{{1\over  2}}

\def\N{{\cal N}}
\def\A{{\cal A}}

\def\L{\Lambda}

\def\eqr{\eqref}

\def\b{\beta}
\def\l{\lambda}

\renewcommand{\Im}{\textrm{Im}\,}
\renewcommand{\Re}{\textrm{Re}\,}

\newcommand\zb{\bar{z}}

\def\rar{\rightarrow}

\def\zb{\overline{z}}

\def\eps{\epsilon}

 \newmuskip\pFqmuskip

\newcommand*\pFq[6][8]{%
  \begingroup % only local assignments
  \pFqmuskip=#1mu\relax
  \mathcode`\,=\string"8000
  \begingroup\lccode`\~=`\,
  \lowercase{\endgroup\let~}\pFqcomma
  {}_{#2}F_{#3}{\left[\genfrac..{0pt}{}{#4}{#5};#6\right]}%
  \endgroup
}
\newcommand{\pFqcomma}{\mskip\pFqmuskip}
\makeatletter
\renewcommand{\@maketitle}{
\newpage
 \begin{center}%
  {\large\bfseries \@title \par}%
 \end{center}%
 \par} \makeatother
\numberwithin{equation}{section}

\begin{document}

\institution{Yale}{Department of Physics, Yale University, New Haven, CT 06511}
\institution{PU}{Department of Physics, Princeton University, Princeton, NJ, 08544}
\institution{CT}{Walter Burke Institute for Theoretical Physics, California Institute of Technology, \cr Pasadena, CA, 91125}

\title{{\bf Beyond ${\boldsymbol{a=c}}$\,:}\\\LARGE Gravitational Couplings to Matter and the ~~~~~~~~~~~~~~~Stress Tensor OPE}

\authors{David Meltzer\worksat{\Yale}, Eric Perlmutter\worksat{\PU, \CT}}

\abstract{We derive constraints on the operator product expansion of two stress tensors in conformal field theories (CFTs), both generic and holographic. We point out that in large $N$ CFTs with a large gap to single-trace higher spin operators, the stress tensor sector is not only universal, but isolated: that is, $\langle TT{\cal O}\rangle=0$, where ${\cal O}\neq T$ is a single-trace primary. We show that this follows from a suppression of $\langle TT{\cal O}\rangle$ by powers of the higher spin gap, $\Delta_{\rm gap}$, dual to the bulk mass scale of higher spin particles, and explain why $\langle TT{\cal O}\rangle$ is a more sensitive probe of $\Delta_{\rm gap}$ than $a-c$ in 4d CFTs. This result implies that, on the level of cubic couplings, the existence of a consistent truncation to Einstein gravity is a direct consequence of the absence of higher spins. By proving similar behavior for other couplings $\langle T{\cal O}_1{\cal O}_2\rangle$ where ${\cal O}_i$ have spin $s_i\leq 2$, we are led to propose that $1/\Delta_{\rm gap}$ is the CFT ``dual'' of an AdS derivative in a classical action. These results are derived by imposing unitarity on mixed systems of spinning four-point functions in the Regge limit. Using the same method, but without imposing a large gap, we derive new inequalities on these three-point couplings that are valid in any CFT. These are generalizations of the Hofman-Maldacena conformal collider bounds. By combining the collider bound on $TT$ couplings to spin-2 operators with analyticity properties of CFT data, we argue that all three tensor structures of $\langle TTT\rangle$ in the free-field basis are nonzero in interacting CFTs.}%: that is, $n_B n_F n_V>0$.}

\date{ }

\maketitle
\setcounter{tocdepth}{2}
\tableofcontents

\sec{Introduction and Summary}

This work aims to explicate properties of the stress tensor, $T_{\mu\nu}$, in conformal field theories. More specifically, we study the operator product expansion (OPE) of two stress tensors, of schematic form
\e{}{T(x) T(0) \sim \sum_\O C_{TT\O} {\O(0)\o x^{2d-\DO}}}
where $\O$ are local operators of conformal dimension $\DO$. 
We will derive new constraints on the coefficients $C_{TT\O}$ by imposing consistency conditions obeyed by all conformal field theories: namely, unitarity and the bounded growth of correlators in the Regge limit. Some of our results apply to all conformal field theories, while some are specialized to holographic conformal field theories with a large gap in operator dimensions. 

\ssec{Motivation}

The $TT$ OPE is central to every CFT$_d$. In $d=2$, the stress tensor generates the Virasoro algebra, a closed subsector parameterized only by the central charge. In $d>2$, because the stress tensor may couple to any operator $\O$ that is a singlet under all global symmetries and respects the Bose symmetry of the $TT$ operator product, the $TT$ OPE is both a challenging and beckoning observable: there are many possibilities and non-universal details, but the $\la TT\O\ra$ three-point couplings should reflect the richness of CFT dynamics.

Study of the $TT$ OPE fits naturally into the context of the AdS/CFT correspondence \cite{Maldacena:1997re,Witten:1998qj,Gubser:1998bc}. There is, in that setting, one main question that motivates us here: what are the sufficient conditions for a CFT to have a weakly coupled, local, Einstein gravity dual, and what would it mean to find them? This question is not new, but nevertheless remains outstanding. It was conjectured in \cite{Heemskerk:2009pn} (henceforth HPPS) that 
\vs

\centerline{Large $N$ + Large higher spin gap ~~$\Rightarrow$  ~~Weakly coupled, local gravity dual.}

\vs

\noindent The gap condition refers to single-trace operators. For future use, we denote the characteristic scale of higher spin operators as $\Dg$, where ``higher-spin'' means spin greater than two.\foot{This refers to symmetric traceless tensors. In $d\geq 4$, there exist operators in mixed symmetry representations of the Lorentz group. For these operators, one way to state the higher spin condition is in terms of the Regge growth: namely, ``higher spin'' means any operator whose contribution to a four-point function grows faster than that of the stress tensor.} The gap condition $\Dg\rar\i$ should be viewed as a proxy for strong coupling, and the two coincide in known examples with marginal couplings.  

This conjecture is remarkable for its reductiveness: if true, all CFT data becomes ``strongly coupled'' thanks to a single spectral condition. HPPS demonstrated a one-to-one correspondence between solutions to crossing symmetry at leading-order in $1/N$ perturbation theory around generalized free scalar fields with no single-trace cubic couplings, and local quartic AdS vertices bounded in derivatives.\foot{HPPS had in mind the stronger form of locality, in which a CFT$_d$ has a local AdS$_{d+1}$ dual, not a local AdS$_{d+1}\times {\cal M}$ dual. 
This distinction, though interesting, plays no role in this paper. For other works on the implications of large gap, see e.g. \cite{Heemskerk:2010ty, ElShowk:2011ag, Fitzpatrick:2012cg, Belin:2016yll}.}  It was only recently shown by Caron-Huot (modulo some low-lying exceptions) that higher-derivative quartic vertices are suppressed by powers of $\Dg$, reproducing the prediction of bulk effective field theory in which heavy higher spin exchanges are integrated out \cite{Caron-Huot:2017vep}. The HPPS counting argument survives the addition of cubic couplings -- for every bulk cubic vertex, dual to a single-trace three-point function, there is another family of solutions to crossing symmetry \cite{Alday:2017gde} -- but the counting arguments do not determine the structure of these couplings. In particular, the gravity dual of a prototypical holographic CFT is not only weakly coupled, local and without higher spin fields, but also obeys the property that its gravitational dynamics are those of {\it Einstein} gravity. That this, too, follows from the higher spin gap was shown by \cite{Camanho:2014apa} (henceforth CEMZ), who studied the graviton three-point coupling dual to the CFT three-point function $\la TTT\ra$. In $d\geq 3$, there are three independent tensor structures,
\e{ttt2}{\la TTT\ra = \la TTT\ra_{\rm Einstein} + \a_2\la TTT\ra_{R^2} + \a_4 \la TTT\ra_{R^3}}
which we have parameterized in terms of the gravity theories which activate them. (In $d=3$, one of these is replaced by a parity-odd structure.) It is a statement of universality in CFTs dual to Einstein gravity that  $\a_2=\a_4=0$ \cite{henningsonskenderis}. CEMZ showed that, indeed,
\e{a2a4}{\a_n \lesssim \Dg^{-n}}
In $d=4$, the constraint on $\a_2$ translates into a parametric bound on the difference in conformal anomaly coefficients, $a-c$:
\e{}{{|a-c|\o c} \lesssim \Dg^{-2}}
This was later derived purely from CFT in \cite{Afkhami-Jeddi:2016ntf,Costa:2017twz,Afkhami-Jeddi:2017rmx}. Thus, on the level of the stress tensor three-point coupling, the higher spin gap guarantees the existence of general relativity in AdS$_{d+1}$, a remarkable result.

\vs
Still, the $a-c$ bound is unsatisfactory, for two reasons. 

First, there is no known explicit example in which $a-c$ saturates the bound, even parametrically. In superconformal field theories, neither $a$ nor $c$ is a function of $\Dg$ at all \cite{grisaru}. Instead, $a-c$ obeys the stronger bound
\e{}{{|a-c|\o c} \lesssim N^{-\#}}
where, roughly speaking, $\#=1$ for an open string dual and $\#=2$ for a closed string dual (e.g. \cite{Buchel:2008vz, szep, Beccaria:2014xda}). On the other hand, without supersymmetry, $a$ is independent of $\Dg$ because of the $a$-theorem, and it is not known whether $c$ is a function of exactly marginal couplings, or whether fixed lines even exist. We thus seek a more robust observable which is sensitive to $\Dg$. In this discussion we are viewing $\Dg$ as a modulus parameterizing a family of large $N$ CFTs, as in the familiar examples, as opposed to a parameter in an isolated CFT.
%\vs
%

Second, the works of HPPS and CEMZ study either the matter sector or the gravity sector, but not their couplings to each other. How are these constrained? What is the low-energy imprint of the decoupled higher spin fields on these couplings? The motivation for answering this question ties to a grander hope, that by studying the fine structure of CFT, we can discover the necessity of string/M-theory in the bulk. Two stringy signatures are especially important in the present context: the existence of Regge trajectories and of a gap scale ($\ell_s$) that controls all particle masses. Classifying what matter we can consistently couple to quantum gravity -- and how we can couple it -- is exactly the holographic dual of determining the allowed CFT local operator spectra and three-point couplings, particularly the couplings between the stress tensor $T$ and other operators $\O$.

\ssec{Summary of results}
To address these issues, consider the following schematic form of the prototypical bulk Lagrangian that appears in AdS/CFT at low energies:  
\e{}{{\cal L}_{\rm bulk} = R+2\Lambda+ \p_\mu\phi^i\p^{\mu}\phi^i + \l_{ijk}\phi^i\phi^j\phi^k + \l_{ijkl}(\p)\phi^i\phi^j\phi^k\phi^l +\ldots}
where $\ldots$ represents higher-point vertices. The $\phi^i$ are matter fields of spins $s\leq 2$, while the $\l_{ijk}$ and $\l_{ijkl}(\p)$ are three- and four-point vertices, where the latter may carry derivatives distributed among the $\phi$'s. This Lagrangian has an obvious but important property: the gravity sector is not only {\it universal}, but {\it isolated}. That is, there exists a consistent truncation to the Einstein gravity sector, whereupon setting $\phi^i=0$ is consistent with the classical equations of motion: at tree-level, only a graviton can decay into gravitons. On the level of three-point couplings, this follows from the absence of a classical coupling between two gravitons and a $\phi^i$. In CFT terms, to leading order in $1/N$,
\e{}{\la TT\O\ra=0}
where $\O$ is a light, single-trace operator not equal to $T$. We emphasize that $\O$ is an operator that survives the low-energy limit, not a heavy field that decouples. 

This poses a natural question:
\e{}{\text{Is }\la TT\O\ra \sim \Dg^{-\#}~~\text{for some }\#>0?}
An affirmative answer would demonstrate, from CFT, that the existence of a consistent truncation to Einstein gravity is a direct consequence of the absence of higher spin particles. Also, $\la TT\O\ra$ can be a function of marginal couplings, and hence $\Dg$, in a 4d SCFT, even when $\O$ is protected by supersymmetry.

In AdS$_{d+1}$, the first coupling of two gravitons to a scalar field has four derivatives, and can be cast in the following form: 
\e{}{\l_{TT\O}\int d^{d+1}x \sqrt{g}\,\phi\,C_{\mu\nu\rho\sigma}^2}
where $C_{\mu\nu\rho\sigma}$ is the Weyl tensor. It is therefore obvious that a two-derivative action in AdS cannot give rise to a $\la TT\O\ra$ coupling for scalar $\O$. What is not obvious is what it means, in CFT, to count bulk derivatives in a classical action. The intuition explained above, combined with dimensional analysis, suggests that
\e{ttogap}{\l_{TT\O}\sim M_{\rm HS}^{-2}\quad \longleftrightarrow \quad \la TT\O\ra \sim \Dg^{-2}}
where $M_{\rm HS}\sim \Dg$ is the mass scale of higher-spin particles. We will prove this below.

In fact, we want to argue for the following more general avatar of ``stringiness'' in CFT. In string theory, $\Dg$ uplifts to ten dimensions, where the low-energy limit yields a two-derivative action. This should be a generic consequence of large gap: in the spirit of HPPS and CEMZ, we make a
\e{proposal}{\text{\it{Proposal:}}~~\text{Counting AdS derivatives }~~ \longleftrightarrow~~\text{ Counting CFT powers of }\Dg} 
That is, the holographic dual of a bulk derivative is an inverse power of $\Dg$; more precisely, at the level of the classical action in a weakly coupled theory of gravity, all bulk derivatives beyond the two-derivative level are suppressed by powers of $M_{\rm HS}$. We offer this as a sharpened definition of what it means to show that, following \cite{Heemskerk:2009pn, Camanho:2014apa,Caron-Huot:2017vep}, the higher spin gap condition
does indeed guarantee the emergence of local, Einstein gravity in the bulk. We will prove \eqr{proposal} for a variety of three-point functions $\la T\O_1\O_2\ra$ where $\O_i$ are symmetric traceless tensors of spin $s\leq 2$. For $d>3$, we can also have operators of mixed symmetry appearing in the $TT$ OPE (e.g. \cite{Costa:2016hju}), which we do not address here.

  \begin{figure}
    \centering
       \includegraphics[width = .5\textwidth]{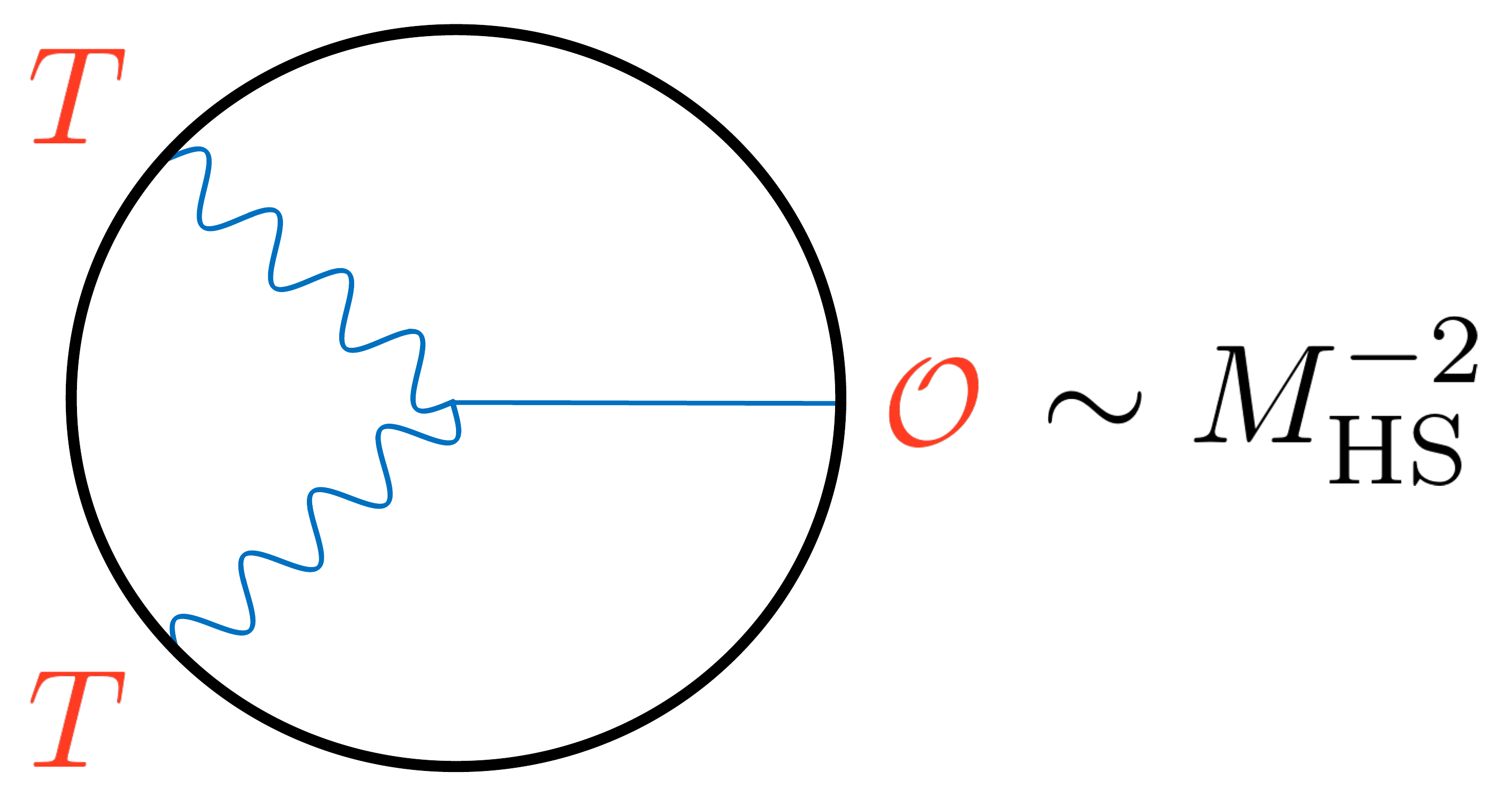}
         \caption{
         In AdS, the cubic coupling of a scalar to two gravitons is suppressed as $M_{\rm HS}^{-2}$, where $M_{\rm HS}\sim \Dg$ is the mass scale of higher spin particles.}\label{fig1}
\end{figure}

Our general strategy combines several ingredients: we study the Regge limit of mixed systems of four-point functions of spinning operators. 
We employ conformal Regge theory \cite{Costa:2012cb}, which is designed to compute the complete contribution of the leading Regge trajectory to CFT four-point functions. This method is not restricted to holographic CFTs: we will also derive rigorous inequalities for $\la T\O_1\O_2\ra$ couplings, valid in {\it any} CFT. In particular, we derive the generalization of the conformal collider bounds of \cite{Hofman:2008ar} on $\la TTT\ra$ couplings. (See \cite{Hofman:2009ug, Hofman:2016awc, Faulkner:2016mzt, Zhou:2016kcz, Chowdhury:2016hjy} for further developments.) These include $\la TT\O\ra$ bounds for various $\O$, thus directly constraining the $TT$ OPE.

\cline

In {\bf Section 2}, we briefly review the salient features of conformal Regge theory for spinning operators, and the unitarity condition that we will employ. Our approach throughout closely follows that of \cite{Costa:2017twz}. A point on the Regge trajectory is parameterized by $\nu$, where \\ $i\nu=(\D-d/2)\geq 0$ is a spectral parameter, and the corresponding spin is given by the function $j(\nu)$. A key point is that the stress tensor lives on this trajectory, at $j(-id/2)=2$. 

In {\bf Section 3}, we summarize our main idea, which we sketch here. Consider a four-point function 
\e{}{\la \Psi\phi\phi\Psi\ra}
where $\phi$ is a scalar primary, and $\Psi$ is a linear combination of primaries:
\e{}{\Psi = a_1\O_1+a_2\O_2}
for some constants $a_i$. This gives rise to a two-by-two matrix of correlators, whose form we study in the Regge limit. Viewed in the $\O_i\O_j\rar j(\nu)\rar\phi\phi$ channel, where $j(\nu)$ stands for the Reggeon, this becomes, up to an overall factor, the following matrix of three-point couplings:
\e{}{\left( 
\begin{array}{cc}
\la \O_1\O_1j(\nu)\ra &\la \O_1\O_2j(\nu)\ra\\
\la \O_2\O_1j(\nu)\ra&\la \O_2\O_2j(\nu)\ra
\end{array} \right)}
For spinning $\O_i$, these entries are functions of polarizations. Unitarity in the Regge limit implies that this matrix is negative semi-definite. So far, this applies everywhere on the Regge trajectory. But by sitting on the $j=2$ stress tensor point of the trajectory, the above constraints become constraints on $\la \O_i\O_j T\ra$ couplings. In particular, positivity of the determinant gives {\it upper} bounds on the off-diagonal couplings. We will be almost entirely interested in the case where $\O_1=T_{\mu\nu}$. 
Then positivity implies bounds of the form
\e{}{\la TT\O\ra^2 \leq \la TTT\ra \la T\O\O\ra}
This is schematic in several ways (e.g. the three-point functions have an independent OPE coefficient for each tensor structure), but will be made precise in what follows. Without imposing a higher spin gap, this yields conformal collider bounds -- that is, inequalities for $\la TT\O\ra$ couplings -- valid in any CFT.\foot{To derive conformal collider bounds from Regge physics, one considers a limit of cross-ratios (impact parameter) in which these matrix elements become exactly those of the null energy operator, which is subject to the average null energy condition \cite{Hofman:2008ar}. Therefore, as in \cite{Costa:2017twz}, they are valid in generic CFTs.} Upon imposing a higher spin gap, a simple argument leads to the suppression with powers of $\Dg$ suggested in \eqr{proposal}. The remaining sections are devoted to implementing this procedure.

In {\bf Section 4}, we study this setup with $\O_1=T_{\mu\nu}$ and where $\O_2=\O$ is a scalar primary. At large gap, a short sequence of steps proves that $\la TT\O\ra \sim \Dg^{-2}$. We explain how to translate this into a bound on the AdS coupling, $\l_{TT\O}$, for all scalar masses, thus proving \eqr{ttogap}. This relation involves a subtlety of ``extremal'' three-point functions in AdS/CFT \cite{DHoker:1999jke}. We discuss further holographic implications of this result, including a parametric no-go result on consistent truncation to Gauss-Bonnet gravity. We identify some promising candidate CFTs in which to holographically compute the leading $1/\Dg$ correction to $\la TT\O\ra$ using string theory. Such CFTs must possess a neutral KK scalar $\O$. Fortunately, this includes the familiar conifold CFT in $d=4$ \cite{Klebanov:1998hh} and the $\N=6$ ABJM theories in $d=3$ in the strongly coupled, `t Hooft limit \cite{Aharony:2008ug}. 

Next, we extract the conformal collider bound on $\la TT\O\ra$. The result can be found in \eqr{TTO-Collider-Even}. This was recently derived using the average null energy condition (ANEC) in \cite{Cordova:2017zej}. As one application of its utility, we infer bounds on OPE coefficients in the stress tensor four-point function, $\la TTTT\ra$, in the mean field theory limit of infinite central charge $C_T$ (see \eqr{attmft}--\eqr{attmftbound}).

In {\bf Section 5}, we extend both the large gap and conformal collider analyses to $\la T\O_1\O_2\ra$ correlators where the $\O_i$ can have spin $\leq 2$. We re-derive the collider bounds of \cite{Cordova:2017zej} involving parity-odd structures, and derive new bounds, including a bound on the parity-odd $\la TTV\ra$ coupling in $d=4$, where $V$ is a non-conserved vector operator, and on $\la TTM\ra$ couplings in all $d$, where $M$ is a non-conserved spin-2 operator. A summary of our main CFT results, along with the relevant AdS vertices whose suppression we derive, is given in the following table: 

{\tabulinesep=1.1mm
\begin{table}[h]
\begin{center}
\begin{tabu}{|c|c|c|c| }
\hline
{\bf Correlator} & $\Dg\rar\i$ {\bf bound}& {\bf Collider bound} & {\bf AdS interaction }\\
\hline
$\<TT\O\>$ &  \eqr{ttogap}  & \eqr{TTO-Collider-Even} & $\f C_{\mu\nu\rho\sigma}^2$ \\
\hline
$\<TT\O\>_{\rm odd}$ & \eqr{eqn:TTO-Holo-Odd} & \eqr{eqn:TTO-Collider-Odd}  & $\f \widetilde C_{\mu\nu\rho\sigma}C^{\mu\nu\rho\sigma}$ \\
\hline
$\<TTV\>_{\rm odd}$ & \eqr{eqn:TTV-Holo} & \eqr{eqn:TTV-Collider}\! , \eqr{eqn:TTJ-Collider}  & $A\wedge R\wedge R$ \\
\hline

$\<TTM\>$ & \eqr{eqn:TTM-Holo}  & \eqr{eqn:TTM_Bound_1}--\eqr{eqn:TTM_Bound_3} & \cite{Camanho:2014apa} \\
\hline
$\<JJT\>_{\rm odd}$ & \eqr{eqn:JJT-Holo-Odd} & \cite{Chowdhury:2017vel}  & $\widetilde R_{~~\kappa\delta}^{\rho\sigma}F^{\mu\kappa}F^{\nu\delta}$ \\
\hline
$\<TTT\>_{\rm odd}$ &  \eqr{eqn:TTT-Holo-Odd}  & \cite{Chowdhury:2017vel}  & $\widetilde{R}_{\mu\nu\sigma\delta}R^{\sigma\delta\rho\gamma}R_{\rho\gamma}\hspace{.02cm} ^{\mu\nu}$ \\
\hline
\end{tabu}
\end{center}

\end{table}}
\noindent Here and elsewhere, we use the notation
\e{}{\O:~\text{spin-0,}~ ~~V:~\text{spin-1,}~~~ M:~\text{spin-2}}
where $V$ and $M$ may, but need not, be conserved. The AdS vertices are those of lowest derivative order that produce the indicated three-point couplings, modulo field redefinitions; the statement is that the vertices are suppressed by the powers of $M_{\rm HS}$ implied by dimensional analysis. (Tilde'd curvatures are made via contraction with the $(d+1)$-dimensional $\epsilon$-tensor.) We have also derived bounds on other three-point functions, including the case of parity-even and parity-odd $\la TV\O\ra$ couplings (see Section \ref{sec:TVO}), which are technically simple but, being highly constrained by stress tensor conservation, physically peculiar. 

Among the collider bounds derived here involving spin $s>0$ operators, the $\la TTM\ra$ bounds seem especially powerful. By combining them with analyticity properties of OPE data as a function of operator spin \cite{Caron-Huot:2017vep}, we give a novel argument that in any interacting CFT, all three tensor structures of $\la TTT\ra$ in the free-field basis are nonzero. In the notation of \eqr{tttfree},
\e{}{n_Bn_Fn_V>0~.}
This was conjectured in \cite{zhib}.

In {\bf Section 6} we briefly conclude.

The {\bf Appendices} contain many details, including our conventions for embedding space computations; various explicit changes of bases needed to relate couplings in the Regge limit to the more familiar bases of spinning correlators \cite{Costa:2011mg}; and reproductions and minor extensions of some results already in the literature using the language of our calculation.

\sec{Spinning conformal Regge theory}
Our goal in this section is to set up the problem of computing systems of mixed spinning correlators in the Regge limit. The requisite technology, that of conformal Regge theory \cite{Costa:2012cb}, has recently been nicely reviewed in \cite{KPZ2017,Li:2017lmh,Afkhami-Jeddi:2017rmx,Costa:2017twz}, so we will keep our discussion streamlined. For experts or casual readers, we have boxed some important equations.% that are most important for future sections.

\begin{figure}
    \centering
       \includegraphics[width = .6\textwidth]{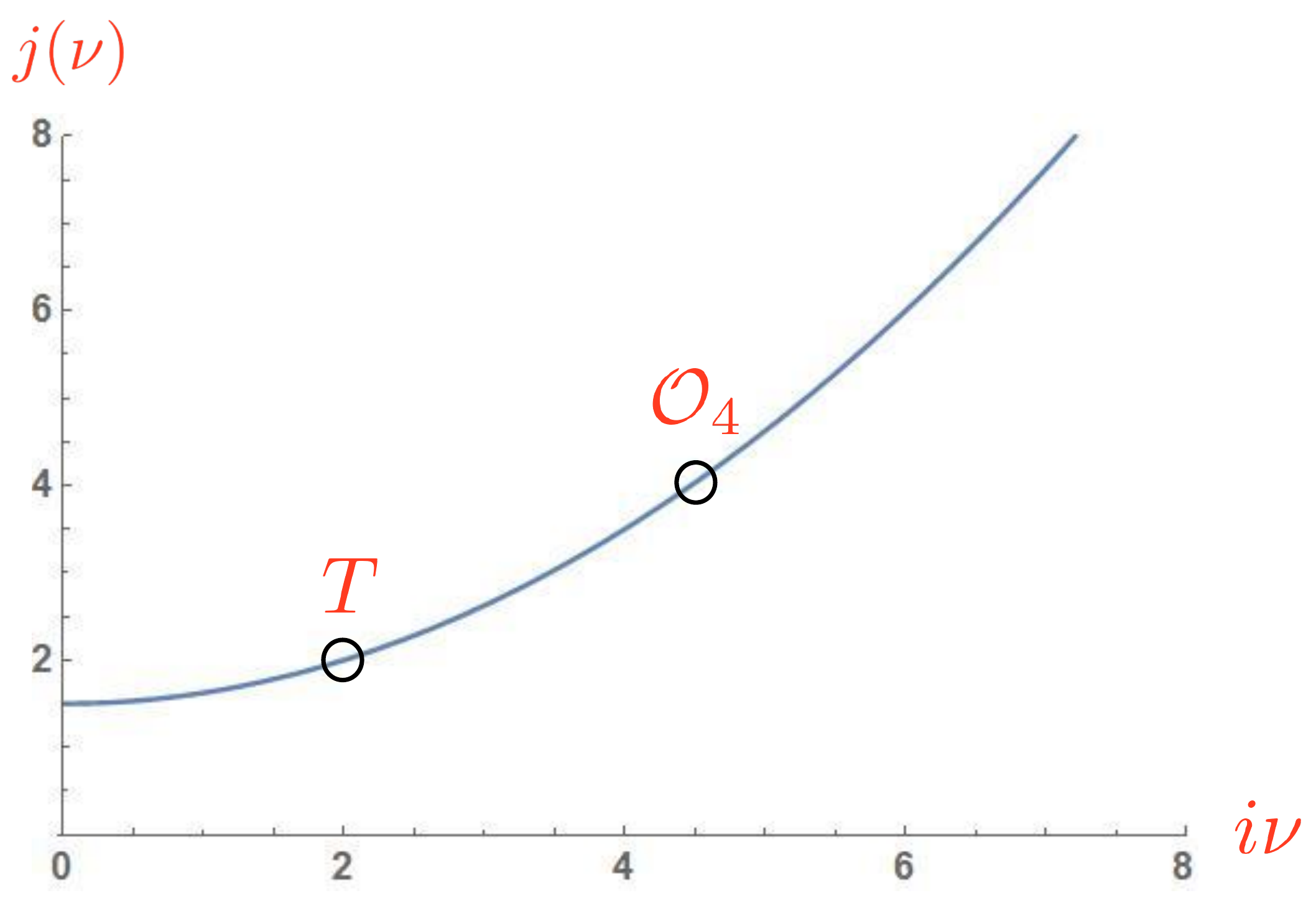}
         \caption{A prototypical leading Regge trajectory, here depicted in $d=4$. Recall that $i\nu\equiv \D-{d\o2}$. We have indicated the locations of the stress tensor, $T$, and the lightest spin-four operator, $\O_4$ which defines $\Dg$ through $j(-i(\Dg-2))=4$.}\label{fig2}
\end{figure}

\ssec{Regge trajectory}
The leading Regge trajectory of a large $N$ CFT may be defined as the set of conformal primary operators of lowest twist for a given spin $s\geq2$, analytically continued to complex spin. Defining conformal dimensions as
\e{dhnu}{\D = h+i\nu~, ~~ \text{where}~{h\equiv {d\o2}}}
the Regge trajectory is defined by the analytic function $j(\nu)=j(-\nu)$. If we define $\nu_s$ as the value of $\nu$ for which the spin-$s$ operator achieves its dimension according to \eqr{dhnu}, then $j(\nu_s)=s$. There are two especially important points on the trajectory:
\es{}{j(0):&\quad \text{Intercept}\\
j(-ih)=2:&\quad \text{Stress tensor}}
We depict a typical leading Regge trajectory in Figure \ref{fig2}. 

The leading Regge trajectory in a finite $N$ theory is known to be convex and monotonic. That is the leading trajectory obeys:
\e{}{0<-\ij'(\nu)<1~, ~~ j''(\nu)<0}
The upper bound on $-ij'(\nu)$ comes from unitarity, $\D=h+i\nu\geq 2h-2+j(\nu)$, while convexity was proven in \cite{Costa:2017twz}. At the intercept, $j'(0)=0$. We will sometimes use the symbol $j(\nu)$ inside correlation functions, as in $\la \O\O j(\nu)\ra$, where it denotes the contribution of the leading Regge trajectory at some value $\nu$.\foot{It is important to distinguish between the exact, leading Regge trajectory at finite $N$ and the leading Regge trajectory of single-trace operators in large $N$ CFTs. The latter are not in general required to be convex, but we will assume the same conditions for the large $N$ trajectory in a neighborhood of the intercept that includes the stress-tensor point. This has passed some consistency checks \cite{Costa:2012cb,KPZ2017,Costa:2017twz}.}

In Regge kinematics, recalled below, large $N$ CFT four-point functions are dominated by the leading Regge trajectory of single trace operators. Its contribution may be resummed into the exchange of an the effective field $j(\nu)$; this is achieved by the program of conformal Regge theory, developed in \cite{Costa:2012cb}. A four-point function of conformal, possibly spinning, primaries is fixed by conformal invariance up to a reduced amplitude $\A(z,\zb)$, where the conformal cross-ratios are
\e{}{u=z\zb={x_{12}^2x_{34}^2\o x_{13}^2x_{34}^2}~, \quad v=(1-z)(1-\zb) = {x_{14}^2x_{23}^2\o x_{13}^2x_{34}^2}}
with $x_{ij} =x_i-x_j$. The Regge limit is reached by analytically continuing $\zb$ around 1 while keeping $z$ fixed, and then taking the limit
\e{}{z,\zb\rar 0~, ~~ {z\o \zb}~\text{fixed}.}
In taking this limit, it is also convenient to use the parameterization
\e{}{z=\s e^\rho~, ~~ \zb=\s e^{-\rho}}
whereupon the Regge limit is $\s\rar0$ with $\rho$ fixed. For scalars $\O_i$, the Regge correlator was derived in \cite{Costa:2012cb}. We will be interested in the analogous result for spinning operators, studied for specific cases in \cite{Costa:2017twz}. In any $d$, four-point functions of symmetric traceless tensors can be written, in a conformal block decomposition, as weighted sums of differential operators acting on scalar conformal blocks \cite{Costa:2011dw,Karateev:2017jgd}. The full four-point function is thus expressed as a sum over a finite number of distinct such operators, each one representing a unique tensor structure. In this ``differential basis,''  
\e{adiff}{\A(\s,\rho) = \sum_\O\sum_{k=1}^{N_{\rm struc}}d^{(k)}_{12\O}d^{(k)}_{34\O}\mathrm{D}_k G_{\D,\ell}(\s,\rho)}
where $G$ is the scalar conformal block for dimension-$\D$, spin-$\ell$ exchange, $N_{\rm struc}$ is the number of independent tensor structures, and $d^{(k)}_{12\O}d^{(k)}_{34\O}$ is the product of OPE coefficients associated to each structure in the exchange of $\O$. We will employ a standard index-free notation \cite{Costa:2011mg} in which Lorentz indices are contracted with null polarization vectors $z_\mu$:
\e{}{\O(x,z) \equiv \O_{\mu_1\ldots \mu_s}(x)z^{\mu_1}\ldots z^{\mu_s}}
where $z^2=0$. In this case, the differential operators $\mathrm{D}_k$ are functions of positions and polarizations. 

\ssec{Spinning conformal Regge theory}
We will be focused on a four-point function of two scalar primaries $\phi$ and two spinning primaries $\O_{1,2}$ of spins $\ell_{1,2}$:
\e{}{\la \O_1(x_1,z_1)\O_2(x_2,z_2)\phi(x_3)\phi(x_4)\ra  }
In this case, in the $\O_1\O_2\rar\O\rar\phi\phi$ channel, $N_{\rm struc}$ is given by the number of independent three-point tensor structures in $\la\O_1\O_2\O\ra$. For even parity, generically we have \cite{Costa:2011mg},
\e{}{N_{\rm struc} = {1\o6}(\ell_1+1)(\ell_1+2)(3\ell_2-\ell_1+3)}
Imposing permutation symmetry or conservation reduces this number. Taking the Regge limit of the reduced amplitude ${\cal A}(\s,\rho)$ in position space, one finds
\e{aregdiff}{ \A^{\rm Regge}(\sigma,\rho)  = (\text{Disconnected})+\int_{-\i}^\i d\nu \sum_{k=1}^{N_{\rm struc}} \a^{(k)}_{12j(\nu)} D_k\,\s^{1-j(\nu)}\Omega_{i\nu}(\rho)}
We leave the $(z_1,z_2)$-dependence of $\A$ implicit. The ingredients here are as follows. First, $\Omega_{i\nu}(\rho)$ is the $\mathbb{H}^{d-1}$ harmonic function over a geodesic distance $\rho$, for scalar dimension $\D=h-\half+i\nu$. In terms of the $\mathbb{H}^{d-1}$ bulk-to-bulk propagator $\Pi_{i\nu}(\rho)$,
\e{eq:Omega}{\Omega_{i \nu} (\rho) = \frac{i \nu}{2 \pi}\, \big(\Pi_{i\nu}(\rho) - \Pi_{-i\nu}(\rho)\big)\,,}
%\label{eq:Omega}
%\eeq
%
where
\e{prop}{\Pi_{i\nu}(\rho) = {\pi^{1-h}\o 2}{\G(h-1+i\nu)\o \G(1+i\nu)}e^{(1-h-i\nu)\rho}{}_2F_1(h-1,h+i\nu-1,i\nu+1,e^{-2\rho})}
Next, the $\lbrace D_k\rbrace$ form a basis of Regge differential operators, and are functions of positions and polarizations. Finally, $\a^{(k)}_{12j(\nu)}$ are proportional to the squared OPE coefficients for Reggeon exchange of the $k$'th structure in the differential basis,
\e{alphadef}{\a^{(k)}_{12j(\nu)} = X(\nu)\g(\nu)\g(-\nu)d^{(k)}_{12j(\nu)}d_{\phi\phi j(\nu)}K_{h+i\nu,j(\nu)}}
where $\g(\nu), X(\nu)$ and $K_{h+i\nu,j(\nu)}$ are defined in Appendix \ref{app:ChangeBases}.  Since we always study $\la \O_1\O_2\phi\phi\ra$ correlators in this paper, we will label squared OPE coefficients only by the operators $\O_1$ and $\O_2$. We will henceforth ignore the disconnected piece of $\A^{\rm Regge}$, which corresponds to identity exchange. 

Following \cite{Costa:2017twz}, we will eventually be imposing unitarity on the Regge limit of the correlator in impact parameter space. Applied to $\A^{\rm Regge}$, the Fourier integral
\e{}{\A(\s,\rho) = (-1)^{-\frac{1}{2}(\D_1+\D_2)-\D_\f}\int dp\, d\pb\, e^{-2ip\cdot x -2i \pb\cdot \xb}{\Bc(p,\pb)\o (-p^2)^{h-\frac{1}{2}(\D_1+\D_2)}(-\pb^2)^{h-\D_\f}}}
where $\s^2 = x^2\xb^2$ and $\s\cosh\rho = -2x\cdot \xb$, defines the impact parameter representation of the Regge correlator,
\e{breglong}{\Bc^{\rm Regge}(S,L) = \int_{-\i}^\i d\nu S^{j(\nu)-1} \sum_{k=1}^{N_{\rm struc}}\beta^{(k)}_{12j(\nu)} \hat D_k \Omega_{i\nu}(L)}
where the impact parameter variables $(S,L)$ are
\e{}{S = 4|p||\pb|~, ~~ \cosh L = -{p\cdot \pb\o |p||\pb|}.}
The form is essentially identical to \eqr{aregdiff}. In these variables, $\Bc^{\rm Regge}(p,\pb)$ is the $S\rar\i$, fixed $L$ limit of $\Bc(p,\pb)$. The $\lbrace \hat D_k\rbrace$ form a basis of Regge differential operators in impact parameter space. 
The $\beta^{(k)}_{12j(\nu)}$ are proportional to the product of OPE coefficients in this basis, which we call $B$:
\e{betakdef}{\beta^{(k)}_{12j(\nu)} = X(\nu)B^{(k)}_{12j(\nu)}B_{\phi\phi j(\nu)}}
The $B^{(k)}_{12j(\nu)}$ basis of OPE coefficients is linearly related to the $d^{(k)}_{12j(\nu)}$ basis.

Let us write \eqr{breglong} in a compact fashion, 
\e{breg}{\boxed{\Bc^{\rm Regge}(S,L) = {i\o\pi}\int_{-\i}^\i d\nu\, \nu\,S^{j(\nu)-1}X(\nu)\Dc(\nu) \,\Pi_{i\nu}(L)}}
The real operator $\Dc$ is a sum over impact parameter space differential structures in the Regge limit,\foot{The relation to \cite{Costa:2017twz} is $[X(\nu)\Dc(\nu)]_{\rm here} = \mathfrak{D}_{\rm there}$. Note that in passing from \eqr{alphadef} to \eqr{betakdef}, we have absorbed a factor of $K_{h+i\nu,j(\nu)}$, which can have zeroes, in the definition of $B_{12j(\nu)}^{(k)}$. This will be convenient when relating our large gap bounds to AdS physics. The $B_{\phi\phi j(\nu)}$ coupling is unique, fixed by a Ward identity, and is never suppressed at large gap.}
\e{}{\boxed{\Dc(\nu) \equiv B_{\phi\phi j(\nu)}\sum_{k=1}^{N_{\rm struc}} B^{(k)}_{12j(\nu)}\hat D_k}}
We have used the evenness of the integrand to trade $\Omega$ for $\Pi$ using \eqr{eq:Omega}. We emphasize that the coefficients $B^{(k)}_{12j(\nu)}$ are defined with respect to a basis of tensor structures in the Regge limit: changing bases can introduce various kinematic factors, as we will see later. 

\sssec{Sliding along the Regge trajectory}
Now we evaluate the integral by the saddle point approximation (recall that $S\rar\i$). The saddle $\nu_0$ is defined by
\e{saddle}{-ij'(\nu_0)={L\over \log S}}
Note that $\nu_0$ is a function of $L$, and that $-ij'(\nu_0)>0$. Expanding $\Bc^{\rm Regge}$ around the saddle point,
\e{bregsaddle}{-i\Bc^{\rm Regge}(S,L) \approx S^{j(\nu_0)-1}\left({\nu_0\o \pi}\sqrt{2\pi\o -j''(\nu_0)\log S}X(\nu_0)\right) \Dc(\nu_0) \,\Pi_{i\nu_0}(L)}
The $S^{j(\nu_0)-1}$ is the hallmark growth, with Reggeon spin $j(\nu_0)$. By dialing $L$, we move the saddle point and access different points on the Regge trajectory.

Expanding \eqr{saddle} around $\nu_0=0$, using the fact that $j'(0)=0$, means keeping $L$ fixed at large $S$. The resulting bounds are optimized by taking $L\ll1$.

To expand \eqr{saddle} at large gap we will make the general ansatz:
\bea\label{jexp}
j(\nu)=2-\sum_{n=1}^\infty j_{n}(\nu^{2})\Delta_{gap}^{-2n}.
\eea
for some degree-$n$ polynomials $j_n(\nu^2)$, about which we will say more in the next section. This is the most general form for $j(\nu)$ such that it remains finite in the limit $\Delta_{gap}\rightarrow\infty$, up to exponential corrections. Then for $|\nu_0|\ll\Dg^2$,
\e{}{-ij'(\nu_0) \sim {1\o\Dg^2}+\ldots}
Therefore, we are still in the regime of fixed $L\ll \log S$. In particular, we may access the stress tensor point $j=2$ at $\nu_0=-ih$. 

In order to sit on the stress tensor point without assuming large gap, we must be a bit more careful. In general, $0<-i j'(\nu_0)<1$, so the saddle point equation is generically satisfied only for $L\sim \log S$. The conditions under which the unitarity condition evaluated in the saddle point approximation is valid, even when we take $\nu$ to the stress tensor point, were explored in Section 4.5 of \cite{Costa:2017twz}. The prescription for deriving conformal collider bounds is to scale $L$ to infinity while sitting at $j=2$. 

\sssec{Unitarity bound}
With all pieces in order, let us introduce the essential constraint. In impact parameter space, $\Bc$ obeys a unitarity condition \cite{KPZ2017,Costa:2017twz}. In a large $N$ theory, where $\Bc$ represents the connected Regge correlator to leading order in $1/N$,
\e{unit}{\Im(-i \Bc^{\rm Regge}(S,L))\geq 0}
This follows from writing $\Bc$ in terms of a phase shift $\chi$ as $\Bc = e^{i\chi}$, and imposing the Cauchy-Schwarz inequality on a pair of states whose inner product defines the Regge correlator; this leads to $\Im(\chi(S,L))\geq 0$, which becomes \eqr{unit} to leading order in the $1/N$ expansion. 

Noting that $\nu_0$ is negative imaginary, and that $\Re(X(\nu))<0$, the unitarity condition applied to  the saddle point result \eqr{bregsaddle} implies
\e{unitd}{\boxed{ \Dc(\nu_0) \,\Pi_{i\nu_0}(L)\geq 0}}
This should be viewed as a bound on the OPE coefficients $B^{(k)}_{12j(\nu_0)}$, analytically continued to the saddle point $\nu_0$. By tuning $\nu_0$, and the polarizations in the definition of $\Dc(\nu_0)$, this can be used to derive constraints on the OPE coefficients at various points along the Regge trajectory.\foot{One can either think of $\Dc$ as a function of polarizations, which can be tuned to yield multiple constraints, or as a matrix in the space of polarizations. We will often employ the former. The latter is useful in making contact with conformal collider bounds, for example, where decomposition into $SO(d-2)$ representations neatly organizes the bounds.} 

\sec{Main idea: $\la TT\O\ra$ bounds from mixed correlators}

 \begin{figure}
    \centering
       \includegraphics[width = .4\textwidth]{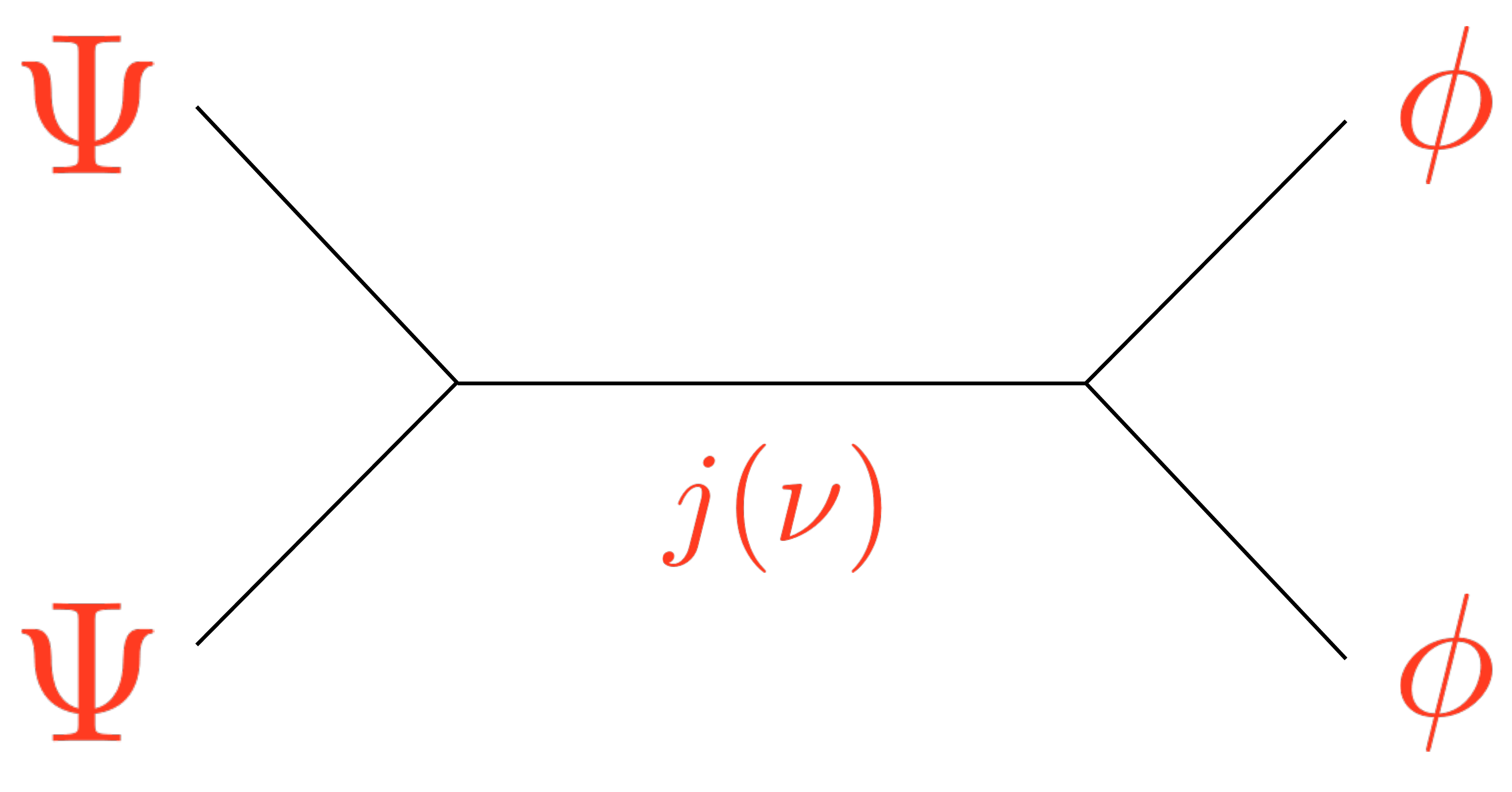}
         \caption{In the Regge limit, the connected four-point function $\la \Psi\phi\phi\Psi\ra$ is dominated by Reggeon exchange in the $\Psi\Psi \rar \phi\phi$ channel. Unitarity bounds the three-point couplings.}\label{fig3}
\end{figure}
Consider now the four-point function
\e{}{\la \Psi\phi\phi\Psi\ra}
where $\Psi$ is a linear combination of spinning primaries:
\e{}{\Psi = a_1\O_1+a_2 \O_2}
for constants $a_i$, where $\O_i$ are spinning operators dotted into polarization vectors. We now have a matrix of correlators,
\e{}{\la \Psi\phi\phi\Psi\ra = a^\dag \cdot \left( 
\begin{array}{cc}
\la \O_1\phi\phi\O_1\ra &\la \O_1\phi\phi\O_2\ra\\
\la \O_2\phi\phi\O_1\ra&\la \O_2\phi\phi\O_2\ra
\end{array} \right)\cdot  a}
In the Regge limit, the unitarity condition on $\la \Psi\phi\phi\Psi\ra$ becomes a positive semi-definite constraint on the differential operator $\Dc$, now viewed as a symmetric matrix in the space of operators:
\e{matrix}{\Dc(\nu_0)\Pi_{i\nu_0}(L)= \left( 
\begin{array}{cc}
\Dc_{11}(\nu_0) &\Dc_{12}(\nu_0)\\
\Dc_{12}(\nu_0)&\Dc_{22}(\nu_0)
\end{array} \right)\Pi_{i\nu_0}(L) \succeq 0}
Its components $\Dc_{ij}(\nu_0)$ are the correlators $\la \O_i\phi\phi\O_j\ra$ in the Regge limit \eqr{breg}, expanded in the $\O_i\O_j\rar j(\nu_0)\rar\phi\phi$ channel of Figure \ref{fig3}:
\e{}{\Dc_{ij}(\nu_0) = B_{\phi\phi j(\nu)}\sum_kB^{(k)}_{\O_i\O_j j(\nu_0)}\hat D_k~.}
Positive semi-definiteness is equivalent to non-negativity of all principal minors of $\Dc$: 
\es{minors}{\Dc_{11}\geq 0~, ~~ \Dc_{22}\geq 0 ~, ~~  \Dc_{11}\Dc_{22}-\Dc_{12}^2\geq 0}
The upper bound on $\Dc_{12}^2$ implies upper bounds on $B_{12j(\nu_0)}^{(k)}$.

In this paper, we will mostly be concerned with linear combinations of stress tensors with other primaries:
\e{tpluso}{\Psi = a_T T + a_\O \O}
Then 
\e{}{\Dc_{12}(\nu_0)\propto B_{ T\O j(\nu_0)}}
At $\nu_0=-ih$ where $j(-ih)=T$, unitarity \eqr{minors} implies upper bounds on $\la TT\O\ra$ couplings. 

We will derive two different types of bounds. The first apply to theories with $\Dg\gg1$. The second are conformal collider bounds, which apply in general. We explain our approach to each of these in turn. 

\ssec{Large gap CFTs}
The scale $\Dg$ is defined by
\es{}{j(-i(\Dg-h))=4}
The spin $j(\nu)$ has an expansion in $1/\Dg$ given in \eqr{jexp}. As higher spin (single-trace) operators decouple, the Regge trajectory flattens out. Note that stress tensor conservation implies $j(-ih)=2$ to all orders in $1/\Dg$. Likewise, all OPE data along the trajectory admits a similar expansion. 

The essential argument for bounding $\la TT\O\ra$ at large gap, inspired by \cite{Costa:2017twz}, is the following. For the spin-four operator, $i\nu \approx \Dg$, so the existence of a finite large gap limit requires OPE coefficients $\la \O_1\O_2 j(\nu)\ra$ to be finite in the limit
\e{}{|\nu|\rar\i~, ~~ \Dg\rar\i~, ~~{|\nu|\o\Dg}~\text{fixed}}
On the other hand, regarded as functions of $\nu$, the $\la \O_1\O_2 j(\nu)\ra$ admit an expansion in $1/\Dg$. This implies
\e{}{\la \O_1\O_2 j(\nu)\ra \approx \la \O_1\O_2 j(0)\ra +\sum_{n=1}^\i {P_n(\nu^2)\o \Dg^{2n}}}
where $P_n(\nu^2)$ is an even, degree-$n$ polynomial in $\nu$ that obeys $P_n(0)=0$, but is otherwise theory-dependent. It follows that suppression of $\la \O_1\O_2 j(\nu)\ra$ at large gap, for any nonzero $\nu$, can be diagnosed by looking at the intercept. In particular, evaluated at the stress tensor point $\nu=-ih$,\foot{We explained above that this is still a large $S$, fixed $L$ regime.} it follows that
\e{}{\la \O_1\O_2 j(0)\ra = 0 \quad \Rightarrow \quad \la \O_1\O_2 T\ra\sim {\Dg^{-2}}}
This suggests a strategy to derive bounds on $\la TT\O\ra$ couplings: using the linear combination \eqr{tpluso} to generate the matrix of correlators \eqr{matrix}, apply the unitarity constraint \eqr{minors} to show that the off-diagonal components $\Dc_{12}(\nu_0)$ must vanish at the intercept, $\nu_0=0$. We emphasize that $\O$ is a light single-trace\foot{Double-trace $\O$ appear in the $TT$ OPE suppressed by an extra power of $1/\sqrt{C_T}$, nor is it clear whether conformal Regge methods apply straightforwardly when the external operator is multi-trace.} operator, whose dimension does not scale with $\Dg$.
 
Now let us explain why $\Dc_{12}(0)$ would vanish, and how to diagnose the degree of suppression with $\Dg$. To optimize the bounds, we take $L\ll1$ as we approach the intercept, as explained below \eqr{saddle}. A key point is that the AdS propagator $\Pi_{i\nu_0}(L)$ diverges as a power law at small $L$ in $d>3$, and logarithmically in $d=3$:
\e{propsmall}{\Pi_{i\nu_0}(L\ll1) \approx \begin{dcases} 
L^{3-d}\,\frac{\pi^{1-d\o2}}{4}  \Gamma \left(\frac{d-3}{2}\right)~, & d>3\,, \\
-{\log L\o 2\pi}, & d =3\,,
\end{dcases}}
Looking back at the form of \eqr{matrix}, and the unitarity constraint $\Dc_{12}^2 \leq \Dc_{11}\Dc_{22}$, this implies that every extra derivative in $\Dc_{12}$, relative to the diagonal terms $\Dc_{ii}$, must have a coefficient with one more power of $\Dg^{-1}$ as we approach the intercept.  
In other words, to count powers of $\Dg$ in the suppression of $\la\O_1\O_2 j(\nu)\ra$ for any $\nu$, we just need to count derivatives in the off-diagonal structure $\la \O_1\O_2 j(\nu_0)\ra$ as we approach $L=0$: in the impact parameter space $B$-basis,
\e{}{\boxed{\Dc(0)\propto \left( 
\begin{array}{cc}
B_{11 j(0)} &\sum\limits_k B^{(k)}_{12j(0)}{\p_L^{n_k}}\\
\sum\limits_k B^{(k)}_{12j(0)}{\p_L^{n_k}}&B_{22j(0)}
\end{array} \right) \quad \Rightarrow \quad B^{(k)}_{12j(\nu)} \sim  {\Dg^{-2\lfloor{n_k+1\o2}\rfloor}}}}
The diagonal $B_{iij(0)}$ always have derivative-free terms, so we have shown only these because they are unsuppressed. (The derivative terms can be set to zero by considering the single correlator $\la \O_i\O_i\phi\phi\ra$ in the Regge limit, rather than the mixed system.) Evaluating this at $\nu=-ih$, the stress tensor point implies the scaling 
\e{}{B_{12T}^{(k)} \sim \Dg^{-2\lfloor{n_k+1\o2}\rfloor}}
for the $k$'th OPE coefficient in $\la \O_1\O_2 T\ra$. The floor symbol implements the evenness in $\nu$ of the OPE coefficients along the Regge trajectory, which buys an extra power of $\Dg$ suppression for odd $n$. In the following sections, we implement this strategy for a host of off-diagonal spinning three-point functions involving stress tensors.

\ssec{Collider bounds for all CFTs}
We also use this setup to derive off-diagonal conformal collider bounds. Here we do not assume that the CFT has a higher spin gap. Instead we implement the unitarity condition \eqr{unitd} in the $L\gg1$ limit, at the stress tensor point, $j(-ih)=2$. In this regime, the AdS propagator has exponential behavior,
\e{}{\Pi_{i\nu}(L\gg1) \approx {\pi^{1-h}\o 2}{\G(h-1+i\nu)\o \G(1+i\nu)}e^{(1-h-i\nu)L}}
which implies that all matrix elements of $\Dc$ are comparable. Then the upper bound on $\Dc_{12}(\nu_0)$ at $\nu_0=-ih$ becomes an upper bound on $\la TT\O\ra$, of the schematic form
\e{}{\la TT\O\ra^2 \leq \la TTT\ra \la T\O\O\ra}
$\la T\O\O\ra$ is essentially inert in this calculation. For instance, for scalar $\O$, $\la T\O\O\ra$ has a unique tensor structure and OPE coefficient fixed by Ward identities, which in our conventions in embedding space (see Appendix \ref{app:Embedding}) reads
\e{}{\la T(P_1;Z_1)\O(P_2)\O(P_3)\ra = -{d\D_\O\o (d-1)\sqrt{C_T}}{V_1^2\o P_{12}^{d+2\o2}P_{13}^{d+2\o2}P_{23}^{2\DO-d-2\o2}}}
where $C_T$ is defined via the stress tensor two-point function,
\e{}{\la T(P_1;Z_1)T(P_2;Z_2)\ra = {C_T}{H_{12}^2\o P_{12}^{d+2}}}
The actual bounds depend on the spin and conformal dimension of $\O$, and will involve some subset of the three $\la TTT\ra$ couplings. Compared to the large gap considerations, deriving the collider bounds requires more precise information about the matrix $\Dc$. 

\sssec*{Review: $\la TTT\ra$}
Before implementing this strategy, we need to summarize the constraints on $\la TTT\ra$, first derived in \cite{Camanho:2014apa} and recapitulated in Regge language in \cite{KPZ2017, Costa:2017twz,Afkhami-Jeddi:2017rmx}. Detailed formulas can be found in \cite{Costa:2017twz}. 

After imposing permutation symmetry and conservation, there are three conformal structures in the $\la TTT\ra$ correlator\cite{Osborn:1993cr}. We may parameterize them using the free-field basis,
\es{tttfree}{\<TTT\>&=n_{B}\<TTT\>_{B}+n_{F}\<TTT\>_{F}+n_{V}\<TTT\>_{V}}
The $n_B$ and $n_F$ structures are those of $n_B$ free bosons and $n_F$ free fermions, respectively; the $n_V$ structure is that of a free ${d-2\o 2}$-form in even dimensions, but may be viewed more generally as a label for the third structure in any dimension. In $d=3$, there are only two independent parity-even structures, and a new parity-odd structure whose form we will recall in due course.

Another convenient parameterization was given in \eqr{ttt2}. The latter parameterization is convenient because $\a_2$ and $\a_4$ are suppressed by the expected powers of $\Dg$ shown in \eqr{a2a4} \cite{Camanho:2014apa, Afkhami-Jeddi:2016ntf,Afkhami-Jeddi:2017rmx, Costa:2017twz, KPZ2017}. This suppression is precisely because the tensor structures in the differential basis have extra derivatives relative to the Einstein structure. $\la TT j(\nu)\ra$ has three conformal structures for general $j(\nu)$, which we write as
\e{ttjnu}{\la TT j(\nu)\ra = \la TT j(\nu)\ra_{\rm Einstein} + \a_2(\nu)\la TT j(\nu)\ra_{R^2} + \a_4(\nu)\la TT j(\nu)\ra_{R^4}}
where the notation means analytic continuation along the Regge trajectory, i.e. $\la TTj(-ih)\ra = \la TTT\ra$ with the parameterization \eqr{ttt2}. The suppression \eqr{a2a4} follows from the analysis of $\la TT j(\nu)\ra$ at the intercept, as articulated in \cite{Costa:2017twz} and above. 
Analyzing \eqr{ttjnu} at large $L$ and $j(-ih)=2$, with no gap assumption, yields the conformal collider bounds of \cite{Hofman:2008ar}. Doing so involves a judicious choice of graviton polarizations.

\section{$\<TT\O\>$ bounds for scalar $\O$}

In this section we will derive bounds on $\<TT\O\>$, where $\O$ is a scalar operator. To do this, we consider the four-point function $\<\Psi\Psi\f\f\>$ where
\bea
\Psi=a_{\O}\O+a_{T}z_{1,\mu}z_{1,\nu}T^{\mu\nu}
\eea
for some constants $a_{\O,T}$. 

The off-diagonal terms in $\Dc$ are proportional to the three-point structure
\e{}{\la T\O \O_{\D,j(\nu)}\ra \equiv z_{1}^{\mu}z_{1}^{\nu}\la T_{\mu\nu}(x_1)\O(x_2)\O_{\D,j(\nu)}(x_3,z_3)\ra,}
In the standard basis,
\e{eqn:TTO_Std_Basis}{\la T\O \O_{\D,J}\ra  \propto c_1 V_{1}^{2}V_{3}^{J}+c_2V_{1}H_{13}V_{3}^{J-1}+c_3H_{13}^{2}V_{3}^{J-2}}
with the denominator fixed by conformal invariance (see Appendix \ref{app:Definitions}). Conservation for the stress tensor yields:
\begin{align}
c_2&= \frac{2 c_1 \left((d-1) \Delta -(d-1) \Delta _\O+J\right)}{(2-d) \left(d-\Delta +J+\Delta _\O\right)},
\\
c_3&= \frac{c_1 \left(-(d-1) \Delta ^2+J (d+J-2)+(d-1) \left(2 \Delta -\Delta _\O\right) \Delta _\O\right)}{(2-d) \left(d-\Delta +J+\Delta _\O-2\right) \left(d-\Delta +J+\Delta _\O\right)}.
\end{align}
Thus, there is a unique structure and we will write $C_{TT\O}=c_1$. We are ignoring a parity-odd structure that exists only in $d=3$, which we will treat separately in section \ref{sec:TTO_odd}.

We can construct $\Dc$ by writing a basis of differential operators directly in the Regge limit \cite{Costa:2017twz}. For a general spinning three-point function $\la \O_1\O_2\O_{\D,J}\ra$, the basis elements are given by degree-$s_i$ monomials built from $z_i\cdot \hat x$ and $z_i\cdot \nabla$, where $i=1,2$ and  $\nabla$ is the covariant derivative on $\mathbb{H}^{d-1}$ constructed using $\hat{x}$ and defined in \eqr{b8}. So for $\la T\O\O_{\D,J}\ra$, a basis is
\begin{align}
{D}_{1}&=(z_{1}\cdot\hat{x})^{2}, \label{eqn:ReggeDIff_TTO_1}
\\
{D}_{2}&=(z_{1}\cdot \hat{x})(z_{1}\cdot\nabla),  \label{eqn:ReggeDIff_TTO_2}
\\
{D}_{3}&=(z_{1}\cdot\nabla)^{2},  \label{eqn:ReggeDIff_TTO_3}
\end{align}
The corresponding impact parameter space basis is given by replacing $\hat{x}\rightarrow \hat{p}$. Details for all the change of bases are given in Appendix \ref{app:TTO_Even}. We will label the coefficient of the unique impact parameter differential operator by $B_{T\O j(\nu)}$.

\ssec{Large gap}
We first apply the arguments of the previous section at the intercept, $\nu_0=0$, to extract bounds at large gap. In the differential basis \eqr{eqn:ReggeDIff_TTO_1}--\eqr{eqn:ReggeDIff_TTO_3} or its impact parameter space counterpart, $\la T\O\O_{\D,J}\ra$ has a unique conformal structure. Moreover, the unique impact parameter space differential operator has two derivatives. The diagonal correlators $\la TT\O_{\D,J}\ra$ and $\la \O\O\O_{\D,J}\ra$ include structures with no derivatives. Therefore, at the intercept, $B_{T\O j(0)}$ vanishes, leading to the desired suppression of $B_{T\O T}$. 

This completes the argument, but let us be more explicit. 
For generic polarizations the scaling of the matrix $\Dc(\nu_0)$ at $\nu_0\approx 0$ takes the form
\e{}{ \Dc(0) \approx \left( 
\begin{array}{cc}
B_{TT j(0)} &B_{T\O j(0)}{\p^2\o \p L^2}\\
B_{T\O j(0)}{\p^2\o \p L^2}&B_{\O\O j(0)}
\end{array} \right)}
Acting on $\Pi_0(L\ll1)\sim L^{3-d}$, the leading behavior is\foot{We leave the logarithmic beahvior in $d=3$, shown in \eqr{propsmall}, understood.}
\e{}{\Dc(0)\Pi_0(L\ll1) \propto L^{3-d} \left( 
\begin{array}{cc}
B_{TTj(0)} &B_{T\O j(0)}L^{-2}\\
B_{T\O j(0)}L^{-2}&B_{\O\O j(0)}
\end{array} \right)}
The off-diagonal terms dominate, which violates positivity unless $B_{T\O j(0)}=0$. Therefore, we conclude that, to leading order in $L\ll1$,
\e{}{B_{T\O j(0)}= 0~.}
The $L^{-2}$ scaling implies that, following our previous arguments, 
\e{ttogap}{B_{TT\O}\sim \Dg^{-2}~.}
This is the desired result.

\ssec{Holographic interpretation}

The result \eqr{ttogap} translates into a suppression of the AdS$_{d+1}$ three-point coupling $\l_{TT\O}$ between two gravitons and a scalar field $\phi$ of mass $m$, where
\e{}{m^2L_{\rm AdS}^2=\D(\D-d)}
After field redefinitions (e.g. \cite{Buchel:2008vz}), we may parameterize this coupling as
\e{phicc}{\l_{TT\O}\int d^{d+1}x \sqrt{g}\, \phi \,C_{\mu\nu\rho\sigma}^2~,}
where $C_{\mu\nu\rho\sigma}$ is the Weyl tensor. 
Since $\Dg \sim M_{\rm HS}$, the mass scale of higher spin fields in AdS, the bulk statement of \eqr{ttogap} is a suppression of $\l_{TT\O}\sim M_{\rm HS}^{-2}$. 

This should hold for all values of the scalar mass $m$. Proving this $m$-independence using conformal Regge theory involves an interesting subtlety. It is well-known \cite{DHoker:1999jke} that for cubic scalar couplings, the proportionality factor between CFT OPE coefficients $C_{123}$ and local AdS couplings $\l_{123}$ vanishes linearly for $\D_3=\D_1+\D_2+2n$ with $n\in \Z_{\geq 0}$:
\e{}{C_{123}\propto \G\left({\D_1+\D_2-\D_3\o2}\right)\l_{123}\quad\quad(\text{scalars})}
This is the statement that for an ``extremal'' three-point function $\la \O_1\O_2\O_3\ra$, the local bulk couplings $\l_{123}=0$. The same is true for spinning correlators (see e.g. \cite{Sleight:2017fpc}), so one might worry that $\l_{TT\O}$ is suppressed only for certain values of $m$. However, it turns out that $B_{TT\O}$ is an especially nice basis in this regard: it is proportional to $\l_{TT\O}$ without any vanishing factors. 
The relation of $B_{TT\O}$ to the standard basis, $C_{TT\O}$, is given in Appendix \ref{app:ChangeBases}. It takes the form
\e{btoc}{B_{TT\O} = g(\D_\O) C_{TT\O}~, ~~\text{where}~~g(\D_\O)\propto \G^{-1}\left(d-{\DO\o2}\right)}
We see that $g(\D_\O)$ has simple zeroes at $\D_\O=2d+2n$. Therefore, for these values of $\D_\O$, there is no bound when expressed in the standard basis. However, $C_{TT\O}$ is related to $\l_{TT\O}$ by a different function, with simple poles at precisely these locations \cite{Sleight:2017fpc, Cordova:2017zej}:
\e{ctol}{C_{TT\O} = f^{-1}(\D_\O)\l_{TT\O}~, ~~\text{where}~~f(\D_\O)\propto \G^{-1}\left(d-{\DO\o2}\right)}
These are the spinning extremal zeroes alluded to above.

Therefore, \eqr{ttogap}, \eqr{btoc}, and \eqr{ctol} imply that for all values of the scalar mass $m$,
\e{ltto}{{\l_{TT\O} \sim M_{\rm HS}^{-2}}}
As discussed in the introduction, this, together with \cite{Camanho:2014apa}, implies that on the level of cubic couplings, in a theory of gravity coupled to a scalar, the existence of a consistent truncation to Einstein gravity is a direct consequence of the absence of higher spin fields. In this way, there is a direct correspondence between a CFT derivation of suppression at large $\Delta_{\rm gap}$ and derivatives in AdS effective actions. We have demonstrated this here for the $\la TT\O\ra$ coupling for scalar $\O$; in the following sections, we support this with several more calculations involving fields of spin $s\leq 2$. The overall picture is that a consistent truncation to Einstein gravity exists in any theory without fields of spin $s> 2$. 

Let us make some comments.

\sssec*{No consistent truncation to Gauss-Bonnet}
Consider a theory of AdS gravity coupled to a scalar field. Through four-derivative order, the most general form of the Lagrangian, up to field redefinitions, is
\e{4deriv}{{\cal L}= R+2\L+\half(\p\phi)^2 +{1\o2}m^2\phi^2+ \l_{GB}(R_{\mu\nu\rho\sigma}^2-4R_{\mu\nu}^2+R^2) + \l_{TT\O}\,\phi\,C_{\mu\nu\rho\sigma}^2}
We have brought the $R^2$ terms into Gauss-Bonnet form using field redefinitions. Regarding \eqr{4deriv} as a classical action, the coefficients obey the parametric constraint
\e{}{\l_{GB}\sim M_{\rm HS}^{-2}\sim \l_{TT\O}}
This implies that no consistent truncation to Gauss-Bonnet gravity is allowed when coupling gravity to scalars, not even at fixed order in low-energy perturbation theory: $\l_{GB}$ and $\l_{TT\O}$ are controlled by the same scale. This result does not follow from CEMZ alone: the $\la TT\O\ra$ coupling is required to rule out the truncation. Stated another way, it is a fine-tuning to couple a free scalar to Gauss-Bonnet gravity, or to add the $\phi R^2$ term without the Gauss-Bonnet term. These statements formalize the naturalness expectation from bulk effective field theory. Our argument here does not apply if either $\l_{GB}$ or $\l_{TT\O}$ is generated solely by loops, which generates a parametric separation. This can indeed happen, as in SUSY AdS$_5$ compactifications.

\sssec*{Computing $\la TT\O\ra$ in top-down examples}

It would be worthwhile to actually compute the leading term in $\la TT\O\ra$ in a CFT with a large gap. This can be done holographically. We are looking for AdS$_{d+1}\times {\cal M}$ string compactifications that include a KK scalar on ${\cal M}$ which is uncharged under all global symmetries. Computing the leading term in $\la TT\O\ra$ in a $1/\Dg$ expansion amounts to deriving the cubic coupling $\l_{TT\O}$ of the $\alpha'$-corrected effective action on AdS$_{d+1}$. 

One familiar example is the conifold theory, a 4d $SU(N)\times SU(N)$ gauge theory with $\N=2$ SUSY and an $SU(2)\times SU(2)\times U(1)_B$ global symmetry. At strong coupling, the gravity dual is type IIB string theory on AdS$_5\times T^{1,1}$. Examination of the KK spectrum \cite{Gubser:1998vd, Ceresole:1999zs, Ceresole:1999rq} reveals the existence of two singlet conformal primary scalars. The first is the superconformal primary of the multiplet containing the Betti current that generates the $U(1)_B$ symmetry; by SUSY, the scalar thus has $\D=2$. The second is an unprotected scalar with $\D=6$, namely, the $k=0$ member of the $Q^k$ tower of operators in \cite{Ceresole:1999zs};\foot{It is a remarkable fact that the $T^{1,1}$ spectrum includes operators which are not protected by SUSY, but nevertheless acquire irrational, order one anomalous dimensions at strong coupling. The $\D=6$ operator we mention in fact acquires no anomalous dimension at all!} this multiplet is the supersymmetrization of an $F^4$ term. Notably, $Q^{k=0}$ descends from an admixture of the 10d metric and four-form potential; this is fortuitous, because the leading $\a'^3$ correction to the type IIB supergravity action in the metric and five-form sector is known explicitly \cite{paulos}. Reducing it to five dimensions, while keeping the KK scalar fluctuations $Q^{k=0}$, would yield the desired cubic coupling; it would be very interesting to carry this out. Other Sasaki-Einstein compactifications AdS$_5\times SE_5$ should have singlet scalars as well. 

There are also some cases in AdS$_4$ that admit singlet KK scalars.\foot{The case of AdS$_4\times M^{1,1,1}$ is quite similar to AdS$_5\times T^{1,1}$: it also contains singlet scalars, including a Betti scalar with $\D=1$, and one unprotected scalar, this time with $\D=4$. (In \cite{Fabbri:1999mk}, this is a $W$ long vector multiplet with $M_1=M_2=J=0$; see p.16.) Being an M-theory example, however, suppression of $\la TT\O\ra$ scales with an inverse power of $N$.} In fact, this includes one of the most well-studied theories, namely, the $\N=6$ ABJM theories in the `t Hooft limit with $\l\rar\i$, with type IIA dual AdS$_4\times \mathbb{CP}^3$. From Table 1 of \cite{Nilsson:1984bj}, one finds $SU(4)_R$ singlet KK scalars of $\D=4,5,6$. 

In both the conifold and ABJM cases, the gap scale is $\Dg \sim \l^{1/4}$. The first corrections to the type II supergravity actions appear at $O(\alpha'^3)$, which implies that $\la TT\O\ra \lesssim \l^{-3/2}$ for these cases. In the ABJM case, it would be nice to check this prediction/bound for using integrability \cite{Minahan:2008hf, Gromov:2008qe, Klose:2010ki}. We emphasize that there is no analog of this computation in 4d $\N=4$ super-Yang-Mills, where the $TT$ OPE contains no scalar singlets to any order in perturbation theory around $\l=\i$.

\sssec*{Changing $C_T$ under marginal deformations}

If $\O$ is exactly marginal, then $C_{TT\O}$ controls the first-order change in $C_T$ along the line of fixed points,
\e{}{\delta C_T \propto \int d^dx \la T(x_1)T(x_2)\O(x)\ra}
In $d=4$, $C_T\propto c$ cannot vary as a function of exactly marginal couplings in a supersymmetric CFT \cite{grisaru}; but in the absence of supersymmetry, it is not known whether $c$ can change. 
Either way, the suppression $C_{TT\O}\sim \Dg^{-2}$ shows that, in a CFT with a large gap, any possible change of $C_T$ is highly suppressed. See \cite{Nakayama:2017oye} for an example where $\la TT\O\ra$ was used to generate a change in $C_T$ in a bottom-up holographic setting.

\sssec*{Model-building applications}
$\la TT\O\ra$ couplings have various phenomenological and cosmological applications, of which we mention a few here. In \cite{Myers:2016wsu, Lucas:2017dqa}, the coupling \eqr{phicc} was used to generate a vev for $\phi$ in a black hole background. Our result \eqr{ltto} implies that without higher spin fields, this vev must be parametrically small. This in turn implies bounds on viscosity and transport coefficients computed from this coupling (see e.g. 4.44 of \cite{Lucas:2017dqa}). The $TT$ OPE also controls the Renyi entropy under second-order shape deformations (e.g. \cite{Lewkowycz:2014jia, Bianchi:2015liz}). Finally, the inflationary observables discussed in \cite{Cordova:2017zej}, including the tensor tilt of  \cite{Baumann:2015xxa} and scalar-tensor-tensor non-gaussianities of the CMB, can now be linked to the higher spin scale, as for $\la TTT\ra$ in \cite{Camanho:2014apa}. We emphasize that the $\la TT\O\ra$ coupling is simpler than the $\la TTT\ra$ couplings. 

\sssec*{Extremal correlations with the Reggeon}

The above zeroes and poles involved in translating between CFT bases and AdS vertices are examples of a more general statement that holds anywhere along the Regge trajectory: transforming our bounds in the $B$-basis to the standard $C$-basis, and then to analytically continued AdS vertices,
\es{extremal}{B_{T\O j(\nu)} &= g_{\nu}(\DO)C_{T\O j(\nu)}\\
C_{T\O j(\nu)} &= f^{-1}_{\nu}(\DO)\l_{T\O j(\nu)}}
where
\e{}{g_{\nu}(\DO)\propto \G^{-1}\left({3h+i\nu-\DO\o2}\right)~,~~f^{-1}_{\nu}(\DO)\propto \G\left({3h+i\nu-\DO\o2}\right)}
$g_{\nu}(\DO)$ is given in Appendix \ref{app:ChangeBases}. These are nothing but the usual zeroes of extremal correlators at $\DO = \D_T+\D(\nu)$, with one operator analytically continued in spin. So, for instance, in the standard basis, the bound on the anaytically-continued OPE coefficient $C_{T\O j(0)}$ vanishes at $\DO = 3h+2n$ for $n\in \Z_{\geq 0}$.

\ssec{Collider bound}
Next, let us study the stress tensor point, $j(-ih)=2$, without taking large gap, which entails taking $L\sim \log S\gg1$. To extract the conformal collider bounds we will need more information about the diagonal matrix elements at the stress-tensor point and the scaling of phase shift matrix elements at large $L$. To start, we should recall that in impact parameter space the conservation condition for $\<T^{\mu\nu}\O \f\f\>$ becomes
\bea
p_{\mu}\mathcal{B}^{\mu\nu}(p,\bar{p})=0,
\eea
There is a similar condition for $\<T^{\mu\nu}T^{\rho\sigma}\f\f\>$. We will use this condition to work directly in $\mathbb{H}^{d-1}$, which plays the role of transverse space in the dual AdS experiment. To perform this projection, we follow \cite{Costa:2017twz} and write $p=Ee$ and $\bar{p}=\bar{E}\bar{e}$, where $E$, $\bar{E}\geq0$ and 
\bea
e=\frac{1}{r}(1,r^{2}+e_{\perp}^{2},e_{\perp}), \hspace{.5cm} \bar{e}=\frac{1}{\bar{r}}(1,\bar{r}^{2}+\bar{e}_{\perp}^{2},\bar{e}_{\perp}), \hspace{1cm} e,\bar{e}\in \mathbb{H}^{d-1}.
\eea
where we use the hyperbolic metric $ds^{2}=r^{-2}(dr^{2}+de^{2}_{\perp})$. We can transform to the coordinate system given by $p^{\mu}=(E,r,e_{\perp})$ and then conservation implies $\mathcal{B}^{E\mu}(p,\bar{p})=0$. Therefore, for conserved operators it is natural to set $E=1$ and restrict to the transverse space parametrized by $p^{\hat{\mu}}=(r,e_{\perp})$.

In practice, this projection is implemented by acting with the operators given in (\ref{eqn:ReggeDIff_TTO_1}) - (\ref{eqn:ReggeDIff_TTO_3}) and making the following replacements: $z\cdot \hat{p}\rightarrow 0$ and $z\cdot\hat{\bar{p}}\rightarrow -\sinh(L) z\cdot n$ where
\bea\label{nvec}
n_{\hat{\mu}}=\frac{1}{r\bar{r}\sinh(L)}(r-\bar{r}\cosh(L),e_{\perp}-\bar{e}_{\perp}).
\eea
The exact form of $n$ is not important, it will play the same role as the position of the detector operator on the sphere at infinity in the corresponding conformal collider set-up\cite{Hofman:2008ar}. That is, after we restrict the polarizations to the transverse space, $z_{E}=0$, we will vary $z$ relative to $n$, which remains fixed, in order to find the optimal bounds. 

Furthermore, we will also make the replacement $z_{\hat{\mu}}z_{\hat{\nu}}\rightarrow \epsilon_{\hat{\mu}\hat{\nu}}$, where $\epsilon_{\hat{\mu}\hat{\nu}}$ is a symmetric, traceless, transverse tensor, to simplify the presentation. After we have done all of this, we find the diagonal terms are \cite{Costa:2017twz}:

\es{eqn:TTdiagonal}{
\mathcal{D}_{TT}(\nu_0)\Pi_{i\nu_0}(L)\big|_{\nu_0=-ih}=B_{\phi\phi T}B^{(1)}_{TTT}\,\Pi_{h}(L)|\epsilon|^{2}\bigg[1&+t_2\left(\frac{n\cdot\epsilon^{*}\cdot\epsilon\cdot n}{|\epsilon|^{2}}-\frac{1}{d-1}\right)\\&+t_4\left(\frac{|n\cdot\epsilon\cdot n|^{2}}{|\epsilon|^{2}}-\frac{2}{d^{2}-1}\right)\bigg],\\
~~\mathcal{D}_{\O\O}(\nu_0)\Pi_{i\nu_0}(L)\big|_{\nu_0=-ih}=B_{\phi\phi T}B_{\O\O T}\,\Pi_{h}(L),~\quad~~ &}
where
\begin{align}
\frac{B_{\phi\phi T}B^{(1)}_{TTT}}{\hat\chi(\nu)\zeta(\nu,5)}\Big|_{\nu=-ih}&=C_T {\Gamma(2h+6)\Gamma \left(h+6\right)\o \pi^h (2h+1)\Gamma(2h)} \label{eqn:beta_TT_1}
\\
\frac{B_{\phi\phi T}B_{\O\O T}}{\hat\chi(\nu)\zeta(\nu,0)}\Big|_{\nu=-ih}&=C_{\O \O T}\label{eqn:beta_TT_2}
\end{align}
where $\hat\chi$ and $\zeta$ defined in Appendix \ref{app:Definitions} and we have suppressed their dependence on the external operator dimensions. The new off-diagonal term is given by:
\begin{align}
\mathcal{D}_{T\O}(\nu_0)\Pi_{i\nu_0}(L)\big|_{\nu_0=-ih}=B_{\phi\phi T}B_{TT\O}\Pi_{h}(L)[d(d-1)n\cdot\epsilon\cdot n]. \label{eqn:TTO_Even_Off_Diagonal}
\end{align}
In writing down these expressions we have implicitly taken $L$ large so that we can access the stress-tensor point. To derive the optimal bounds we can then decompose $\Dc$ with respect to the $SO(d-2)$ transverse rotational group that leaves $n$ invariant. Given the form of (\ref{eqn:TTO_Even_Off_Diagonal}) we can see it is a singlet under $SO(d-2)$ and therefore is bounded by the effective number of scalars in $\<TTT\>$. It will also be convenient to write
\bea\label{epsrep}
\epsilon_{\hat{\mu}\hat{\nu}}=\frac{1}{2}\left(\epsilon_{1,\hat{\mu}}\epsilon_{2,\hat{\nu}}+\epsilon_{2,\hat{\mu}}\epsilon_{1,\hat{\nu}}\right)-\frac{1}{d-1}g_{\hat{\mu}\hat{\nu}}\epsilon_{1}\cdot\epsilon_{2}~.
\eea 
We derive the bound by setting $\epsilon_1=\epsilon_2=n$. In the standard basis (\ref{eqn:TTO_Std_Basis}) the result is:
\e{ttosingle}{{C_{TT\O}^2\o C_\O}f(\D) \leq n_B}
where $C_\O$ is the norm of $\O$, $f(\D)$ is defined as
\e{fscal}{f(\D) = {\pi^{4h}(2h-1)^3 \Gamma\left(h+1\right)\Gamma\left(2h+1\right)\Gamma\left(\D\right)\Gamma\left(\D-h+1\right) \o 2(h-1)^2 \Gamma^4\left(2+{\D\o2}\right)\Gamma^2\left(h+{\D\o2}\right)\Gamma^2\left(2h-{\D\o2}\right)}}
and $n_B$ is the coefficient of the bosonic structure in the free-field representation of $\la TTT\ra$ in \eqr{tttfree}. The non-trivial, positive function $f(\D)$ arises from the various factors in \eqr{eqn:beta_TT_1}--\eqr{eqn:beta_TT_2}, which come from the transformation from the Regge differential basis to the standard basis and the Fourier integration.

The explicit calculation above was for a single $\O$, but we may generalize to the four-point function $\la \Psi\Psi\phi\phi\ra$ where $\Psi$ includes a sum over all scalar primaries of the theory,
\e{}{\Psi= \sum_i a_i \O_i + a_T T}
At $\nu=-ih$, the phase shift matrix $\Dc$ takes the simple form
\e{}{\Dc(-ih)= \left( 
\begin{array}{ccccc}
\Dc_{TT}(-ih) &\Dc_{T\O_1}(-ih)&\Dc_{T\O_2}(-ih)&\Dc_{T\O_3}(-ih)&\cdots\\
\Dc_{T\O_1}(-ih)&\Dc_{\O_1\O_1}(-ih)&0&0&\cdots\\
\Dc_{T\O_2}(-ih)&0&\Dc_{\O_2\O_2}(-ih)&0&\cdots\\
\Dc_{T\O_3}(-ih)&0&0&\Dc_{\O_3\O_3}(-ih)&\cdots\\
\vdots&\vdots&\vdots&\vdots&\ddots
\end{array}\right)}
The diagonality of the $\Dc_{\O_i\O_j}$ sub-matrix follows from the fact that, for scalar primaries $\O_i$ in an orthogonal basis, $\la T\O_i\O_j\ra \propto \d_{ij}$. Demanding the positivity of each $2\times 2$ principal minor of $\Dc(-ih)$ acting on $\Pi_{h}(L)$ yields \eqr{ttosingle} for each individual $\O$; positivity of the full matrix determinant yields the stronger bound
\e{TTO-Collider-Even}{\sum_\O {C_{TT\O}^2\o C_\O}f(\D) \leq n_B}\label{ttocoll} 

This result was recently derived using the ANEC in a mixed state in \cite{Cordova:2017zej}. There, $f(\D)$ arises from performing integrals over the sphere at null infinity. The fact that Regge constraints at large impact parameter include ANEC constraints is well-known \cite{Costa:2017twz}.\foot{One may also state this directly on the level of the $\phi\phi$ OPE \cite{Afkhami-Jeddi:2017rmx}: in the Regge limit, the leading correction to the identity is a ``shockwave operator,'' which has support on a $(d-1)$-dimensional ball and must be positive when evaluated in perturbative states. In the lightcone limit, this ball localizes on the null line, the shockwave operator becomes the null energy, and Regge positivity becomes the ANEC.}
\sssec{Comments}\label{431}
The bound \eqr{TTO-Collider-Even} was nicely analyzed in \cite{Cordova:2017zej}. So as to avoid redundancy, let us make just a few new comments. 

\sssec*{Zeroes of $f(\D)$}

First, $f(\D)$ has double zeroes at $\D=4h+2n=2d+2n$.  As noted in \cite{Cordova:2017zej}, the zero is required due to the existence of scalar double-trace operators
\e{}{[TT]_n\equiv :T_{\mu\nu}\p^{2n}T^{\mu\nu}:}
which are present in the $TT$ OPE of any large $C_T$ CFT. These operators have dimension
\e{ttnops}{\D_{TT}(n) = 2d+2n+{\g_{TT}(n)\o C_T}+\ldots}
This implies that without these zeroes, the bound \eqr{TTO-Collider-Even} would be violated in a theory with $C_T\sim n_B\rar\i$, because ${C_{TT[TT]_n}^2/ C_{[TT]_n}}\sim C_T^2$. 

Alternatively, one can consider a theory of $n_F\gg1$ free fermions, which has $n_B=0$ but still contains these double-trace operators. Yet another natural way to understand these zeroes, including the fact that $f(\D)$ has a {\it double} zero, was explained earlier using the AdS interpretation of the extremal $\la TT\O\ra$ correlator. 

\sssec*{Large $n$ scaling of mean field theory OPE coefficients}

If we expand \eqr{TTO-Collider-Even} around $C_T=\i$ and sum over operators $[TT]_n$, we find
\e{ttoexp}{\sum_{n=0}^{\Lambda}a_{TT}^{\rm MFT}(n)f''(2d+2n)\g_{TT}^2(n)\leq 2n_B}

where 
\e{attmft}{a_{TT}^{\rm MFT}(n) \equiv {(C_{TT[TT]_n}^{\rm MFT})^2\o C_{[TT]_n}}}
is the normalized squared OPE coefficient for the operators \eqr{ttnops} in the mean field theory (MFT) of stress tensors. The sum is cut off at $n_{\rm max}= \L$, which scales with some power of $C_T$ due to perturbative unitarity \cite{Fitzpatrick:2010zm}. While $a_{TT}^{\rm MFT}(n)$ can be obtained in principle by decomposing the MFT result for $\la TTTT\ra$ into confomal blocks, it is not yet known explicitly; but we can bound its large $n$ growth by expanding the rest of the summand of \eqr{TTO-Collider-Even} at $n\gg1$ and demanding consistency. The idea is to consider a CFT with $C_T\gg1$ but finite $n_B$. An example is a theory of $n_F\gg1$ generalized free spin-1/2 fermions, dual to $n_F$ free fermions in AdS. Ignoring overall constants and further subleading terms,
\e{}{f''(2d+2n\gg1)\sim {16^n n^{-{7d\o2}-4}}}
The large $n$ scaling of  $\g_{TT}(n)$ depends on the details of the CFT. The fastest growth, which will give us the strongest bound, happens when stress tensor exchange dominates the $TT$ interaction, as in holographic $\Dg\gg1$ theories. In this case, \cite{Cornalba:2006xm,Cornalba:2007zb,Kaviraj:2015cxa,Alday:2017gde,Li:2017lmh}
\e{}{\g_{TT}(n\gg1) \sim n^{d-1}}
Then approximating the sum as an integral and demanding finiteness bounds the large $n$ MFT OPE coefficients as
\e{attmftbound}{a_{TT}^{\rm MFT}(n\gg1) \lesssim 16^{-n} n^{{3d\o2}+5}}
This takes a similar parametric form as the known asymptotic behavior of the squared OPE coefficients for the scalar double-trace operators $:\!\phi\p^{2n}\phi:$ in a MFT of scalars $\phi$, of conformal dimension $\D$ \cite{Fitzpatrick:2011dm}:
\e{}{a_{\phi\phi}^{\rm MFT}(n\gg1) \sim 16^{-n} n^{-{3d\o2}+4\D}}
The explicit $a_{TT}^{\rm MFT}(n)$ may in principle be derived using the weight-shifting operators of \cite{Karateev:2017jgd}; it would be nice to carry this out. Furthermore, once we know $a_{TT}^{\rm MFT}(n)$, we can also use this bound to derive constraints on the anomalous dimensions generated from more general solutions to crossing at large $N$.

Note also that \eqr{ttoexp} implies that in a large $C_T$ CFT, $n_B=0$ is only possible if $\g_{TT}=0$. For generic CFTs, where the notion of double-trace is ill-defined, $n_B=0$ implies that the only scalar operators which appear in the $TT$ OPE must have the dimensions of double-trace operators. Presumably, in either case, this is only possible if the theory is free.\foot{We note in passing the similarity of $f(\D)$ to the function appearing in the scalar sum rule of \cite{Gillioz:2016jnn}. Whether those results can be unified with ours deserves to be understood.}
%%%%%%%%%%%%%%%%%%%%%%%%%%%%%%%%%%%%%%%%%%%%%%%%%%%%%%%%%%%%%%%%%%%%%%%%%%%%%%%%%%%%%%%%%%%%%%%%%%%%%%%

%%%%%%%%%%%%%%%%%%%%%%%%%%%%
\section{Holographic and collider bounds on $\la T\O_1\O_2\ra$}
\label{sec:MoreSystems}
%%%%%%%%%%%%%%%%%%%%%%%%%%%%
In this section we will apply our method to other systems of correlators; derive new bounds on three-point couplings for theories with a higher spin gap; and derive conformal collider bounds for general CFTs. Throughout we will assume that the stress tensor Regge trajectory is dominant in the limit $S\rightarrow\infty$ with $L$ held fixed and will use the same logic as in the previous section to derive the bounds.

First, a word on changes of bases and AdS couplings. In every case to follow, we derive bounds on the $\b$-basis of impact parameter space, Regge differential operators. When translated into the standard position space $C$-basis of conformal structures, our bounds become trivial at certain values of $\Delta_\O$, due to the presence of zeroes in the basis change. Likewise, when translating the $C$-basis to the basis of local AdS vertices $\l$, there are poles at these locations. In particular, in computing $\la T \O j(\nu)\ra$ for a traceless, symmetric operator $\O$ of spin-$\ell_\O$ and twist $\t_\O = \DO-\ell_\O$, a slightly modified version of \eqr{extremal} holds for parity-even couplings: 
\es{}{B^{(k)}_{T\O j(\nu)} &= g^{(k)}_{\nu}(\DO)C^{(k)}_{T\O j(\nu)}\\
\l^{(k)}_{T\O j(\nu)} &= f^{(k)}_{\nu}(\DO)C^{(k)}_{T\O j(\nu)}}
where $k$ indexes the independent structures, and
\e{}{g^{(k)}_{\nu}(\DO)\propto \G^{-1}\left({3h+i\nu-\t_\O\o2}\right)~,~~f^{(k)}_{\nu}(\DO)\propto \G^{-1}\left({3h+i\nu-\t_\O\o2}\right)}

This is an exact analog of the situation in the $\ell_\O=0$ case. For parity-odd couplings, the poles are shifted by one, $\t_\O \rar \t_\O-1$, in the above formulas. 
 \begin{figure}
    \centering
       \includegraphics[width = .4\textwidth]{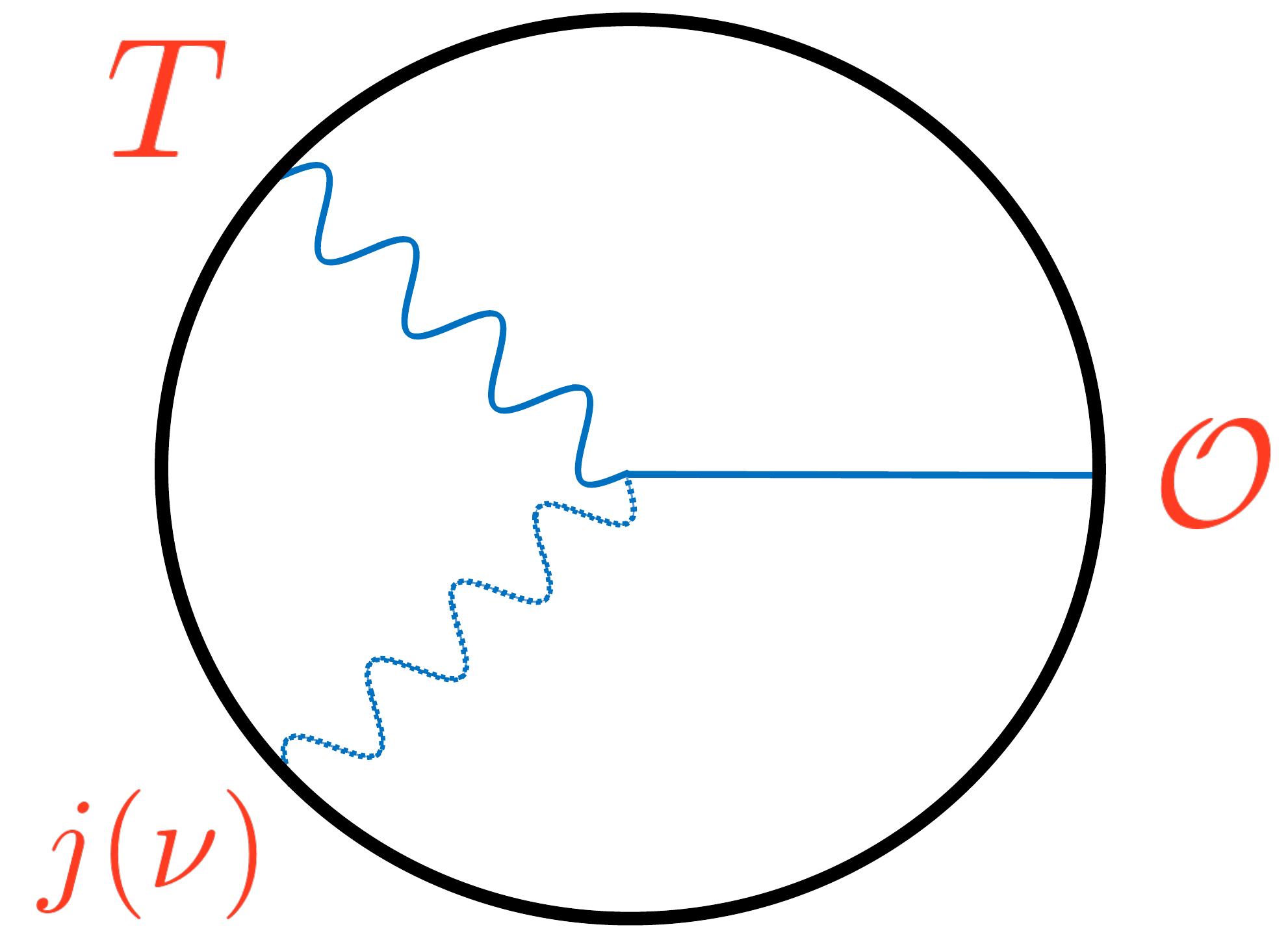}
         \caption{CFT three-point functions $\la{T\O j(\nu)}\ra$, computed holographically from AdS three-point diagrams like the one shown, are ``extremal'' when $\tau_\O = \D_T+\D(\nu)+2n$.}\label{fig4}
\end{figure}

When $\nu=-ih$ and this becomes a $\la TT\O\ra$ correlator, these zeroes/poles at $\t_\O = 4h+2n$ can be understood as arising from consistency of the bounds with the MFT of stress tensors, which contains totally-symmetric spin-$\ell_\O$ double-trace operators comprised of two stress tensors.\foot{For general spin-$\ell_\O$, the structure of poles may depend on the index $k$. For instance, at $\ell_\O=2$, there are two families of totally symmetric double-trace operators, whose conformal dimensions differ by two:

\es{}{
:\!T_{\rho\sigma}\p^{2n}\p_{(\mu}\p_{\nu)}T^{\rho\sigma}\!:~, \quad :\!T_{\rho(\mu}\p^{2n}T^{\rho}_{\nu)}\!:}
Which conformal structures these operators turn on -- and, in turn, for which $k$ there are poles/zeroes in the corresponding functions $f^{(k)}(\DO)$ and $g^{(k)}(\DO)$ -- is a matter of computation. We return to this issue in Section \ref{ttmcoll}.}

In the examples considered below, we will give the explicit form of the relevant double-trace operators. 

%%%%%%%%%%%%%%%%%%%%%%%%%%%%
\subsection{$\<TT\O\>_{\rm odd}$ in $d=3$}
\label{sec:TTO_odd}
%%%%%%%%%%%%%%%%%%%%%%%%%%%%
Here we will again consider the four-point function $\<\Psi\Psi\phi\phi\>$ with

\e{}{\Psi=a_{\O}\O+a_{T}z_{1,\mu}z_{1,\nu}T^{\mu\nu},}
However, now we will focus on the parity-odd part of the three-point function $\<T\O\O_{\Delta,J}\>$, which can only exist in $d=3$. In the standard embedding space basis there are two possible structures:
\e{}{\<T\O\O_{\Delta,J}\> \propto \epsilon(P_{1},P_{2},P_{3},Z_{1},Z_{3})\left(c_{1}V_{1}V_{3}^{J-1}+c_{2}H_{13}V_{3}^{J-2}\right),}
where $\epsilon$ is the 5d Levi-Civita symbol in embedding space. Conservation implies 
\begin{eqnarray}
c_1= c_2\frac{\left(\Delta_\O-\Delta+J+2\right)}{\Delta_\O-\Delta}.
\end{eqnarray}
To match the conventions of \cite{Cordova:2017zej} we define
\bea
C^{\rm odd}_{TT\O}=\frac{1}{4}(c_{2}-c_{1}).
\eea

When the third operator is the stress-tensor ($\Delta=3$ and $J=2$), the above solution automatically guarantees conservation at $P_3$.

The Regge differential operators are 
\begin{align}
\hat{D}_{1}&=\epsilon(z_{1},\hat{p},\nabla) z_{1}\cdot \hat{p},
\\\hat{D}_{2}&=\epsilon(z_{1},\hat{p},\nabla)z_{1}\cdot\nabla,
\end{align}
and we will label their coefficients as $\widetilde{B}^{(i)}_{T\O j(\nu)}$. Conservation of the external stress tensor implies $\widetilde{B}^{(1)}_{T\O j(\nu)}=0$.

\sssec{Large gap}
To derive the holographic bounds, we count derivatives in $\Dc$.  There are two powers of $\nabla$ in $\hat{D}_{2}$, which indicates that at $S\rightarrow \infty$ with $L$ held fixed, the off-diagonal matrix elements grow faster than the diagonal terms by a factor of $L^{-2}$. This implies:
\bea
\widetilde{B}_{T\O j(0)}=0\quad\Rightarrow\quad \widetilde{B}_{T\O T}\sim {\Delta_{gap}^{-2}}
\eea
In terms of the standard basis, this becomes:
\bea
C^{\rm odd}_{T\O j(0)}=0 \hspace{.5cm} \text{except for  } \Delta_{\O}= \frac{11}{2}+2n 
\label{eqn:TTO-Holo-Odd}
\eea

Holographically, the result is that coefficient of the parity-odd coupling $\int \phi \eps_{\mu\nu\lambda\eta}C_{~~\rho\sigma}^{\lambda\eta}C^{\mu\nu\rho\sigma}$ is bounded to scale as $M_{\rm HS}^{-2}$, as indicated in the table in the introduction. There are many known parity-violating $d=3$ CFTs with a large gap. These include the ABJ theories \cite{Aharony:2008gk} of $U(N)_k\times U(M)_{-k}$ Chern-Simons gauge fields coupled to bifundamental matter, and the Gaiotto-Tomasiello theories \cite{Gaiotto:2009mv} that are instead based on $U(N)_{k_1}\times U(N)_{-k_2}$ with $k_1\neq k_2$. Both have AdS$_4\times \mathbb{CP}^3$ supergravity duals in the `t Hooft limit, with $\Dg \sim \l^{-1/4}$.\foot{In the Gaiotto-Tomasiello theory, there are two `t Hooft couplings, $\l_i = N/k_i$. In the approximation in which the zero-form flux $F_0$ is small, $k_1 -k_2\ll k_1+k_2$, so $\Dg\sim \l_1^{-1/4} \sim |\l_2|^{-1/4}$.}

\sssec{Collider bound}
To derive the conformal collider bounds we instead set $\nu=-\frac{3i}{2}$ and consider the full phase shift matrix. To simplify this analysis we will perform the replacement $z_{\hat{\mu}}z_{\hat{\nu}}\rightarrow \epsilon_{\hat{\mu}\hat{\nu}}$, where $\epsilon_{\hat{\mu}\hat{\nu}}$ was defined in \eqr{epsrep}. The optimal bounds are found by choosing $\epsilon_{1}=n$ and $\epsilon_{1}\perp \epsilon_{2}$. This picks out the free fermion structure in $\<TTT\>$. 
In terms of this coefficient, we find
\e{eqn:TTO-Collider-Odd}{(C^{\rm odd}_{TT\O})^{2}f_{\rm odd}(\Delta_\O)\leq n_{F}}
where
\e{}{f_{\rm odd}(\Delta_\O)=\frac{4608\, \pi ^6 \,\Gamma \left(2 \DO-1\right)}{\Gamma^2 \left(\frac{7-\DO}{2} \right) \Gamma^2 \left(\frac{\DO+1}{2}\right) \Gamma^2 \left(\DO+3\right)}}
The bound has zeroes at $\DO=7+2n$, due to the existence of the scalar double-trace operators 
\e{}{[TT]_{n,0}^{\rm odd}=\epsilon_{\mu\rho\sigma}:\!T^{\mu\nu}\p^{2n}\partial^{\rho}T^{\sigma}_{~~\nu}\!:}
in parity-violating large-$C_T$ theories in $d=3$. 
%%%%%%%%%%%%%%%%%%%%%%%%%%%%%%%
\subsection{$\<TTV\>_{\rm odd}$ in $d=4$}
\label{sec:TTV}
%%%%%%%%%%%%%%%%%%%%%%%%%%%%%%%

Here we will study the four-point function $\<\Psi\Psi\phi\phi\>$ with

\e{}{\Psi=a_{T}z_{1,\mu}z_{1,\nu}T^{\mu\nu}+a_{V}z_{2,\rho}V^{\rho},}
where $V$ is an arbitrary vector operator in $d=4$. For the three-point function $\<TV\mathcal{O}_{\Delta,J}\>$, there are two possible parity-odd embedding space structures:
\bea
\<TV\mathcal{O}_{\Delta,J}\>\propto \epsilon(P_1,P_2,P_3,Z_1,Z_2,Z_3)(c_{1}V_1 V_{3}^{J-1}+c_2 H_{13}V_{3}^{J-2}).
\eea
Conservation for the stress tensor implies:
\bea
c_1= c_2 \left(\frac{J+3}{\Delta _V-\Delta}+1\right). \label{eqn:consTTV}
\eea
If we set $\Delta=4$ and $J=2$ the above solution guarantees conservation at $P_3$ as well. Following the conventions of \cite{Cordova:2017zej}, we define the unique OPE coefficient $C_{TTV}$ as
\bea
c_2=\frac{C_{TTV}}{2 \pi ^6}.
\eea
If $V=J$ where $J$ is a conserved, Abelian current, then imposing conservation at $P_2$ does not yield any new conditions. 

There are two Regge differential operators:
\begin{align}
\hat{D}_{1}&=\epsilon(z_1,z_2,\hat{p},\nabla) z_{1}\cdot \hat{p},
\\
\hat{D}_{2}&=\epsilon(z_1,z_2,\hat{p},\nabla) z_1\cdot\nabla.
\end{align}
Conservation in impact parameter space implies that $\widetilde{B}^{(1)}_{TVj(\nu)}=0$.

\sssec{Large gap}

To proceed we need some knowledge about the diagonal phase shift matrix element $\<V|\chi|V\>$, for which we present the relevant details in Appendix \ref{app:VVT}. To derive bounds in holographic theories we will simply need to know that all impact parameter space operators for $\<VVT\>$ with covariant derivatives will be suppressed by $\Delta_{gap}$.

From the above analysis, we also know the only allowed differential operator, $\hat{D}_{2}$, is quadratic in $\nabla$. We therefore find the off-diagonal terms grow faster than the diagonal terms $\<T| \chi |T\>$ and $\<V| \chi |V\>$ by a factor of $L^{-2}$ as $L\rightarrow 0$. This implies
\bea
\widetilde{B}^{(2)}_{TVj(0)}=0\quad\Rightarrow\quad \widetilde{B}^{(2)}_{TTV}\sim{\Delta_{gap}^{-2}}~.
\eea
In terms of the standard OPE basis we have:
\bea
C_{TVj(0)}=0 \hspace{.5cm}\text{except at  } \Delta_{V}=7+2n \label{eqn:TTV-Holo}
\eea

The $\Dg$ suppression of $\la TTV\ra$ and $\la TTJ\ra$ translate, holographically, into a suppression of the coefficient of the AdS$_5$ mixed gauge-gravitational Chern-Simons term,
\e{}{\l_{ARR}\int A\wedge R \wedge R~,}
with the higher spin scale:
\e{}{\l_{ARR} \sim M_{\rm HS}^{-2}}
When $A$ is a gauge field dual to a conserved current $J$, this term captures the mixed anomaly in the dual CFT \cite{AGW, Witten:1998qj},
\e{}{\nabla_{\mu}J^{\mu} \propto \l_{ARR} \eps_{\mu\nu\rho\sigma}R^{\mu\nu}_{~~\eta\zeta}R^{\eta\zeta\rho\sigma}}
If $V$ is not conserved, then $A$ is massive, and this term no longer represents a mixed anomaly in the dual CFT, but may still be present in the bulk. For instance, a bulk gauge symmetry may be broken in the $1/N$ expansion, where a gauge field acquires a mass due to loop corrections. 

It was argued in \cite{Bhattacharyya:2016knk} using a flat space shockwave calculation, following \cite{Camanho:2014apa}, that this coupling violates causality when $A$ is a $U(1)$ gauge field. See \cite{Landsteiner:2011iq,Azeyanagi:2015uoa, Iqbal:2015vka, Landsteiner:2017lwm, Landsteiner:2011cp, Azeyanagi:2015gqa, Landsteiner:2011iq, Nishioka:2015uka} for further applications of this mixed Chern-Simons term to holographic entanglement and chiral vortical transport. Note that if the bulk has $\N=2$ supersymmetry, the $A\wedge R\wedge R$ term is a superpartner of the $\phi R^2$ term \cite{Hanaki:2006pj}, where $A$ is the graviphoton that generates the $U(1)_R$; therefore, in this case, $\l_{TTV}\sim \Dg^{-2}$ follows from $\l_{TT\O}\sim \Dg^{-2}$. 

\sssec{Collider bounds}
To derive the conformal collider bounds we set $\nu=-2i$ and make the same replacement $z_{\hat{\mu}}z_{\hat{\nu}}\rightarrow \epsilon_{\hat{\mu}\hat{\nu}}$, where $\epsilon_{\hat{\mu}\hat{\nu}}$ was defined in \eqr{epsrep}. 

We find the optimal bound by choosing $\epsilon_{1}=n$ and $\epsilon_{2}\perp n$. In terms of the free-field basis for $\la TTT\ra$ introduced in \eqr{tttfree}, and the basis for $\la TVV\ra$ given in \eqr{vvostruc}, the bound may be written as
\e{eqn:TTV-Collider}{{C_{TTV}^{2}\o \left(3-a_{2,V}\right) C_{VVT}}f_1(\Delta_V)\leq n_{F}}
where
\e{eqn:TTV_Collider}{f_1(\Delta_V)=\frac{3^2\, 4^{\Delta _V+4} \Gamma^2 \left(\frac{\Delta _V}{2}+1\right)}{\pi ^7 \left(\Delta _V-4\right){}^2 \Delta _V \left(\Delta _V+1\right) \Gamma^2 \left(\frac{9-\Delta _V}{2}\right) \Gamma^2 \left(\frac{\Delta _V+1}{2}\right) \Gamma^2 \left(\frac{\Delta _V+5}{2}\right)}}
We call this function $f_1(\D_V)$ to indicate its role as the spin-1 version of the function $f(\D)$ entering the scalar $\la TT\O\ra$ bound. The coefficients $C_{VVT}$ and $a_{2,V}$, defined in Appendix \ref{app:VVT}, are linear combinations of the OPE coefficients which appear in $\<VVT\>$. 

The function $f_1(\D_V)$ has similar behavior as $f(\D)$, given in \eqr{fscal}. Here we mention a few properties. At large $\D_V$, ignoring overall factors and a multiplicative oscillating factor accounting for the double-trace zeroes,
\e{}{f_1(\D_V\gg1) \sim {4^{\D_V} \D_V^{-15}}~.}
As in our discussion of Section \ref{431}, this implies a bound on the large $n$ scaling of the MFT OPE coefficients for the parity-odd double-trace vectors,
\e{}{[TT]_{n,1}^{\rm odd}=\epsilon_{\mu\rho\sigma\delta}:\!T^{\mu\nu}\p^{2n}\partial^{\rho}T^{\sigma}_{~~\nu}\!:~,}
though we leave the algebra implicit. The double zeroes at $\Delta_{V}=9+2n$ are the expected double-trace zeroes, due to the existence of $[TT]_{n,1}^{\rm odd}$ in parity-violating large-$C_T$ theories in $d=4$.\foot{The zero at $\Delta_{V}=4$ is not physical: we note, from (\ref{eqn:consTTV}), that at that point we instead have $c_{2}=C_{TTV}=0$ and need to use $c_{1}$ to parametrize $\<TTV\>$. } Finally, let us also quote the expansion around the unitarity bound $\D_V=3$,
\e{}{f_1(3+\eps) \approx {48\o \pi^6}(1+3\eps+O(\eps^2))}

We can specialize to the case where $V$ is a conserved current $J$, and parametrize $\<JJT\>$ by its free field structures
\bea
\<JJT\>=Q^{2}_{WF}\<JJT\>_{WF}+Q^{2}_{CB}\<JJT\>_{CB},
\eea
where, following \cite{Cordova:2017zej}, the subscripts ``$WF$" and ``$CB$" refer to the three-point function structures found in free field theories of Weyl fermions and free complex bosons, respectively. Then the bound above becomes
\bea
C_{TTV}^{2}\leq n_{WF}Q^{2}_{WF}. \label{eqn:TTJ-Collider}
\eea
where $n_{WF} = {n_F/2}$ parameterizes the $\la TTT\ra$ structure in a theory of Weyl fermions. This matches the result found in \cite{Cordova:2017zej}. One can think of \eqr{eqn:TTV-Collider} as a generalization of that result. This becomes relevant when, for example, $J$ is approximately conserved, but becomes non-conserved in perturbation theory in some parameter. The example of $1/N$ gauge symmetry-breaking was mentioned above. In these situations, \eqr{eqn:TTV-Collider} is the appropriate bound on the cubic couplings $\la TTV\ra$. 

%%%%%%%%%%%%%%%%%%%%%%%%%%%%%%%
\ssec{$\<TV\O\>$}
\label{sec:TVO}
%%%%%%%%%%%%%%%%%%%%%%%%%%%%%%%

We will now consider the stranger case of the four-point function $\<\Psi\Psi\phi\phi\>$ where 
\e{}{\Psi=a_{V}z_{1,\mu}V^{\mu}+a_{\O}\O~.}
At a technical level this is the simplest case to consider. On the other hand, $\<TV\O\>$ is constrained to vanish for generic operator dimensions $\D_\O$ and $\D_V$, as we show below. Nevertheless, we analyze it for the sake of completeness.

%%%%%%%%%%%%%%%%%%%%%%%%%%%%%%%
\sssec{Parity-even}
\label{sec:TVO_Even}
%%%%%%%%%%%%%%%%%%%%%%%%%%%%%%%
$\<V\O\mathcal{O}_{\Delta,J}\>$ has two independent parity-even structures
\bea
\<V\O\mathcal{O}_{\Delta,J}\>\propto c_{1}V_{1}V_{3}^{J}+c_{2}H_{13}V_{3}^{J-1}.
\eea
When $\mathcal{O}_{\Delta,J}=T$, conservation for the stress tensor implies
\bea
c_1&=&\frac{1}{2} c_2 \left(-d \Delta _V+d \Delta _\O+2\right),
\\
\Delta_{\O}&=&\Delta_{V}\pm 1.\label{dvdo}
\eea
We will henceforth rename $c_2=C_{V\O T}$.

The typical interacting CFT will not possess two operators $\O$ and $V$ which satisfy the requirement \eqr{dvdo}, but let us proceed.\foot{In \cite{Giombi:2011rz}, it was shown that found that a similar three-point function $\la \widehat TV\O\ra$ is non-zero in a free theory of $U(N)$ bosons or fermions, where $ \widehat T$ is a conserved spin-2 operator that is charged under the global symmetry. However, here we are considering $\la TV\O\ra$, which was shown in \cite{Giombi:2011rz} to vanish in these free theories. There may also be other interesting cases where a similar analysis to what follows is useful, e.g. if $V$ and $\O$ are charged under some global symmetry and couple to a subleading Regge trajectory.} Furthermore, if $V$ is a conserved current $J$, only the solution $\Delta_{\O}=d-2$ is allowed. We do not know of any interacting CFT where $\la TJ\O\ra\neq 0$ -- indeed, there is an argument in \cite{Bzowski:2013sza} that $\la TJ\O\ra$  vanishes identically.

These caveats aside, we find the following Regge impact parameter space operators:
\begin{align}
\hat{D}_{1}=&z_{1}\cdot \hat{p},
\\
\hat{D}_{2}=&z_{1}\cdot\nabla.
\end{align}

Since the external operators do not satisfy any conservation conditions, we have to be more careful about the conservation conditions in the $B$ basis. That is, we can only relate $B^{(1)}_{V\O j(\nu)}$ and $B^{(2)}_{V\O j(\nu)}$ at the stress tensor point $\nu= -id/2$. For generic $\nu$ there are in principle no relations between different $B^{(i)}_{V\O j(\nu)}$. Therefore, this case requires a slightly different procedure. First we can bound $B^{(2)}_{V\O j(\nu)}$ at the intercept point $\nu=0$, because $\hat{D}_{2}$ is linear in $\nabla$. This implies
\e{eqn:TVO-Holo-Even}{B^{(2)}_{V\O j(0)}=0 \quad \Rightarrow \quad C_{V\O j(0)}=0}
Since $j(\nu)$ is even in $\nu$, this implies that at the stress-tensor point we have $B^{(2)}_{V\O T}\sim \Delta_{gap}^{-2}$. In this case there are no subtleties with zeroes when we convert to the standard basis for both solutions to conservation. At the stress-tensor point, $B^{(1)}_{V\O j(\nu)}$ and $B^{(2)}_{V\O j(\nu)}$ are proportional to each other, so this is sufficient to show  $B^{(1)}_{V\O j(\nu)}$ is also suppressed.

To derive the conformal collider bounds we set $\nu=-id/2$ and choose $z=n$. When we choose the solution corresponding to $\Delta_{\O}=\Delta_{V}+1$, the collider bound is:
\bea
C_{V\O T}^2\leq \frac{4 \pi ^{-\frac{d}{2}} \Gamma \left(\frac{d+2}{2}\right) C_{VVT} \left((d-2) a_{2, V}+d-1\right)}{(d-1)^4 \left(\Delta _V+1\right)}, \label{eqn:TVO-Collider-Even_pt1}
\eea
where $a_{2,V}$ and $C_{VVT}$ are the same linear combination of OPE coefficients for $\<VVT\>$ which appeared in (\ref{eqn:TTV_Collider}) and were defined in Appendix \ref{app:VVT}.

When we choose the solution corresponding to $\Delta_{\O}=\Delta_{V}-1$ the collider bound is
\bea
C_{V\O T}^2\leq \frac{4 \pi ^{-\frac{d}{2}} \Gamma \left(\frac{d}{2}+1\right) \Delta _V C_{VVT} \left((d-2) a_{2, V}+d-1\right) \left(d-2 \Delta _V\right)}{(d-1)^4 \left(\Delta _V^2-1\right) \left(d-2 \Delta _V-2\right)}.  \label{eqn:TVO-Collider-Even_pt2}
\eea
When $V$ is actually a conserved current $J$ this bound becomes:
\bea
C_{J\O T}^2\leq \frac{d \pi ^{-2 d} Q^{2}_{F} \Gamma \left(\frac{d}{2}+1\right) \Gamma \left(\frac{d}{2}\right)^3}{4 (d-1)^2}.  \label{eqn:TJO-Collider}
\eea
where $Q^{2}_{F}=Q^{2}_{WF}/2$ gives the contribution of the effective number of Dirac fermions to $\<JJT\>$.
%%%%%%%%%%%%%%%%%%%%%%%%%%%%%%%
\sssec{Parity-odd}
\label{sec:TVO_Odd}
%%%%%%%%%%%%%%%%%%%%%%%%%%%%%%%
We will now repeat the above analysis in $d=3$ where we can also write down the parity-odd structure:
\e{}{\<V\O\mathcal{O}_{\Delta,J}\>_{\rm odd}\propto \widetilde{C}_{V\O \O_{\Delta,J}}\epsilon(P_1,P_2,P_3,Z_1,Z_3)V_{3}^{J-1}.}
%\eea
%
If $\mathcal{O}_{\Delta,J}=T$ then conservation implies $\Delta_{V}=\Delta_{\O}$. There is also a unique Regge differential operator in impact parameter space:
\bea
\hat{D}=\epsilon(z_{1},\hat{p},\nabla).
\eea
Since the only allowed differential operator is linear in $\nabla$, we see that in the limit $L\rightarrow 0$, the off-diagonal terms will grow faster by a factor of $L^{-1}$. This immediately implies:
\bea
\widetilde{B}_{V\O j(0)}=0 \quad \rightarrow \quad \widetilde{C}_{V\O j(0)}=0.  \label{eqn:TVO-Holo-Odd}
\eea
At the stress tensor point, $\widetilde{B}_{V\O T}\sim \Delta_{gap}^{-2}$. Once again, here there are no subtleties when going between the bases. 

To derive the conformal collider bound we choose $z_{1}\perp n$ and find:
\bea
\widetilde{C}_{V\O T}^{2}\leq \frac{3 \Delta _V \left(2-a_{2,V}\right) C_{VVT}}{8 \pi  \left(\Delta _V+1\right) \left(2 \Delta _V-1\right)}.  \label{eqn:TVO-Collider-Odd}
\eea
If $V=J$ is a conserved current, the bound becomes
\bea
\widetilde{C}_{J\O T}^{2}\leq \frac{9 Q^{2}_{F}}{128 \pi ^4}.  \label{eqn:TJO-Collider-Odd}
\eea

%%%%%%%%%%%%%%%%%%%%%%%%%%%%%%%
\subsection{$\<TTM\>$ for spin-2 $M$}
\label{sec:TTM}
%%%%%%%%%%%%%%%%%%%%%%%%%%%%%%%

Finally we study the four-point function $\la \Psi\Psi\phi\phi\ra$ with
\e{}{\Psi=a_{T}z_{1,\mu}z_{1,\nu}T^{\mu\nu}+a_{M}z_{2,\rho}z_{2,\sigma}M^{\rho\sigma},}
where $M$ is a general spin-2 operator of dimension $\D_M\geq d$, that is a singlet under all global symmetries. This will yield bounds on three-point functions $\la TTM\ra$. 

When studying $TM\rightarrow \mathcal{O}_{\Delta,j(\nu)}\rightarrow \phi\phi$, there are at least two Regge trajectories of which to be aware. One is the stress tensor trajectory, whose OPE coefficients we want to bound. The other is the $M$-trajectory, which will generically appear in the above channel. We will assume that the stress tensor trajectory is dominant in the limit $S\rightarrow\infty$ with $L$ fixed.

In the standard basis, there are 11 structures for $\la TMT\ra$:
\bea
\<TMT\>\propto \underset{n_1,n_2,n_3}{\sum}C^{(n_1,n_2,n_3)}_{TMT}V_1^{2-n_{2}-n_{3}}V_{2}^{2-n_{1}-n_{3}}V_{3}^{2-n_{1}-n_{2}} H_{23}^{n_1} H_{13}^{n_2}H_{12}^{n_{3}},
\eea
where $n_{i}+n_{j}\leq 2$ for $i\neq j$. After imposing conservation we find two solutions, which we can parametrize by $C^{(0,0,0)}_{TMT}$ and $C^{(1,0,1)}_{TMT}$. Our method involves studying $\la TM\O_{\D,J}\ra$, for which there are 14 structures for general $J$; in impact parameter space, these are spanned by
\bea
\hat{D}_{n_1,n_2,n_{12}}=(z_1\cdot z_2)^{n_{12}}(z_1\cdot \hat{p})^{2-n_{12}-n_{1}}(z_2\cdot \hat{p})^{2-n_{12}-n_2}(z_1\cdot\nabla)^{n_1}(z_2\cdot\nabla)^{n_2},
\eea
where $n_{1}+n_{12}\leq2$ and $n_{2}+n_{12}\leq2$. When $J\leq4$, the $\hat{D}_{n_1,n_2,n_{12}}$ do not all create linearly independent structures.

\sssec{Large gap}

We start by considering a large gap theory where $\D_M$ is finite in the limit $\Delta_{gap}\rightarrow\infty$. The resulting bounds were derived in flat space in \cite{Camanho:2014apa} by studying $2\rightarrow2$ graviton scattering.\foot{See also \cite{KPZ2017} for similar bounds on $\<JJM\>$ derived by studying $\<JJJJ\>$.} The analogous analysis in the CFT would entail studying $\<TTTT\>$. One advantage of the present approach is that it treats bounds on $\<TTM\>$ on an equal footing with all the other bounds: in all cases, the $\phi$ operators produce the phase shift operator $\chi$, whose matrix elements we are directly constraining. 

For simplicity we impose the unitarity condition in impact parameter space and restrict to transverse polarizations for both $T$ and $M$. Transversality involves taking $z_1\cdot \hat p \rar 0$ and $z_2\cdot \hat p\rar0$, so only the differential operators $\hat{D}_{0,0,2}$, $\hat{D}_{1,1,1}$, and $\hat{D}_{2,2,0}$ can produce a non-zero phase shift. Because $\hat{D}_{1,1,1}$, and $\hat{D}_{2,2,0}$ have two and four derivatives, respectively, the unitarity condition at $L\ll1$ implies
\begin{align}
B^{(2,2,0)}_{TMj(0)}=0 \hspace{1cm} B^{(1,1,1)}_{TMj(0)}=0,
\end{align}
As in previous sections, we are using the fact that the diagonal three-point couplings $B_{TTj(0)}$ and $B_{MMj(0)}$ have terms with no derivatives.\foot{For details on $\Dg$ scaling of $\<MMT\>$ structures at large gap, see Appendix \ref{app:MMT}.} Further imposing conservation on $T$ fixes the other differential operators in terms of these two couplings, thus implying that all $B^{(i,j,k)}_{TMj(0)}=0$ at $\Dg\gg1$.  In terms of the OPE coefficients these bounds become:
\bea
C^{(n_1,n_2,n_3)}_{TMj(0)}=0 \hspace{.5cm}\text{except at   }\Delta_{M}=\frac{3d}{2}+2+2n \text{, with } n\in\mathbb{Z}_{n\geq0}. \label{eqn:TTM-Holo}
\eea 
At the stress-tensor point, this implies
\begin{align}\label{553}
B^{(1,1,1)}_{TMT}\sim{\Delta_{gap}^{-2}~,\hspace{1cm}B^{(2,2,0)}_{TMT}\sim{\Delta_{gap}^{-4}} } ~.
\end{align}
This matches the result of \cite{Camanho:2014apa}. In $d=3$, the $B^{(1,1,1)}_{TMT}$ structure is not allowed, and we find the same constraint on $B^{(2,2,0)}_{TMT}$.

\sssec{Collider bound}\label{ttmcoll}
Next, we will derive collider bounds on $\<TTM\>$. Without imposing large gap, the matrix $\Dc$ is rather complicated, involving several couplings on both the diagonal and off-diagonal. However, we will take a technical shortcut to derive some bounds, postponing a fuller approach to future work.

The shortcut is to only consider transverse polarizations for $M$. This is, of course, not the most general configuration. Choosing the polarization tensors $\eps_{1,2}$ to be real and transverse, we can immediately write our matrix $\Dc$ as
\es{}{&\mathcal{D}_{ij}(\nu_0)\Pi_{i\nu_0}(L)\big|_{\nu_0=-ih}=B_{\phi\phi T}B^{(1)}_{ij T}\,\Pi_{h}(L)\epsilon_i\cdot\epsilon_j \times\\
&\bigg(\delta^{ij}+t^{(ij)}_2\left(\frac{n\cdot\epsilon_i\cdot\epsilon_j\cdot n}{\epsilon_i\cdot\epsilon_j}-\frac{1}{d-1}\right)+t^{(ij)}_4\left(\frac{n\cdot\epsilon_i\cdot n \ n\cdot\epsilon_j\cdot n}{\epsilon_i\cdot\epsilon_j }-\frac{2}{d^{2}-1}\right)\bigg),}
where $i,j=T$ or $M$. In the above expression we have implicitly taken a large $L$ limit, so the coefficients $t_{2,4}^{(ij)}$ are independent of $L$. 

The simplicity follows from our choice of transverse polarization for $M$: on the level of this matrix, we are essentially ignoring its non-conservation. $\Dc_{TT}$ was given in \eqr{eqn:TTdiagonal}, and $\Dc_{MM}$ has an identical functional form: while $\la TMM\ra$ has six linearly independent tensor structures after imposing permutation symmetry of $M$ and conservation of $T$, imposing transversality for $M$ ensures that only three structures participate. Finally, $\Dc_{TM}$ and $\Dc_{MT}$ have only two structures, and no diagonal piece because $\la T(P_1;Z_1)M(P_2;Z_2)\ra=0$.\foot{The coefficients $t_2^{(ij)}$ and $t_4^{(ij)}$ can be related, if desired, to the standard basis. We provide this (long) relation for $t_2^{(TM)}$ and $t_4^{(TM)}$ in Appendix \ref{app:TTM}, as well as our normalization conditions for $B^{(1)}_{TMT}$ and $B^{(1)}_{MMT}$.}

We now impose non-negativity of the principal minors of $\Dc$. $\Dc_{TT}\geq 0$ and $\Dc_{MM}\geq 0$ yields collider bounds in the pure states $\Psi=T$ \cite{Hofman:2008ar} and $\Psi=M$ \cite{Komargodski:2016gci}, respectively. The new bounds come from non-negativity of the full matrix determinant, $\det \Dc\geq 0$. After setting some normalizations (see Appendix \ref{app:TTM}), this yields three constraints, one for each structure in $\la TTT\ra$:

\begin{align}
\frac{\left((d+1)(d-3) t_{2}^{(TM)}+((d-1) d-4) t_{4}^{(TM)}\right)^2}{\left((d+1) ((d-3) t_{2}^{(MM)}+(d-1))+((d-1) d-4) t_{4}^{(MM)}\right)E^{(MM)}}\, f_{2}(\Delta_{M}) \leq& \ n_{B}  \label{eqn:TTM_Bound_1} \\ \nonumber \\
%%%%%%%%%%%%%%%%%%%%%%%%%%%%%%%%%%%%%%%
\frac{(d+1)\left((d+1)(d-3) t_{2}^{(TM)}-4 t_{4}^{(TM)}\right)^2 }{(d-1)\left((d+1)((d-3) t_{2}^{(MM)}+2 (d-1))-4 t_{4}^{(MM)}\right)E^{(MM)}
}\,f_{2}(\Delta_{M})\leq& \ n_{F}  \label{eqn:TTM_Bound_2} \\ \nonumber \\
%%%%%%%%%%%%%%%%%%%%%%%%%%%%%%%%%%%%%%%
\frac{(d-3)\left((d+1) t_{2}^{(TM)}+2 t_{4}^{(TM)}\right)^2 }{(d-1)\left((d+1) (-t_{2}^{(MM)}+(d-1))-2 t_{4}^{(MM)}\right)E^{(MM)}
}\,f_{2}(\Delta_{M})\leq& \ n_{V}  \label{eqn:TTM_Bound_3}
\end{align}
where 
\e{}{f_{2}(\Delta_{M})=\frac{4 \pi ^{\frac{3 d}{2}} \Gamma ^2\left(\frac{d}{2}+1\right) \Gamma (d+1) \Gamma (\Delta_{M}+5) \Gamma \left(-\frac{d}{2}+\Delta_{M}+5\right)}{(d-2) (d+1) (d+5) (d+10) \Gamma^2 \left(\frac{\Delta_{M}}{2}+5\right) \Gamma ^2\left(\frac{\Delta_{M}}{2}+1\right) \Gamma^2 \left(d-\frac{\Delta_{M}}{2}+1\right) \Gamma^2 \left(\frac{d+\Delta_{M}}{2}+5 \right)}\nonumber}
We call this function $f_2(\D_M)$ to indicate its role as the spin-2 version of the function $f(\D)$ entering the scalar $\la TT\O\ra$ bound. It obeys $f_2(\D_M)\geq 0$ for all unitary $\D_M\geq d$. In these bounds, we have parametrized $\<TTT\>$ by its free field structures \eqr{tttfree}. The factors in parenthesis in the denominators of \eqr{eqn:TTM_Bound_1}--\eqr{eqn:TTM_Bound_3} are constrained to be non-negative, due to $\Dc_{MM}\geq 0$. We have included an overall constant $E^{(MM)}$, which fixes the norm of $M$ and is thus constrained to be positive for all unitary $\D_M\geq d$; we find it convenient to keep this factor explicit. 

First, we note the consistency check that the last bound disappears when $d=3$, as expected: there is no $n_{V}$ structure in $d=3$. In addition, in the remaining bounds, the $t^{(ij)}_{2}$ drop out, leaving us with just $t^{(ij)}_{4}$, correctly reflecting the reduction in the number of tensor structures.

Let us now analyze the (double) zeroes of $f_{2}(\D_M)$. These occur at $\D_M=2d+2+2n$, the conformal dimensions of MFT double-trace operators of the form
\e{}{\!:T^{\rho\sigma}\partial^{\mu}\partial^{\nu}\partial^{2n}T_{\rho\sigma}:\!}
On the other hand, MFT also has a spin-2 double-trace operator of $\D_M=2d$, of the form 
\e{tt2d}{\!:T^{\rho(\mu}T_{~~\rho}^{\nu)}\!:}
As in previous examples, the left-hand sides of the bounds scale as $C_T^2$, while the right-hand sides scale as $C_T$ at most; so $f_{2}(2d)\neq 0$ seems to present a contradiction. A resolution can be seen by passing to the standard basis using \eqr{t2tm}: at $\D_M=2d$, the coefficients $t_{2,4}^{(TM)}$ are functions of only one of the two independent $\la TMT\ra$ OPE coefficients, namely, $C^{(0,0,0)}_{TMT}$. The second coefficient $C^{(1,0,1)}_{TMT}$ remains unconstrained. This implies that in MFT, the operator \eqr{tt2d} must have vanishing $C^{(0,0,0)}_{TMT}$, but $C^{(1,0,1)}_{TMT}$ can be nonzero. It would be worthwhile to confirm this explicitly by performing an OPE decomposition of the MFT result for $\<TTTT\>$.

It is also interesting to study what happens for free theories. Consider free bosons for concreteness, with $n_F=n_V=0$. All operators $M$ must obey
\e{freecond}{\D_M=2d+2+2n~~~~~\text{and/or}~~~~~\la TTM\ra=0~.}
Even though $n_B\neq 0$, the bosonic bound \eqr{eqn:TTM_Bound_1} will still be saturated if the ANEC is saturated in the pure $\Psi=M$ state (i.e. $\Dc_{MM}=0$). This is still consistent with a nonzero result for $\la TMM\ra$ for the $\D_M\neq 2d+2+2n$ operators, because the denominator of \eqr{eqn:TTM_Bound_1} depends on a linear combination of three $\la TMM\ra$ OPE coefficients. Compared to the saturation of the scalar $\la TT\O\ra$ bound \eqr{TTO-Collider-Even}, in which both sides are nonzero, this is a novel mechanism.

We emphasize that we have restricted to transverse polarizations for $M$. We may also expect to derive interesting new bounds from the ANEC by considering longitudinal polarizations \cite{Komargodski:2016gci,Hartman:2016lgu}. We hope to return to this problem in future work \cite{WIP}.

\ssec{Application: $n_Bn_Fn_V>0$ in interacting CFTs}
We now use these bounds to argue that
\e{}{n_Bn_Fn_V>0} 
in interacting CFTs. This was conjectured by \cite{zhib}, who argued using different methods that at least two of $n_B, n_F$ and $n_V$ must be nonzero in interacting CFTs.

It is clear that if even a single spin-2 operator $M$ gives a nonzero contribution to the left-hand side of one of the bounds \eqr{eqn:TTM_Bound_1}--\eqr{eqn:TTM_Bound_3} the corresponding parameter $n_B, n_F$ or $n_V$ must be nonzero. Therefore, proving that $n_Bn_Fn_V>0$ in a given CFT boils down to proving the necessity of such contributions to the $TT$ OPE. 

To warm up, let us set two of the parameters to zero. As noted above, every operator $M$ must then obey \eqr{freecond}. These conditions are hallmarks of free theories: the typical CFT spectrum is irrational, and the condition that $\la TTM\ra=0$ for all operators is highly non-generic. We have reproduced the conclusion of \cite{zhib} with this argument; in a moment, we will give a stronger one.

Let us now set only one parameter to zero, say, $n_F=0$ for concreteness. Then every operator $M$ must obey
\e{nf0}{\D_M=2d+2+2n~~~~~\text{and/or}~~~~~t_2^{(TM)} = {4\o (d+1)(d-3)}\,t_4^{(TM)}}
This, too, is highly non-generic.

To strengthen what ``generic'' means, and to establish that there must exist non-conserved spin-2 operators appearing in the $TT$ OPE, we now appeal to the analyticity in spin of CFT operator data. In \cite{Caron-Huot:2017vep}, it was shown that the OPE data appearing in a scalar OPE is an analytic function of spin $J$ for all $J>1$. More precisely, a scalar four-point function $\la \phi\phi\phi'\phi'\ra$, is characterized by an ``OPE function'' $c(J,\D)$, an analytic function of $J$ which is meromorphic for real integer $J$ and has poles at physical operator dimensions with residues
\e{}{c(J,\D)\Big|_{\O_{\D_\O,J}} \sim {C_{\phi\phi\O_{\D,J}}C_{\phi'\phi'\O_{\D_\O,J}}\o \D-\D_\O}}
This implies the remarkable fact that operators contributing to a given four-point function can be organized in families, extending from asymptotically large $J$ all the way down to (but not including) $J=1$. In the proof of \cite{Caron-Huot:2017vep}, the bound $J>1$ follows from the statement that a conformal block with $J\leq 1$ does not grow in the Regge limit. (See also \cite{ssw}.)

Every CFT contains infinite towers of multi-twist operators at asymptotically large spins \cite{light1,light2}. One can label families of CFT operators by their large spin representatives.\foot{See \cite{dsdi} for a hands-on example of this approach in the 3d Ising model, and \cite{Alday:2016jfr, Alday:2016njk} for a closely related approach.} Analyticity in spin implies that these families extend all the way down through $J=2$. This implies that an infinite number of spin-2 operators appears in a $\phi\phi$ OPE.\foot{We are making an assumption here: that an infinite number of families do not decouple from the OPE as the spin is decreased from infinity. This seems impossible, but would not violate analyticity. For instance, an infinite number of residues could vanish. The full range of allowed behaviors is not yet understood. Note that this does not happen for free theories. We thank David Simmons-Duffin and Simon-Caron-Huot for discussions on this.} 

This discussion applied to the $\phi\phi$ OPE, but the same physics applies to the $TT$ OPE. An analogous formula can be derived for spinning correlators \cite{dsdwip}, including for $\la TTTT\ra$. Following the logic of \cite{Caron-Huot:2017vep}, the analyticity of the $TT$ OPE data will extend to spins $J\geq J'$, where $J'$ is the lowest spin for which the spinning conformal blocks grow in the Regge limit. This growth is determined by the spin of the {\it internal} operator, so $J'=1$. (For an explicit example, see e.g. Appendix B of \cite{Afkhami-Jeddi:2016ntf} for the Regge limit of the $T$ exchange block in a $\la TT\phi\phi\ra$ correlator.) In particular, this includes $J=2$. 

We conclude that an infinite number of operators $M$ appears in the $TT$ OPE. Therefore, a condition like \eqr{nf0} would require fixing an infinite set of OPE data. We take this to imply that $n_Bn_Fn_V>0$ in any interacting CFT. In $d=3$, where only $n_B$ and $n_F$ bounds survive and $t_4^{(TM)}$ is the lone structure in $\la TTM\ra$, the argument is even stronger: a theory with $n_Bn_F=0$ necessarily obeys the ``free conditions'' \eqr{freecond} for all operators $M$.\foot{We note that the numerical data of \cite{krav} supports this conclusion -- see Figure 13.}

%%%%%%%%%%%%%%%%%%%%%%%%%%%%%%%
\section{Final thoughts}
%%%%%%%%%%%%%%%%%%%%%%%%%%%%%%%
\label{sec:Discussion}

Overall, our results at large higher spin gap, following up on \cite{Heemskerk:2009pn, Camanho:2014apa,Caron-Huot:2017vep, KPZ2017, Costa:2017twz, Afkhami-Jeddi:2017rmx}, 
constitute a significant step toward proving the sufficiency of the gap condition for holographic emergence. We have proven this for all three-point functions $\la T\O_1\O_2\ra$ when the $\O_i$ are scalars, vectors or the stress tensor. We also proved bounds on $\<TTM\>$ when $M$ is a non-conserved spin-2 operator. A full proof would also include mixed symmetry tensor fields (in $d>3$), and the expected suppression of all higher-derivative contact terms of the low-spin fields. It would be satisfying to prove the correspondence between AdS derivatives and CFT powers of $\Dg$ more abstractly. It would be also be interesting to understand in more detail the structure of the leading Regge trajectory beyond the stress tensor point. In particular, the bound derived at the intercept can constrain not only the stress tensor, but also the spin-four operator on the same trajectory \cite{Costa:2017twz}. Given the importance of both the gap scale in deriving bounds on the $TT$ OPE and the role of higher-spin operators in restoring causality and unitarity, we can also expect to derive powerful new constraints on this operator and possibly the entire trajectory.

We believe the conformal collider bounds presented here are likely to be the tip of an iceberg in deriving universal constraints on CFT data. The centrality of the $TT$ OPE in CFT encourages further study. In this paper we have focused on operators of low spin, in part to understand holographic CFTs, but there remains a trove of new bounds to be discovered for higher spin operators in general. We plan to return to this analysis \cite{WIP} in order to further delineate the space of CFTs and the structure of the $TT$ OPE.

\sec*{Acknowledgments}
We thank Clay Cordova, Simone Giombi, Diego Hofman, Petr Kravchuk, Juan Maldacena, Joao Penedones, David Simmons-Duffin,  Kostas Skenderis and Joaquin Turiaci for helpful discussions. We also thank Joaquin Turiaci for comments on a draft. We gratefully acknowledge support from Johns Hopkins University during the March Workshop on Quantum Gravity and the Bootstrap, where this work was initiated; from the Simons Summer Workshop at the Simons Center for Geometry and Physics, Stony Brook University; and from the Princeton Center for Theoretical Science. EP is supported in part by the Department of Energy under Grant No. DE-FG02-91ER40671, and by Simons Foundation grant 488657 (Simons Collaboration on the Nonperturbative Bootstrap). DM is supported by NSF grant PHY-1350180 and Simons Foundation grant 488651.

\appendix

%%%%%%%%%%%%%%%%%%%%%%%%%%%%%%%
\section{Embedding Space}
\label{app:Embedding}
%%%%%%%%%%%%%%%%%%%%%%%%%%%%%%%

We use the standard embedding space formalism as presented in \cite{Costa:2011dw, Costa:2011mg}. We introduce an embedding space $\mathbb{R}^{2,d}$ and lift the vectors $x^{\mu}$ and $z^{\mu}$ as
\bea
P=(P^+,P^-,P^{\mu})=(1,x^2,x^\mu), \hspace{1cm} Z=(0,2x\cdot z,z^\mu),
\eea
with the metric given by $P\cdot P=-P^+P^-+\eta_{\mu\nu}P^{\mu}P^{\nu}$. We will also define $P_{ij}=-2P_{i}\cdot P_{j}$ which projects down as $P_{ij}\rightarrow x_{ij}^{2}$ when going back to physical space.

\ssec*{Standard basis}
The standard parity-even structures are:
\es{}{H_{ij}&=-2[(Z_{i}\cdot Z_{j} P_{i}\cdot P_{j}-P_{i}\cdot Z_{j} P_{j}\cdot Z_{i})]~,
\\ \\
V_{i,jk}&=\frac{Z_{i}\cdot P_{j} P_{i}\cdot P_{k}-Z_{i}\cdot P_{k} P_{i}\cdot P_{i}}{P_{j}\cdot P_{k}}~.}

For most operators we will use the normalization:
\bea
\<\O_{\Delta,J}(P_1;Z_1)\O_{\Delta,J}(P_2;Z_2)\>=\frac{H_{12}^{J}}{P_{12}^{\Delta+J}}. \label{eqn:TwoPtNorm}
\eea
The exceptions will be conserved currents and the stress tensors, where we will define $C_{J}$ and $C_{T}$ as follows:
\bea\label{ttnorm}
\<J(P_1;Z_1)J(P_2;Z_2)\>=C_{J}\frac{H_{12}}{P_{12}^{d}}, \hspace{1cm} \<T(P_1;Z_1)T(P_2;Z_2)\>=C_{T}\frac{H_{12}^{2}}{P_{12}^{d+2}}.
\eea
When discussing three-point functions, for convenience we will write $V_{1}=V_{1,23}$, $V_{2}=V_{2,31}$, and $V_{3}=V_{3,12}$. Parity-even three-point functions then take the form:

\begin{align}
&\<\mathcal{O}_{\Delta_{1},J_{1}}\mathcal{O}_{\Delta_{2},J_{2}}\mathcal{O}_{\Delta_{3},J_{3}}\>=\frac{V_{1}^{m_{1}}V_{2}^{m_{2}}V_{3}^{m_{3}}H_{12}^{n_{12}}
H_{13}^{n_{13}}H_{23}^{n_{23}}}{P_{12}^{h_{123}}P_{13}^{h_{132}}P_{23}^{h_{231}}}~,
\\
&h_{ijk}\equiv\frac{1}{2}\big(\Delta_{i}+J_{i}+\Delta_{j}+J_{j}-\Delta_{k}-J_{k}\big).
\end{align}
The $m_i$ and $n_{ij}$ run over all possible values consistent with 
\e{}{m_i \equiv J_i-n_{ij} -n_{ik}\geq 0~, ~~\text{where  } j,k\neq i~, ~~ (n_{ij}\equiv n_{ji})}
for each $i=1,2,3$. 
To generate parity-odd structures we will use the $d+2$ dimensional Levi-Civita symbol $\epsilon$ in embedding space. Parity-odd three-point function structures in $d=4$ can then be found by multiplying parity-even structures by
\bea
\epsilon^{(4d)}=\frac{1}{\sqrt{P_{12}P_{13}P_{23}}}\epsilon(Z_{1},Z_{2},Z_{3},P_{1},P_{2},P_{3}).
\eea
In $d=3$ the basic parity-odd structure we will need is
\bea
\epsilon_{ij}=\frac{1}{\sqrt{P_{12}P_{13}P_{23}}}\epsilon(Z_{i},Z_{j},P_{1},P_{2},P_{3}).
\eea
There are three such structures, but only two are linearly independent.

\ssec*{Differential basis}
There are four basic parity-even differential operators, given by
\begin{align}
D_{11}\equiv& \left[(P_{1}\cdot P_{2})(Z_{1}\cdot\frac{\partial}{\partial P_{2}})-(Z_{1}\cdot P_{2})(P_{1}\cdot\frac{\partial}{\partial P_{2}})-(Z_{1}\cdot Z_{2})(P_{1}\cdot\frac{\partial}{\partial Z_{2}})+(P_{1}\cdot Z_{2})(Z_{1}\cdot\frac{\partial}{\partial Z_{2}})\right]\Sigma^{1,0}, \nonumber
\\ \nonumber
\\
D_{12}\equiv& \left[(P_{1}\cdot P_{2})(Z_{1}\cdot\frac{\partial}{\partial P_{1}})-(Z_{1}\cdot P_{2})(P_{1}\cdot\frac{\partial}{\partial P_{1}})+(Z_{1}\cdot P_{2})(Z_{1}\cdot\frac{\partial}{\partial Z_{1}})\right]\Sigma^{0,1},
\\\nonumber
\\ D_{22} \equiv&D_{11}|_{1\leftrightarrow 2}~, ~~ D_{21} \equiv D_{12}|_{1\leftrightarrow 2}\nonumber
\end{align}
where $\Sigma^{a,b}$ implements the shift $(\Delta_1,\Delta_2)\rightarrow (\Delta_1+a,\Delta_2+b)$. The basis is then given by
\e{}{H_{12}^{n_{12}}D_{12}^{n_{13}}D_{21}^{n_{23}}D_{11}^{m_1}D_{22}^{m_2}}
acting on a basic (scalar)-(scalar)-(spin-$J_3$) structure \cite{Costa:2011dw, Costa:2011mg}. 

In $d=4$ there is a unique parity-odd differential operator given by:
\bea
\widetilde{D}^{(4d)}=\epsilon\left(Z_1,Z_2,P_1,P_2,\frac{\partial}{\partial P_1},\frac{\partial}{\partial_{P_2}}\right).
\eea
When constructing parity-odd three-point functions in $d=3$ we will use:
\bea
\widetilde{D}_{1}=\epsilon\left(Z_{1},P_{1},\frac{\partial}{\partial P_{1}},P_{2},\frac{\partial}{\partial P_{2}}\right),
\\
\widetilde{D}_{2}=\epsilon\left(Z_{2},P_{2},\frac{\partial}{\partial P_{2}},P_{1},\frac{\partial}{\partial P_{1}}\right).
\eea
There are other parity-odd differential operators one can write down in $d=3$, e.g.
\bea
\widetilde{D}_{3}=\left(Z_1,Z_2,P_1,P_2,\frac{\partial}{\partial P_{1}}\right)
\\
\widetilde{D}_{4}=\left(Z_1,Z_2,P_1,P_2,\frac{\partial}{\partial P_{2}}\right)
\eea
but these operators will not generate new, linearly independent tensor structures for the cases we consider. We will therefore restrict to using $\widetilde{D}_{1}$ and $\widetilde{D}_{2}$ without loss of generality.

%%%%%%%%%%%%%%%%%%%%%%%%%%%%%%%%%%%%%%%%%%%%%%%%%%%%%%%%

%%%%%%%%%%%%%%%%%%%%%%%%%%%%%%%
\section{Change of Bases: Mixed Systems}
\label{app:ChangeBases}
%%%%%%%%%%%%%%%%%%%%%%%%%%%%%%%

%%%%%%%%%%%%%%%%%%%%%%%%%%%%%%%
\subsection{Definitions for Regge Limit}
\label{app:Definitions}
%%%%%%%%%%%%%%%%%%%%%%%%%%%%%%%

In this section we will give the details for how to go from the standard, embedding space basis to the impact parameter space basis in the Regge limit. A more detailed discussion can be found in \cite{Costa:2017twz} and \cite{Cornalba:2009ax}. For convenience we will need to define the following functions, which appear when studying the Regge limit for $\<\O_{1}\O_{2}\O_{3}\O_4\>$:
\begin{align}
X(\nu) =& \left[\left(i\cot\left({\pi j(\nu)\o 2}\right)-1\right)\pi^{h+1}4^{j(\nu)}{(-ij'(\nu))\o 4i\nu}\right]\\
\nonumber
\\\gamma(\nu)=&\Gamma\left(\frac{\Delta_1+\Delta_2+j(\nu)+i\nu-\frac{d}{2}}{2}\right)\Gamma\left(\frac{\Delta_3+\Delta_4+j(\nu)+i\nu-\frac{d}{2}}{2}\right),
\\
\nonumber
\\
\chi(\nu) =& \gamma(\nu) \gamma(-\nu)  \,C_{{\cal O}_3{\cal O}_4j(\nu)}  K_{\frac{d}{2}+ i\nu,j(\nu)} X(\nu),\label{chiform}\\
\nonumber
\\
\hat \chi(\nu) =& \chi(\nu)/X(\nu),\label{chihat}\\
\nonumber
\\
K_{\D,J}=\ & \frac{\Gamma(\Delta+J) \,\Gamma(\Delta-\frac{d}{2}+1) \, (\Delta-1)_J   }{ 
4^{J-1} 
\Gamma\!\left( \frac{\Delta +J+\Delta_{1 2}}{2}\right)
\Gamma\!\left( \frac{\Delta +J-\Delta_{1 2}}{2}\right) 
\Gamma\!\left( \frac{\Delta +J+\Delta_{3 4}}{2}\right)
\Gamma\!\left( \frac{\Delta +J-\Delta_{3 4}}{2}\right)}
\label{KDeltaJ}\\&
\frac{1}{
 \Gamma\!\left( \frac{\Delta_1 +\Delta_{2} -\D+J}{2}\right)
\Gamma\!\left( \frac{\Delta_3 +\Delta_{4} -\D+J}{2}\right)
\Gamma\!\left( \frac{\Delta_1 +\Delta_{2} +\D+J-d}{2}\right)
\Gamma\!\left( \frac{\Delta_3 +\Delta_{4} +\D+J-d}{2}\right)},
 \nonumber
\end{align}
with $\Delta_{ij}=\Delta_i-\Delta_j$. We always choose $\O_3$ and $\O_4$ to be scalars, so we include their OPE coefficients in the definition of $\chi$. The gamma functions in the denominator of the first line of $K_{\Delta,J}$ will be (partially) responsible for the fact that the bounds on the physical OPE coefficients disappear for special values of the external or internal operator dimensions. Recall that $-ij'(\nu)$ is positive and bounded for negative imaginary $\nu$,
\e{}{0\leq -ij'(\nu)\leq 1}
This implies $\Re(X(\nu))<0$. Another function that will appear when doing Fourier transforms is:
\es{zetaform}{\zeta(\nu,n)
= \Bigg[{\pi^{d-2} \o4  }
&\Gamma\!\left( \frac{\Delta_1+\Delta_2+j(\nu) -\frac{d}{2} +i\nu}{2} +n \right)\Gamma\!\left( \frac{\Delta_1+\Delta_2+j(\nu)-\frac{d}{2}-i\nu}{2} +n \right)\\
\times~&
\Gamma\!\left( \frac{\Delta_3+\Delta_4+j(\nu)-\frac{d}{2}+i\nu}{2} \right)\Gamma\!\left( \frac{\Delta_3+\Delta_4+j(\nu)-\frac{d}{2}-i\nu}{2} \right)\Bigg]^{-1}}
Note that $\g(\nu)\g(-\nu)\zeta(\nu,n)$ is independent of $\D_{3},\D_{4}$, and that $\g(\nu)\g(-\nu)\zeta(\nu,0)={4\pi^{2-d}}$.

Finally, we will implement the action of the covariant derivatives as:
\begin{align}\label{b8}
\nabla^{\mu} F^{\mu_{1}...\mu{_\ell}}(\hat{x})&=P^{\mu}_{\nu}P^{\mu_1}_{\nu_1}...P^{\mu_\ell}_{\nu_\ell}\frac{\partial}{\partial \hat{x}_{\nu}}F^{\nu_{1}...\nu_{\ell}}(\hat{x})
\\
P^{\mu}_{\nu}&=(\delta^{\mu}_{\nu}+\hat{x}^{\mu}\hat{x}_{\nu})
\end{align}
where $P^{\mu}_{\nu}$ implements the projection onto $\mathbb{H}^{d-1}$.

In the remaining sections we will adopt the following conventions for labelling OPE coefficients in the various bases:
\es{}{c~\text{or}~C:&\quad \text{standard embedding space basis}\\
d:&\quad \text{differential embedding space basis}\\
B:&\quad \text{Regge differential impact parameter space basis}}
The Regge differential impact parameter space differential operators, which we denote $\hat D_i$, are of the same form as the position space operators, which we denote $D_i$, except with $\hat{x}\rightarrow \hat{p}$. In the $c$ and $d$ bases, $c_i$ or $d_i$ will stand for the coefficients of the $i$'th element. We also use $\a$ and $\beta$ as bases for the phase shift operator for a given four-point function -- see \eqr{alphadef} and \eqr{betakdef}, respectively, for their relation to the $d$ and $B$ coefficients. $\b$ is defined in the same way as in \cite{Costa:2017twz} to facilitate simple comparison.

In the following we will also suppress the dependence of $j$ on $\nu$, the dependence of $\chi$ and $\zeta$ on the external scaling dimensions, and will use $\Delta$ instead of $\frac{d}{2}+i\nu$ to label the point on the exchanged Regge trajectory. We do this to keep the expressions below as compact as possible.
%%%%%%%%%%%%%%%%%%%%%%%%%%%%%%%
\subsection{$\<TT\O\>$}
%%%%%%%%%%%%%%%%%%%%%%%%
\sssec{Parity-even}
\label{app:TTO_Even}
%%%%%%%%%%%%%%%%%%%%%%%%%%%%%%%
When $\O$ is a scalar, the standard basis for $\<T\O\mathcal{O}_{\Delta,J}\>$ is:
\bea
\{V_{1}^{2}V_{3}^{J},V_{1}H_{13}V_{3}^{J-1},H_{13}^{2}V_{3}^{J-2}\}.
\eea
Conservation for the stress tensor yields:
\begin{align}
c_{1}&= \frac{(d-2) c_{3} (d+\Delta_{\O}-\Delta+J-2) (d+\Delta_{\O}-\Delta+J)}{(d-1) (\Delta_{\O}-\Delta)^2-(d-2) J-J^2},
\\
c_{2}&= \frac{2c_{3} ((d-1) (\Delta_{\O}-\Delta)-J) (d+\Delta_{\O}-\Delta+J-2)}{(d-1) (\Delta_{\O}-\Delta)^2-(d-2) J-J^2}. 
\end{align}
The differential, embedding space basis is given by:
\bea
\{D_{12}^{2},D_{11}D_{12},D_{11}^{2}\}.
\eea
The conversion between the two bases is:
\begin{align}
d_{1}&=\frac{(J-1) (c_{1} J-c_{2} (2a+\Delta+J))+c_{3} (2a+\Delta+J-2) (2a+\Delta+J)}{(\Delta-1) \Delta (J-1) J}, \label{eqn:TTO_dToc_1}
\\
d_{2}&= \frac{2 c_{3} (2a-\Delta+J) (2a+\Delta+J-2)+2 (J-1) (c_{1} J-c_{2} (2a+J))}{(\Delta-1) \Delta (J-1) J},
\\
d_{3}&= \frac{(J-1) (c_{2} (\Delta-J-2a)+c_{1} J)+c_{3} (\Delta-J-2a) (\Delta-J+2-2a)}{(\Delta-1) \Delta (J-1) J}. \label{eqn:TTO_dToc_3}
\end{align}
where %
\e{}{a\equiv\frac{1}{2}(\Delta_{\O}-d)~.}
Next we consider the action of these differential operators in the Regge limit. It is convenient to define the differential operators:
\begin{align}
{D}_{1}&=(z_{1}\cdot\hat{x})^{2},  \label{eqn:ReggeOpTO1}
\\
{D}_{2}&=(z_{1}\cdot \hat{x})(z_{1}\cdot\nabla),
\\
{D}_{3}&=(z_{1}\cdot\nabla)^{2}.  \label{eqn:ReggeOpTO3}
\end{align}
In order to find the relation between this basis and the position-space, Regge differential basis, we need to compare the action of both sets of operators when acting on a conformal partial wave in the Regge limit. For the details of this procedure, see appendix C.1 of \cite{Costa:2017twz}. Here we will present the results:

\begin{align}
\frac{\alpha_{1}}{\chi(\nu)}&= \frac{\Gamma \left(\frac{1}{2} (-2 a+\Delta +J)\right) \Gamma \left(\frac{1}{2} (2 a+\Delta +J)\right)}{8 \Gamma \left(\frac{1}{2} (-2 a+\Delta +J+2)\right) \Gamma \left(\frac{1}{2} (2 a+\Delta +J+2)\right)}  \nonumber
\\
&\bigg(d_{1} (2 a+J-1) (2 a+J+1) (2 a-\Delta -J) (2 a-\Delta -J+2)\nonumber
\\
&+d_{2} (2 a-J+1) (2 a+J-1)(2 a+\Delta +J) (-2 a+\Delta +J)
\\
&+d_{3} \left((J-2 a)^2-1\right) (2 a+\Delta +J-2) (2 a+\Delta +J)
\bigg), \nonumber
\\ \nonumber \\
\frac{\alpha_{2}}{\chi(\nu)}&= \frac{\Gamma \left(\frac{1}{2} (-2 a+\Delta +J)\right) \Gamma \left(\frac{1}{2} (2 a+\Delta +J)\right)}{2}  \bigg(\frac{a d_{2} (2 a+\Delta +J)-\frac{1}{2} d_{1} (2 a+J) (2 a-\Delta -J+2)}{\Gamma \left(\frac{1}{2} (-2 a+\Delta +J)\right) \Gamma \left(\frac{1}{2} (2 a+\Delta +J+2)\right)}
\nonumber \\
&+\frac{d_{3} (2 a-J)}{\Gamma \left(\frac{1}{2} (-2 a+\Delta +J+2)\right) \Gamma \left(\frac{1}{2} (2 a+\Delta +J-2)\right)}\bigg),
\\ \nonumber \\
\frac{\alpha_{3}}{\chi(\nu)}&={\Gamma \left(\frac{1}{2} (-2 a+\Delta +J)\right) \Gamma \left(\frac{1}{2} (2 a+\Delta +J)\right) \o2}\bigg(\frac{d_{1} (-2 a+\Delta +J-2)+d_{2} (2 a+\Delta +J)}{2 \Gamma \left(\frac{1}{2} (-2 a+\Delta +J)\right) \Gamma \left(\frac{1}{2} (2 a+\Delta +J+2)\right)}
\nonumber \\
&+\frac{d_{3}}{\Gamma \left(\frac{1}{2} (-2 a+\Delta +J+2)\right) \Gamma \left(\frac{1}{2} (2 a+\Delta +J-2)\right)}\bigg).
\end{align}

Finally we need to perform the Fourier transforms and consider the conformal partial wave in the impact parameter space representation. We will use the same set of differential operators (\ref{eqn:ReggeOpTO1})-(\ref{eqn:ReggeOpTO3}) in impact parameter space. We then find the following relation between the $\beta$ and $\alpha$ basis:
\begin{small}
\begin{align}
\frac{\beta _1}{\zeta(\nu,1)}&= \frac{1}{64} \bigg(16 (\Delta -1) (d-\Delta -1) \big(\alpha _3 \left(-d \Delta+\Delta ^2-2 J-2 \Delta_{\O}-d+1\right)
\nonumber \\ & \hspace{1cm} +2 \alpha _2 \left(-J-\Delta_{\O}+1\right)\big)-16 \alpha _1 \left(-J-\Delta_{\O}-1\right) \big(-J-\Delta_{\O}+1\big)\bigg),
\\
\frac{\beta _2}{\zeta(\nu,1)}&= \frac{1}{4} \bigg(\alpha _3 \left(-d (\Delta  (J-2)+1)+\left(-d \Delta +\Delta ^2+1\right) \Delta_{\O}+\Delta  (\Delta -d)d+\Delta ^2 (J-2)+J+d\right)
\nonumber \\ & \hspace{1cm} +\alpha _2 \left(d \Delta +\Delta_{\O} \left(-2 J-d+2\right)+J (-d-J+2)-\Delta ^2-\Delta_{\O}^2\right)+\alpha _1 \left(J+\Delta_{\O}\right)\bigg),
\\
\frac{\beta _3}{\zeta(\nu,1)}&= \frac{1}{4} \left(-\alpha _1-\alpha _3 \left(J+\Delta_{\O}+d-1\right){}^2+2 \alpha _2 \left(J+\Delta_{\O}+d-1\right)\right).
\end{align}
\end{small}

After imposing conservation for the stress tensor we find 
\begin{align}
\b_2&=0\\
\beta_{3}&=\frac{\beta _1 (d-1)}{(\Delta -1) (d-\Delta -1)}
\end{align}
and
\e{}{\frac{\beta_1}{\chi(\nu)\zeta(\nu,1)}= \frac{c_1 (d-\Delta -1) \left(\Delta +J+\Delta_{\O}-2\right) \left(\Delta +J+\Delta_{\O}\right) \left(d+\Delta +J-\Delta_{\O}-2\right)}{2 (d-2) \Delta  \left(d-\Delta -J-\Delta_{\O}\right)}~.}
%\end{footnotesize}

The term $\left(d+\Delta +J-\Delta_{\O}-2\right) $ in the numerator leads to one of the double-trace zeroes. The remaining zeroes are due to $\chi(\nu)$. A consistency check on these calculations is that this conservation condition agrees with the conservation condition in the embedding space basis.
%%%%%%%%%%%%%%%%%%%%%%%%%%%%%%%
\sssec{Parity-odd}
\label{app:TTO_Odd}
%%%%%%%%%%%%%%%%%%%%%%%%%%%%%%%
For $\<T\O\mathcal{O}_{\Delta,J}\>$ in $d=3$ there are two parity-odd structures:
\bea
\epsilon_{13}\{V_{1}V_{3}^{J-1},H_{13}V_{3}^{J-2}\}.
\eea
Conservation of the stress-tensor implies: 
\bea
c_1= \frac{c_2 \left(\Delta_{\O}-\Delta+J+2\right)}{\Delta_{\O}-\Delta}. \label{eqn:appTTOoddCons}
\eea
Our basis for the differential operators is:
\bea
\{D_{11}\widetilde{D}_{1},D_{12}\widetilde{D}_{1}\}.
\eea
The change of basis between the two is:
\begin{align}
d_1&= \frac{c_2 \left(-3+\Delta_{\O}-\Delta+J-1\right)-c_1 (J-1)}{2 \left(\Delta-1\right) \Delta (J-1) J},
\\
d_2&= \frac{c_2 \left(-3+\Delta_{\O}+\Delta+J-1\right)-c_1 (J-1)}{2 \left(\Delta-1\right) \Delta (J-1) J}.
\end{align}

Similarly, we find two Regge differential operators:
\begin{align}
{D}_{1}&= \epsilon(z_{1},\hat{x},\nabla) z_{1}\cdot \hat{x},
\\
{D}_{2}&=\epsilon(z_{1},\hat{x},\nabla)z_{1}\cdot\nabla.
\end{align}

The $\alpha$ to embedding differential change of basis is given by:
\begin{small}
\begin{align}
&\frac{\alpha _1}{\chi(\nu)}=  \frac{J\Gamma \left(\frac{ -2 a+J+\Delta }{2}\right) \Gamma \left(\frac{2 a+J+\Delta }{2}\right)}{2 \Gamma \left(\frac{-2 a+J+\Delta +1}{2}\right) \Gamma \left(\frac{2 a+J+\Delta +1}{2} \right)} \bigg(d_1 (-2 a+J-1) (2 a+\Delta +J-1) 
\nonumber \\ & \hspace{6.4cm} -d_2 (2 a+J-1) (-2 a+\Delta +J-1)\bigg),
\\ \nonumber\\
&\frac{\alpha _2}{\chi(\nu)}= -\frac{J \left(d_1 (2 a+\Delta +J-1)+d_2 (-2 a+\Delta +J-1)\right) \Gamma \left(\frac{1}{2} (-2 a+J+\Delta )\right) \Gamma \left(\frac{1}{2} (2 a+J+\Delta )\right)}{2 \Gamma \left(\frac{ -2 a+J+\Delta +1}{2}\right) \Gamma \left(\frac{2 a+J+\Delta +1}{2}\right)},
\end{align}
\end{small}

where $a=-\frac{1}{2}(\Delta_{T}-\Delta_{\O})=\frac{1}{2}(\Delta_{\O}-3)$.
\\ \\
Finally, after performing the Fourier transform, we find $\beta$ is related to $\alpha$ by:
\begin{small}
\begin{align}
\frac{\beta _1}{\zeta\left(\nu,\frac{3}{2}\right)}&= -\frac{1}{8} \left(-\Delta +J+\Delta_{\O}+4\right) \left(\Delta +J+\Delta_{\O}+1\right) \left(\alpha _1 \left(J+\Delta_{\O}-1\right)-\alpha _2 (\Delta -3) \Delta \right),
\\
\frac{\beta _2}{\zeta\left(\nu,\frac{3}{2}\right)}&= \frac{1}{8} \left(-\Delta +J+\Delta_{\O}+4\right) \left(\Delta +J+\Delta_{\O}+1\right) \left(\alpha _1-\alpha _2 \left(J+\Delta_{\O}+2\right)\right).
\end{align}
\end{small}
After imposing conservation, we find $\beta_{1}=0$ and 
\begin{small}
\es{}{\frac{\beta_{2}}{\zeta\left(\nu,\frac{3}{2}\right)\chi(\nu)}&=\frac{c_2 \left(-\Delta +J+\Delta_{\O}+2\right) \left(-\Delta +J+\Delta_{\O}+4\right) \left(\Delta +J+\Delta_{\O}-1\right) \left(\Delta +J+\Delta_{\O}+1\right)}{8 (\Delta -1) \Delta  \left(\Delta -\Delta_{\O}\right) \Gamma \left(\frac{J+\Delta -\Delta_{\O}+2}{2}\right) \Gamma \left(\frac{J+\Delta +\Delta_{\O}-2}{2}\right)}\\&\times\Gamma \left(\frac{ J+\Delta -\Delta_{\O}+3}{2}\right) \Gamma \left(\frac{J+\Delta +\Delta_{\O}-3}{2} \right).}
\end{small}

Note the pole at $\Delta=\Delta_{\O}$ is not physical: we can see from (\ref{eqn:appTTOoddCons}) that at this point $c_2=0$ and we should use $c_1$ instead.
%%%%%%%%%%%%%%%%%%%%%%%%%%%%%%%
\subsection{$\<TTV\>$}
\label{app:TTV}
%%%%%%%%%%%%%%%%%%%%%%%%%%%%%%%
The parity-odd, embedding space structures for $\<TV\mathcal{O}_{\Delta,J}\>$ in $d=4$ are given by:
\bea
\epsilon^{(4d)}\{V_{1}V_{3}^{J-1},H_{13}V_{3}^{J-2}\}.
\eea 
Conservation of the stress tensor implies:
\bea
c_1= c_2 \left(\frac{J+3}{\Delta _V-\Delta }+1\right).
\eea
The differential basis is generated by:
\bea
\{D_{12}\widetilde{D}^{4d},D_{11}\widetilde{D}^{4d}\}.
\eea
The conversion between the embedding space bases is:
\begin{align}
d_1&= \frac{c_2 \left(\Delta +J+2a-1\right)+c_1(1-J)}{2 (\Delta -1) \Delta  (J-1) J},
\\
d_2&= -\frac{c_2 \left(\Delta -J-2a+1\right)+c_1 (J-1)}{2 (\Delta -1) \Delta  (J-1) J},
\end{align}
where $a=\frac{1}{2}\left(\Delta_V-4\right)$.

We also find there are two Regge differential operators:
\begin{align}
{D}_{1}&= \epsilon(z_1,z_2,\hat{x},\nabla) z_{1}\cdot \hat{x},
\\
{D}_{2}&=\epsilon(z_1,z_2,\hat{x},\nabla) z_1\cdot\nabla.
\end{align}

The relation between the two position space, differential bases is:
\begin{footnotesize}
\begin{align}
\frac{\alpha _1}{\chi(\nu)}=& \frac{J \left(d_1 (2 a+J-1) (-2 a+\Delta +J-1)+d_2 (2 a-J+1) (2 a+\Delta +J-1)\right) \Gamma \left(\frac{J+\Delta-2a}{2}\right) \Gamma \left(\frac{J+\Delta+2a}{2}\right)}{2 \Gamma \left(\frac{J+\Delta +1-2a}{2}\right) \Gamma \left(\frac{J+\Delta +1+2a}{2}\right)},
\\
\frac{\alpha _2}{\chi(\nu)}=& \frac{J \left(d_1 (-2 a+\Delta +J-1)+d_2 (2 a+\Delta +J-1)\right) \Gamma \left(\frac{J+\Delta-2a}{2}\right) \Gamma \left(\frac{J+\Delta+2a}{2}\right)}{2 \Gamma \left(\frac{J+\Delta +1-2a}{2}\right) \Gamma \left(\frac{J+\Delta +1+2a}{2}\right)}.
\end{align}
\end{footnotesize}

Once again, we perform a Fourier transform and find the $\beta$ to $\alpha$ change of basis is:
\begin{small}
\begin{align}
&\hspace{-.2cm}\frac{\beta _1}{\zeta\left(\nu,\frac{3}{2}\right)}= -\frac{1}{8} \left(-\Delta +J+5+\Delta _V\right) \left(\Delta +J+\Delta _V+1\right) \left(\alpha_2 \Delta  (4-\Delta )+\alpha_1 \left(J+\Delta _V-1\right)\right),
\\
&\hspace{-.2cm}\frac{\beta _2}{\zeta\left(\nu,\frac{3}{2}\right)}= -\frac{1}{8} \left(-\Delta +J+5+\Delta _V\right) \left(\Delta +J+\Delta _V+1\right) \left(\alpha_1-\alpha_2 \left(J+3+\Delta _V\right)\right).
\end{align}
\end{small}

After imposing conservation for the stress tensor at position one, we find $\beta_{1}=0$ and:
\begin{small}
\es{}{
&\hspace{-.3cm} \frac{\beta_2}{\zeta\left(\nu,\frac{3}{2}\right)\chi(\nu)}=\frac{c_2 \left(-\Delta +J+\Delta _V+3\right) \left(-\Delta +J+\Delta _V+5\right) }{8 (\Delta -1) \Delta  \left(\Delta -\Delta _V\right) \Gamma \left(\frac{J+\Delta -\Delta _V+3}{2}\right) \Gamma \left(\frac{J+\Delta +\Delta _V-3}{2}\right)}
\\ & 
\times\bigg(\left(\Delta +J+\Delta _V-1\right) \left(\Delta +J+\Delta _V+1\right) \Gamma \left({J+\Delta -\Delta _V+4\o2}\right) \Gamma \left({J+\Delta +\Delta _V-4\o2}\right)\bigg).}
\end{small}
%%%%%%%%%%%%%%%%%%%%%%%%%%%%%%%
\subsection{$\<TV\O\>$}
\label{app:TVO}
%%%%%%%%%%%%%%%%%%%%%%%%%%%%%%%

%%%%%%%%%%%%%%%%%%%%%%%%%%%%%%%
\subsubsection{Parity-even}
\label{app:TVO_Even}
%%%%%%%%%%%%%%%%%%%%%%%%%%%%%%%
We start with $\<V\O\mathcal{O}_{\Delta,J}\>$. The parity-even embedding space structures are: 
\bea
\{V_{1}V_{3}^{J},H_{13}V_{3}^{J-1}\}.
\eea
When $\mathcal{O}_{\Delta,J}=T$, conservation of the stress tensor implies
\bea
c_1=\frac{1}{2} c_2 \left(-d \Delta _V+d \Delta_{\O}+2\right),
\\
\Delta_{\O}=\Delta_{V}\pm 1.
\eea
If $V$ is a conserved operator only $\Delta_{\O}=\Delta_{V}- 1$ is allowed. The embedding space, differential basis is generated by: 
\bea
\{D_{12},D_{11}\}.
\eea
The conversion between the differential and standard basis is:
\begin{align}
d_1&=\frac{c_1 J-c_2 \left(-\Delta _V+\Delta_{\O}+\Delta+J-1\right)}{\left(\Delta-1\right) J},
\\
d_2&= \frac{c_2 \left(\Delta _V-\Delta_{\O}+\Delta-J-1\right)+c_1 J}{\left(\Delta-1\right) J}.
\end{align}

The two Regge differential operators are:
\bea
{D}_{1}=z_{1}\cdot\hat{x},
\\
{D}_{2}=z_{1}\cdot\nabla.
\eea
After taking the Regge limit we find:
\begin{footnotesize}
\begin{align}
&\frac{\alpha _1}{\chi(\nu)}=\frac{\Gamma \left(\frac{J+\Delta +\Delta_{\O}-\Delta _V}{2}\right) \Gamma \left(\frac{J+\Delta -\Delta_{\O}+\Delta _V}{2}\right)}{4 \Gamma \left(\frac{J+\Delta +\Delta_{\O}-\Delta _V+1}{2}\right) \Gamma \left(\frac{J+\Delta -\Delta_{\O}+\Delta _V+1}{2}\right)}
\\&~~~~~~~\times \bigg(d_1 \left(J+\Delta_{\O}-\Delta _V\right) \left(\Delta +J-\Delta_{\O}+\Delta _V-1\right)
-d_2 \left(\Delta +J+\Delta_{\O}-\Delta _V-1\right) \left(J-\Delta_{\O}+\Delta _V\right)\bigg),\nonumber
\\ \nonumber\\
&\frac{\alpha _2}{\chi(\nu)}= \frac{\left(d_2 \left(\Delta +J+\Delta_{\O}-\Delta _V-1\right)+d_1 \left(\Delta +J-\Delta_{\O}+\Delta _V-1\right)\right)}{4 \Gamma \left(\frac{J+\Delta +\Delta_{\O}-\Delta _V+1}{2}\right) \Gamma \left(\frac{J+\Delta -\Delta_{\O}+\Delta _V+1}{2}\right)}\nonumber\\&~~~~~~~\times \Gamma \left(\frac{J+\Delta +\Delta_{\O}-\Delta _V}{2}\right) \Gamma \left(\frac{J+\Delta -\Delta_{\O}+\Delta _V}{2}\right)
\end{align}
\end{footnotesize}
Doing the Fourier transform yields:
\begin{align}
\frac{\beta _1}{\zeta\left(\nu,\frac{1}{2}\right)}&= \frac{1}{2} \left(\alpha _2 (\Delta -1) (-d+\Delta +1)+\alpha _1 \left(d-J-\Delta_{\O}-\Delta _V\right)\right),
\\
\frac{\beta _2}{\zeta\left(\nu,\frac{1}{2}\right)}&= \frac{1}{2} \left(\alpha _1-\alpha _2 \left(J+\Delta_{\O}+\Delta _V-2\right)\right).
\end{align}

The full change of basis is given by:
\begin{small}
\begin{align}
&\frac{\beta_1}{\chi(\nu)\zeta\left(\nu,\frac{1}{2}\right)}=\frac{\Gamma \left(\frac{J+\Delta +\Delta_{\O}-\Delta _V}{2}\right) \Gamma \left(\frac{J+\Delta -\Delta_{\O}+\Delta _V}{2}\right) }{4 \Gamma \left(\frac{J+\Delta +\Delta_{\O}-\Delta _V+1}{2} \right) \Gamma \left(\frac{J+\Delta -\Delta_{\O}+\Delta _V+1}{2}\right)}\times \nonumber \\ & \hspace{2.5cm} \bigg( 
c_1 \big(\left(\Delta_{\O}-\Delta _V\right) \left(d-J-\Delta_{\O}-\Delta _V\right)-(d-\Delta -1) (\Delta +J-1)\big)
\nonumber \\ & \hspace{2.75cm} +c_2 \left(\Delta +J+\Delta_{\O}-\Delta _V-1\right) \left(-\Delta +J+\Delta_{\O}+\Delta _V-1\right)\bigg),
\\
&\frac{\beta_2}{\chi(\nu)\zeta\left(\nu,\frac{1}{2}\right)}=-\frac{ \Gamma \left(\frac{ J+\Delta +\Delta_{\O}-\Delta _V}{2}\right) \Gamma \left(\frac{J+\Delta -\Delta_{\O}+\Delta _V}{2}\right)}{4 (\Delta -1) \Gamma \left(\frac{J+\Delta +\Delta_{\O}-\Delta _V+1}{2}\right) \Gamma \left(\frac{J+\Delta -\Delta_{\O}+\Delta _V+1}{2}\right)}
\nonumber \\ &\hspace{2.5cm} \bigg(c_1 \big((J-2) (\Delta +J-1)+J \Delta_{\O}+(2 \Delta +J-2) \Delta _V\big)
\nonumber \\ & \hspace{2.5cm} -c_2 \left(\Delta +J+\Delta_{\O}-\Delta _V-1\right) \left(-\Delta +J+\Delta_{\O}+\Delta _V-1\right)\bigg).
\end{align}
\end{small}
%%%%%%%%%%%%%%%%%%%%%%%%%%%%%%%
\subsubsection{Parity-odd}
\label{app:TVO_Odd}
%%%%%%%%%%%%%%%%%%%%%%%%%%%%%%%
For $\<V\f \mathcal{O}_{\Delta,J}\>$ in $d=3$ there is a unique parity-odd structure given by:
\bea
\epsilon_{13}V_{3}^{J-1}.
\eea
with coefficient $\tilde c$. This is generated by the embedding space differential operator $\tilde{D}_{1}$. The unique Regge differential operator is given by:
\bea
\hat{\tilde D}=\epsilon(z_{1},p,\nabla).
\eea
Here the change of basis is trivial, since all coefficients must be proportional to each other:
\begin{align}
&\tilde{d}= -\frac{\tilde{c}}{2J (\Delta-1)},
\\
&\frac{\alpha}{\chi(\nu)} = -2 J  \tilde{d},
\\
&\frac{\beta}{\zeta(\nu,1)} = \frac{1}{16} \left((3-2 \Delta )^2-\left(3-2 \left(J+\Delta_{\O}+\Delta _V\right)\right){}^2\right),
\end{align}
so we have
\bea
\frac{\beta}{\chi(\nu)\zeta(\nu,1)}=\tilde{c}\,\frac{ J \left((3-2 \Delta )^2-\left(3-2 \left(J+\Delta_{\O}+\Delta _V\right)\right){}^2\right)}{32 (\Delta -1)}.
\eea
If the exchanged operator is the stress tensor, conservation implies $\Delta_{\O}=\Delta_{V}$.
%%%%%%%%%%%%%%%%%%%%%%%%%%%%%%%
\subsection{$\<TMT\>$}
\label{app:TTM}
%%%%%%%%%%%%%%%%%%%%%%%%%%%%%%
Here we give some details relevant for the collider calculation in Section \ref{ttmcoll}. 

We choose to define the parameters $B_{TMT}^{(1)}$ and $B_{MMT}^{(1)}$ as
\es{}{
\frac{B_{\phi\phi j(\nu)}B_{TMj(\nu)}^{(1)}}{\hat\chi^{(TM)}(\nu)\zeta^{(TM)}(\nu,4)}\bigg|_{\nu=-i\frac{d}{2}} &= 1\\
\frac{B_{\phi\phi j(\nu)}B_{MMj(\nu)}^{(1)}}{\hat\chi^{(MM)}(\nu)\zeta^{(MM)}(\nu,4)}\bigg|_{\nu=-i\frac{d}{2}}&=E^{(MM)}}
where the functions $\hat\chi$ and $\zeta$ were defined in \eqr{chihat} and \eqr{zetaform}, and the superscript denotes the external operator dimensions $\D_1$ and $\D_2$ in those formulas. We have included an overall constant $E^{(MM)}$, which fixes the norm of $M$ and is thus constrained to be positive for all unitary $\D_M\geq d$; we find it convenient to keep this factor explicit. $\beta^{(1)}_{TTT}$ can be found in \eqr{eqn:beta_TT_1}.

We also give the relation between the $t_{2,4}^{(TM)}$ structures and the standard OPE coefficients. With the overall normalization factor given above, the final result is
\begin{tiny}
\begin{align}\label{t2tm}
&t_{2}^{(TM)}= \frac{(\Delta_{M}+4) (d+\Delta_{M}+2) (d+\Delta_{M}+4)}{16 (\Delta_{M}+2) \left(d^3 (\Delta_{M} (\Delta_{M}+14)+16)-d^2 (\Delta_{M} (\Delta_{M} (\Delta_{M}+15)+42)+112)+3 d (\Delta_{M} (\Delta_{M} (\Delta_{M}+10)+44)+32)-2 \Delta_{M} (\Delta_{M}+4) (\Delta_{M}+8)\right)}
\nonumber \\ &  \big(-C^{(1,0,1)}_{TMT} (2 d-\Delta_{M}) \big(12 \left(-d^3+d+2\right) \Delta_{M}^2+(d-2) (d-1) \Delta_{M}^4+(d-1) (d (d+10)-8) \Delta_{M}^3-4 (d+2) (d (7 d-11)+8) \Delta_{M}
\nonumber \\ & +32 (d-1) d (d+2)\big)-8 C^{(0,0,0)}_{TMT} (2 d^4 (\Delta_{M} (\Delta_{M}+4)+2)-d^3 (\Delta_{M} (\Delta_{M} (2 \Delta_{M}+9)+24)+32)+d^2 (\Delta_{M} (\Delta_{M}+2) (3 \Delta_{M}+17)+4)
\nonumber \\ & -d (\Delta_{M} (\Delta_{M} (\Delta_{M}+2)-22)-24)-2 \Delta_{M} (7 \Delta_{M}+8))\big)
%%%%%%%%%%%%%%%%%%%%%%%%%%%%%%%%%%%%
\\ \nonumber\\
&t_{4}^{(TM)}= \frac{(d+1) (d+2) (\Delta_{M}+4) (d+\Delta_{M}+2) (d+\Delta_{M}+4)}{8 (\Delta_{M}+2) \left(d^3 (\Delta_{M} (\Delta_{M}+14)+16)-d^2 (\Delta_{M} (\Delta_{M} (\Delta_{M}+15)+42)+112)+3 d (\Delta_{M} (\Delta_{M} (\Delta_{M}+10)+44)+32)-2 \Delta_{M} (\Delta_{M}+4) (\Delta_{M}+8)\right)}
\nonumber \\ &
\big(2 C^{(0,0,0)}_{TMT} \left(d^3 (\Delta_{M} (\Delta_{M}+5)+2)-d^2 (\Delta_{M} (\Delta_{M} (\Delta_{M}+5)+18)+20)+d (\Delta_{M}+3) (\Delta_{M} (2 \Delta_{M}+9)+6)-\Delta_{M} (\Delta_{M} (\Delta_{M}+11)+12)\right)
\nonumber \\ &-C^{(1,0,1)}_{TMT} (2 d-\Delta_{M}) \left(d^2 (3 \Delta_{M} (\Delta_{M}+2)-8)-d (\Delta_{M}+4) (3 \Delta_{M}-2)-2 (\Delta_{M}-4) \Delta_{M}\right)\big)
\end{align}
\end{tiny}
As a reminder, we use the basis
\bea
\<TMT\>\propto \underset{n_1,n_2,n_3}{\sum}C^{(n_1,n_2,n_3)}_{TMT}V_1^{2-n_{2}-n_{3}}V_{2}^{2-n_{1}-n_{3}}V_{3}^{2-n_{1}-n_{2}} H_{23}^{n_1} H_{13}^{n_2}H_{12}^{n_{3}}.
\eea
One can verify with a magnifying glass that when $\Delta_{M}=2d$ the dependence on $C^{(1,0,1)}_{TMT}$ disappears in $t_{2,4}^{(TM)}$, as noted in the main text.

\section{Filling in the Gaps}
\label{app:Gaps}
%%%%%%%%%%%%%%%%%%%%%%%%%%%%%%%
In the following sections we will derive new bounds on three point functions in CFTs with $\Dg\gg1$. The bounds on $V$ and $M$ are necessary to prove our earlier holographic bounds on mixed systems involving those operators. The bounds on parity odd couplings involving $J$ and $T$ will be necessary to show the universality of $\<JJT\>$ and $\<TTT\>$ in $d=3$.
%%%%%%%%%%%%%%%%%%%%%%%%%%%%%%%
\subsection{$\<VVT\>$}
\label{app:VVT}
%%%%%%%%%%%%%%%%%%%%%%%%%%%%%%%

In this section we will both derive bounds for $\<VVT\>$ at large gap and define OPE coefficients which appeared in bounds for $\<VT\O\>$ and $\<TTV\>$. When we set $\Delta_{V}=d-1$ we will recover the results of \cite{Costa:2017twz,Cornalba:2009ax} for conserved currents. Here we will only consider parity-even structures.

We start with $\<VV\mathcal{O}_{\Delta,J}\>$. A basis for parity-even structures is:
\bea\label{vvostruc}
\{V_{1}V_{2}V_{3}^{J},H_{23}V_{1}V_{3}^{J-1}+H_{13}V_{2}V_{3}^{J-1},H_{12}V_{3}^{J},H_{13}H_{23}V_{3}^{J-2}\}.
\eea
If $\mathcal{O}_{\Delta,J}=T$, then conservation at $P_3$ implies:
\bea\label{c7}
c_1= \frac{1}{2} c_4 \left(d^2-4\right)+2 c_2.
\eea
Furthermore, the Ward identity for $T$ \cite{Komargodski:2016gci} implies
\bea
c_{3}= \frac{\Delta_{V}}{d-1}c_2+\frac{d^2-d-2 \Delta_{V}}{2 (d-1)}c_4
\eea
The basis for differential operators is given by:
\bea
\{D_{12}D_{21},D_{12}D_{22}+D_{11}D_{21},D_{11}D_{22},H_{12}\}.
\eea
The corresponding change of basis is:
\begin{align}
d_1&= \frac{(J-1) \left(2 c_2 \left(\Delta+J\right)-c_1 J\right)-c_4 \left(J^2+\Delta \left(\Delta+2 J-4\right)\right)}{\left(\Delta-1\right) \Delta (J-1) J},
\\ d_2&= \frac{c_4 \left(\Delta^2-J^2\right)-\left(c_1-2 c_2\right) (J-1) J}{\left(\Delta-1\right) \Delta (J-1) J},
\\ d_3&= -\frac{c_4 \left(J-\Delta\right){}^2+(J-1) \left(2 c_2 \left(\Delta-J\right)+c_1 J\right)}{\left(\Delta-1\right) \Delta (J-1) J},
\\ d_4&= \frac{c_3 \Delta+\frac{c_4 \left(\Delta-J\right)}{J-1}-c_1+2 c_2}{\Delta}.
\end{align}

A basis of Regge differential operators is:
\begin{align}
{D}_{1}&=(z_{1}\cdot \hat{x})( z_{2}\cdot \hat{x}), \\
{D}_{2}&=(z_{1}\cdot \hat{x})( z_{2}\cdot\nabla)+(z_{2}\cdot \hat{x})( z_{1}\cdot\nabla), \\
{D}_{3}&=(z_{1}\cdot\nabla )(z_{2}\cdot \nabla), \\
{D}_{4}&=z_{1}\cdot z_{2}.
\end{align}
The relation between the $\alpha_i$ and $d_i$ basis is:
\es{}{
\frac{\alpha _1}{\chi(\nu)}&= \frac{d_1 \left(J^2-1\right) (\Delta +J)+d_3 \left(J^2-1\right) (\Delta +J)-2 d_2 (J+1)^2 (\Delta +J-2)+8 d_4 (\Delta +J)}{4 (\Delta +J)}, \\
\frac{\alpha _2}{\chi(\nu)}&= \frac{1}{4} \left(d_1-d_3\right) J, \\
\frac{\alpha _3}{\chi(\nu)}&= \frac{\left(d_1+2 d_2+d_3\right) (\Delta +J)-4 d_2}{4 (\Delta +J)}, \\
\frac{\alpha _4}{\chi(\nu)}&= \frac{\Delta  \left(d_1 (J-1)-2 d_2 (J+1)+d_3 (J-1)+4 d_4\right)+\left(d_1-2 d_2+d_3\right) (J-1) J+4 d_4 J+4 d_2}{4 (\Delta +J)}~.\nonumber}
Performing the Fourier transform, we find:
\es{}{
\frac{\beta_1}{\zeta(\nu,1)}&= \frac{1}{4} \bigg((\Delta -1) (d-\Delta -1) \left(\alpha _3 \left(-d \Delta +d+\Delta ^2-2 J-4 \Delta _V+1\right)+2 \alpha _2 \left(d-J-2 \Delta _V+1\right)\right) \nonumber \\ & \hspace{.5cm} -\alpha _1 \left(d-J-2 \Delta _V-1\right) \left(d-J-2 \Delta _V+1\right)\bigg) ,
\\ \frac{\beta_2}{\zeta(\nu,1)}&= \frac{1}{4} \bigg(\alpha _3 \left(-d (\Delta  (J-2)+1)+2 \left(-d \Delta +\Delta ^2+1\right) \Delta _V+\Delta ^2 (J-2)+J\right)
\nonumber \\ &\hspace{.5cm} +\alpha _2 \left(d \Delta +2 \Delta _V \left(d-2 J-2 \Delta _V+2\right)+(J-2) (d-J)-\Delta ^2\right)+\alpha _1 \left(-d+J+2 \Delta _V\right)\bigg),
\\ \frac{\beta_3}{\zeta(\nu,1)}&= \frac{1}{4} \left(\left(J+2 \Delta _V-1\right) \left(2 \alpha _2-\alpha _3 \left(J+2 \Delta _V-1\right)\right)-\alpha _1\right),
\\ \frac{\beta_4}{\zeta(\nu,1)}&= \frac{1}{16} \bigg(8 \alpha _2 (\Delta -1) (d-\Delta -1)-4 \alpha _3 (\Delta -1) (d-\Delta -1) \left(J+2 \Delta _V-1\right) 
\nonumber \\ &\hspace{.5cm}+4 \alpha _1 \left(-d+J+2 \Delta _V+1\right)+4 \alpha _4 \left(-\Delta +J+2 \Delta _V\right) \left(d-\Delta -J-2 \Delta _V\right)\bigg).}
%\end{align}

In order to make our expressions more compact and in line with the conventions of \cite{Costa:2017twz}, we define:
\begin{align}
\beta^{(1)}_{VVj(\nu)}&=\beta_4-\frac{\beta_3 \left(\left(\frac{d}{2}-1\right)^2+\nu ^2\right)}{d-1}, \label{eqn:betaVVT_1} \\
\beta^{(2)}_{VVj(\nu)}&=\beta_{4}-\beta_{1}, \label{eqn:betaVVT_2} \\
\beta^{(3)}_{VVj(\nu)}&=\beta_2, \label{eqn:betaVVT_3} \\
\beta^{(4)}_{VVj(\nu)}&=-\beta_3. \label{eqn:betaVVT_4}
\end{align}

Using these variables, the full change of basis is given by:
\begin{small}
\es{}{
\frac{\beta^{(1)}_{VVj(\nu)}}{\zeta(\nu,1)\chi(\nu)}&= \frac{1}{4 (d-1)^2}\bigg(-2 c_2 \left(d^2+d (\Delta -3)+2 (\Delta -1) (d-\Delta -1) \Delta _V-\Delta ^2+2\right)
\nonumber \\ &
\hspace{-.5cm}+c_4 \big(\Delta _V \left(2 (d-1) \left(d^2-d (\Delta +1)+\Delta ^2\right) \Delta _V+2 (d-3) \Delta ^2-2 (d-3) d \Delta -d (d ((d-4) d+5)+2)+4\right)
\nonumber \\ &
\hspace{-.5cm}-(d-2) \left(-(d+1) \Delta ^2+d (d+1) \Delta -2 d+2\right)\big)+2 c_3 (d-1) \left(\Delta  (d-\Delta )+(d-1) \Delta _V \left(d-2 \Delta _V-2\right)\right)\bigg),
\\ 
\frac{\beta^{(2)}_{VVj(\nu)}}{\zeta(\nu,1)\chi(\nu)}&=\frac{1}{4} \left(d-2 \Delta _V-2\right) \left(\left(2 c_3-c_4 d\right) \left(d-\Delta _V-2\right)+2 c_2-2 c_4\right),
\\ 
\frac{\beta^{(3)}_{VVj(\nu)}}{\zeta(\nu,1)\chi(\nu)}&= \frac{\left(d-2 \Delta _V-2\right) \left((d-1) \left(c_4 d-2 c_3\right)+2 \left(c_2-c_4\right) \Delta _V\right)}{4 (d-1)},
\\ 
\frac{\beta^{(4)}_{VVj(\nu)}}{\zeta(\nu,1)\chi(\nu)}&= \frac{c_4 \left(-d^2-2 \Delta _V \left((d-1) \Delta _V+d-3\right)+d+2\right)+2 c_3 (d-1)-2 c_2 \left(2 \Delta _V+1\right)}{4 (d-1)}.}
%\end{align}
\end{small}

At this point, we note that if we restrict to purely transverse polarizations for the vectors $V$, then the final result for $\<VV\f\f\>$ is identical in form with the result for $\<JJ\f\f\>$. Assuming $z_{1,2}$ are transverse and denoting the full differential operator in impact parameter space as $\mathcal{D}_{VV}(\nu_0)$ we find \cite{Costa:2017twz}:
\bea
\mathcal{D}_{VV}(\nu_{0})\Pi_{i\nu_0}(L) = B_{\phi\phi j(\nu_0)}B^{(1)}_{VVj(\nu_0)}\Pi_{i\nu_0}(L)z_1\cdot z_2 \left(1+\frac{B^{(4)}_{VVj(\nu_0)}}{B^{(1)}_{VVj(\nu)}}a(\nu_0,L)\left(\frac{n\cdot z_1 n\cdot z_2}{z_1\cdot z_2}-\frac{1}{d-1}\right)\right) \label{eqn:VVphaseshift}
\nonumber \\
\eea
The function $a(\nu_0,L)$ is defined in (2.42) of \cite{Costa:2017twz}. When deriving bounds at $\Dg\gg1$, we need that in the limit $L\rightarrow 0$, $a(0,L)$ behaves like $L^{-2}$ for $d\geq4$ and $L^{-2}\log^{-1}(L)$ in $d=3$. Therefore for theories with a large gap we find $B^{(4)}_{VV}(\nu_0)=0$. In terms of the standard basis this yields:
\bea
c_2=((1-d) \Delta_V+1)c_4, \quad c_4\geq0
\eea
Therefore, we can fix $\<VVT\>$ up to a single, positive OPE coefficient in theories with a large gap.

When deriving conformal collider bounds we need, from \cite{Costa:2017twz},
\bea
\underset{L\rightarrow\infty}{\lim}\ a(\nu,L)=-\frac{1}{4}(d+2i\nu)(d+2i\nu-2)
\eea
Then the variables $a_{2,V}$ and $C_{VVT}$ which appear in collider bounds are defined as:
\begin{align}
a_{2,V}&=(1-d)d\frac{B^{(4)}_{VVT}}{B^{(1)}_{VVT}}, \label{a2def}
\\
C_{VVT}&=\frac{\beta^{(1)}_{VVj(\nu)}}{\zeta(\nu,1)\chi(\nu)}\bigg|_{\nu=-\frac{id}{2}}
\end{align}
After setting $\Delta_{V}=d-1$ we have $a_{2,V}\rightarrow a_{2}$ where $a_2$ was the variable first introduced for $\<JJT\>$ in \cite{Hofman:2008ar}.

In generic CFTs, the collider bound on $a_{2,V}$ is identical in form to the bound on $a_{2}$:
\bea
-\frac{d-1}{d-2}\leq a_{2,V}\leq d-1.
\eea
The full conformal collider bounds for non-conserved vectors can be found in \cite{Komargodski:2016gci}.
%%%%%%%%%%%%%%%%%%%%%%%%%%%%%%%
\subsection{$\<MMT\>$}
\label{app:MMT}
%%%%%%%%%%%%%%%%%%%%%%%%%%%%%%%

We now derive bounds on $\<MMT\>$ at $\Dg\gg1$. Here all we need is that the Regge differential operators in impact parameter space for $\<MM\f\f\>$ are of the form
\bea
\hat{D}_{n_1,n_2,n_{12}}=(z_1\cdot z_2)^{n_{12}}(z_1\cdot \hat{p})^{2-n_{12}-n_{1}}(z_2\cdot \hat{p})^{2-n_{12}-n_2}(z_1\cdot\nabla)^{n_1}(z_2\cdot\nabla)^{n_2}. 
\eea
Restricting to transverse polarizations for $M$ involves taking $z_1\cdot \hat p \rar 0$ and $z_2\cdot \hat p\rar0$, as explained above \eqr{nvec}, so only $\hat{D}_{0,0,2}$, $\hat{D}_{1,1,1}$, and $\hat{D}_{2,2,0}$ lead to a non-zero phase shift. Imposing the bounds from unitarity we find:
\bea
B^{(2,2,0)}_{MMj(0)}=0, \hspace{1cm} B^{(1,1,1)}_{MMj(0)}=0.
\eea
The structure $B^{(0,0,2)}_{MMj(0)}$ is still allowed. Therefore we have that at small $L$, assuming transverse polarizations, $\mathcal{D}(\nu_{0})\Pi_{i\nu_0}(L)$ scales like $\Pi_{i\nu_0}(L)$ to leading order. This had to be the case: spin-2 operators are allowed in $\Dg\gg1$ theories, and the Ward identity relates $\la TMM\ra$ to $\la MM\ra$. Combining this with the same fact about $\<TTT\>$ at small $L$ leads to bounds on the size of the coupling $\<TTM\>$, as stated in \eqr{553}.

We can also go further and derive bounds on polarization tensors with longitudinal components. If we consider the polarization tensor 
\bea
\epsilon_{\mu\nu}=\frac{1}{2}(\hat{p}_{\mu}e_{\nu}+\hat{p}_{\nu}e_{\mu})
\eea
where the vector $e_{\mu}$ is transverse -- that is $e\cdot \hat{p}=0$ -- then we find:
\bea
B^{(1,1,0)}_{MMj(0)}=0
\eea
This will be crucial later to derive bounds on the parity odd coupling $\<TTM\>$ in $d=3$. This is also consistent with our expectations that any differential operator in impact parameter space with derivatives is suppressed at the intercept.

%%%%%%%%%%%%%%%%%%%%%%%%%%%%%%%
\subsection{$\<JJT\>_{\rm odd}$}
\label{sec:JJT_Odd}
%%%%%%%%%%%%%%%%%%%%%%%%%%%%%%%
We now derive bounds on the parity-odd part of $\<JJT\>$ in $d=3$ at $\Dg\gg1$. Here we will consider $z_{1,\mu}z_{2,\nu}\<J^{\mu}J^{\nu}\phi\phi\>$ and generalize some results already found in \cite{Costa:2017twz}. This case is also simple enough that we can work directly in the $B$ basis. We find there are two parity-odd Regge differential operators:
\begin{align}
\hat{D}_{1}&=\epsilon(z_1,\hat{p},\nabla)z_2\cdot \hat{p} + \epsilon(z_2,\hat{p},\nabla) z_1\cdot \hat{p}, \\
\hat{D}_{2}&=\epsilon(z_1,\hat{p},\nabla)z_2\cdot\nabla+\epsilon(z_2,\hat{p},\nabla)z_1\cdot\nabla,
\end{align}
and conservation implies $B^{(1)}_{JJj(\nu)}=0$. To simplify the analysis, we choose 
\bea
z_{1}=z_{2}=\sin(\theta)n+\cos(\theta)n_{\perp} \text{   with   } n_{\perp}\cdot n_{\perp}=1  ~,~ ~  n_{\perp}\cdot n=0.
\eea
Then in the small $L$ limit we find:
\bea
\hat{D}_{2}\Pi_{i\nu_0}(L\ll1)\propto -\frac{4\sin(\theta)\cos(\theta)}{\pi L^{2}}.
\eea

The first we thing we should note is that this term vanishes if $\theta=0,\frac{\pi}{2}$ and therefore does not interfere with the derivation of bounds on the parity-even part of $\<JJT\>$. Furthermore, since the parity-even terms grow like $\log(L)$ at small $L$, this term is clearly dominant. Since it is not sign-definite, we must have
\bea
\widetilde{B}^{(2)}_{JJj(0)}=0 \quad \Rightarrow \quad \widetilde{C}_{JJj(0)}=0, \label{eqn:JJT-Holo-Odd}
\eea
where we used that in the standard embedding space basis there is only structure, and there are no extra zeroes when changing bases. In a theory with a large gap we then have $\widetilde{B}^{(2)}_{JJT}\sim\Delta_{gap}^{-2}$, equivalently, $\widetilde{C}_{JJT}\sim\Delta_{gap}^{-2}$. For conformal collider bounds on this OPE coefficient, see \cite{Chowdhury:2017vel}.
%%%%%%%%%%%%%%%%%%%%%%%%%%%%%%%
\subsection{$\<TTT\>_{\rm odd}$}
\label{sec:TTT_Odd}
%%%%%%%%%%%%%%%%%%%%%%%%%%%%%%%

We now derive similar bounds on the parity-odd part of $\<TTT\>$ in $d=3$. Bounds on the corresponding gravitational interaction were first found in \cite{Camanho:2014apa}. We will consider the four-point function  $z_{1,\mu}z_{1,\nu}z_{2,\rho}z_{2,\sigma}\<T^{\mu\nu}T^{\rho\sigma}\phi\phi\>$ and once again work entirely in the $B$ basis. The most general Regge differential operator in impact parameter space is:
\begin{align}
\widehat{{D}}=&\epsilon(z_1,\hat{p},\nabla)\bigg( \ \underset{n_{1},n_{2}}{\sum}c_{n_1,n_2}(z_1\cdot \hat{p})^{1-n_1}(z_2\cdot \hat{p})^{2-n_2}(z_1\cdot\nabla)^{n_1}(z_2\cdot\nabla)^{n_2} +(z_1\cdot z_2)(d_{1}z_2\cdot \hat{p} + d_2 z_2\cdot\nabla)\bigg) \nonumber \\ &\hspace{1in}+(z_1\leftrightarrow z_2), \label{eqn:TTT_odd_Diffs}
\end{align}
where $0\leq n_1\leq 1$ and $0\leq n_2\leq 2$. After imposing conservation there is only one solution parametrized by $c_{1,2}$. The $c_{1,2}$ term will lead to the most divergent contribution to the phase shift in the limit $L\rightarrow 0$, so we will define $\widetilde{B}_{TTj(\nu)}=c_{1,2}(\nu)$ where we made the $\nu$ dependence explicit. When we project down to $\mathbb{H}^{2}$ we will make the replacement \eqr{epsrep}. 
%\bea
%\epsilon^{\mu\nu}\rightarrow\frac{1}{2}\left(e_1^{\hat{\mu}} e_2^{\hat{\nu}}+e_2^{\hat{\mu}} e_1^{\hat{\nu}}\right)-\frac{g^{\hat{\mu}\hat{\nu}}e_1\cdot e_2}{d-1}.
%\eea
%
We will also define the angles $\theta_{i}$ by:
%\bea
\e{}{\eps_{i}\cdot n=\cos\theta_i}
%\eea
The final answer after projecting down to $\mathbb{H}^{2}$ and making these replacements is
\bea
\widehat{{D}} \Pi_{i\nu_0}(L\ll 1)\propto \frac{6}{L^{4}\pi}\sin(2(\theta_1+\theta_2)).
\eea
which is not sign-definite. Moreover, it grows faster than the parity-even terms and therefore 
\bea
\widetilde{B}_{TTj(0)}=0 \quad \Rightarrow \quad \widetilde{C}_{TTj(0)}=0, \label{eqn:TTT-Holo-Odd}
\eea
where we once again used that there is a unique structure in the embedding space basis and there are no subtleties when performing the change of bases. In a theory with a large gap this becomes $\widetilde{B}_{TTT}\sim \Delta_{gap}^{-4}$ or $\widetilde{C}_{TTT}\sim\Delta_{gap}^{-4}$, which matches \cite{Camanho:2014apa}. Conformal collider bounds on this OPE coefficient were also found in \cite{Chowdhury:2017vel}.

We can also note that since the general parity-odd differential operator (\ref{eqn:TTT_odd_Diffs}) necessarily has derivatives, we can derive similar bounds for the parity odd correlator $\<TMT\>$ in $d=3$. That is, since any parity-odd differential operator in impact parameter space for $\<TMT\>$ must contain derivatives and we have already shown such terms in $\<MMT\>$ are suppressed, we must have $\widetilde{B}_{TMj(0)}=0$ in a theory with a large gap.
\bibliography{Biblio}{}
\bibliographystyle{ssg}

\end{document}